%% file: ms_21aefx_jwst.tex
\documentclass{aa}  

\usepackage{graphicx}
\usepackage{xfrac}
\usepackage[varg]{txfonts} 
\usepackage[colorlinks=true,citecolor=blue,linkcolor=blue,urlcolor=blue]{hyperref}
\usepackage[utf8]{inputenc}
\usepackage{amsmath}	
\usepackage{amssymb}	
\usepackage{longtable}
\usepackage{pdflscape}

\usepackage{layouts}
\usepackage{booktabs} 
\usepackage{multicol}
\usepackage{ragged2e} 

\usepackage{cleveref}[2012/02/15]
\crefformat{footnote}{#2\footnotemark[#1]#3}

\newcommand{\snia}{SN~Ia}
\newcommand{\sneia}{SNe~Ia}

\usepackage{scalefnt}
\newcommand{\textmc}[1]{\textsc{\scalefont{1.2}#1}}

\newcommand{\one}{\,\,\textmc{i}}
\newcommand{\two}{\,\,\textmc{ii}}
\newcommand{\three}{\,\,\textmc{iii}}
\newcommand{\four}{\,\,\textmc{iv}}
\newcommand{\five}{\,\,\textmc{v}}

\renewcommand\ion[2]{#1\,\,\textmc{\romannumeral #2}}

\newcommand{\nifs}{\ensuremath{^{56}\rm{Ni}}}
\newcommand{\nife}{\ensuremath{^{58}\rm{Ni}}}
\newcommand{\cofs}{\ensuremath{^{56}\rm{Co}}}

\newcommand{\msun}{\ensuremath{M_{\odot}}}

\newcommand{\mtot}{\ensuremath{M_\mathrm{tot}}}

\newcommand{\kms}{\ensuremath{\rm{km\,s}^{-1}}}

\newcommand{\gcc}{\ensuremath{\rm{g\,cm}^{-3}}}
\newcommand{\ergs}{\ensuremath{\rm{erg\,s}^{-1}}}

\newcommand{\bmv}{\ensuremath{B\!-\!V}}

\newcommand{\ebv}{\ensuremath{E(\bmv)}}

\newcommand{\mch}{\ensuremath{M_{\rm Ch}}}

\newcommand{\micron}{\ensuremath{\mu\rm{m}}}

\usepackage{color}

\makeatletter
\newcommand{\vast}{\bBigg@{4}}
\newcommand{\Vast}{\bBigg@{5}}
\makeatother

\begin{document} 

   \title{Nebular spectra from Type Ia supernova explosion models
     compared to JWST observations of SN~2021aefx\thanks{Full
       Tables~\ref{tab:nkiiilev} and \ref{tab:nkiiicol} are available
       at the CDS via anonymous ftp to
       \url{cdsarc.cds.unistra.fr} (\url{ftp://130.79.128.5})
       or via
       \url{https://cdsarc.cds.unistra.fr/viz-bin/cat/J/A+A/678/A170}}}
   \titlerunning{Non-LTE nebular \snia\ spectra from the optical to the MIR}
   
   \author{
     S.~Blondin\inst{\ref{lam}}
     \and
     L.~Dessart\inst{\ref{iap}}
     \and
     D.~J.~Hillier\inst{\ref{upitt}}
     \and
     C.~A.~Ramsbottom\inst{\ref{qub}}
     \and
     P.~J.~Storey\inst{\ref{ucl}}
   }
   \authorrunning{S. Blondin et al.}
   
   \institute{
     Aix Marseille Univ, CNRS, CNES, LAM, Marseille, France\\
     \email{stephane.blondin@lam.fr}
     \label{lam}
     \and
     Institut d'Astrophysique de Paris, CNRS-Sorbonne Université, 98
     bis boulevard Arago, 75014, Paris, France
     \label{iap}
     \and
     Department of Physics and Astronomy \& Pittsburgh Particle
     Physics, Astrophysics, and Cosmology Center (PITT PACC),\\
     University of Pittsburgh, 3941 O’Hara Street, Pittsburgh, PA
     15260, USA
     \label{upitt}
     \and
     Astrophysics Research Centre, School of Mathematics and Physics,
     Queen's University Belfast, Belfast BT7 1NN,\\Northern Ireland, UK
     \label{qub}
     \and
     Department of Physics and Astronomy, University College London,
     Gower Street, London WC1E 6BT, UK
     \label{ucl}
   }

   \date{Received 10 June 2023 / Accepted 25 August 2023}

  \abstract
   {Recent JWST observations of the Type Ia supernova (\snia)
     2021aefx in the nebular phase have paved the way for late-time
     studies covering the full optical to mid-infrared (MIR) wavelength
     range, and with it the hope to better constrain \snia\ explosion
     mechanisms.}
   {We investigate whether public \snia\ 
     models covering a broad range of progenitor scenarios and
     explosion mechanisms (Chandrasekhar-mass, or \mch, delayed detonations,
     pulsationally assisted gravitationally confined detonations,
     sub-\mch\ double detonations, and violent mergers)
     can reproduce the full optical-MIR spectrum of SN~2021aefx at
     $\sim$270 days post explosion.}
   {We consider spherically averaged 3D models available from the
     Heidelberg Supernova Model Archive with a \nifs\ yield in the
     range 0.5--0.8\,\msun. We performed 1D steady-state non-local
     thermodynamic equilibrium simulations with the radiative-transfer
     code CMFGEN, and compared the predicted spectra to SN~2021aefx.}
   {The models can explain the main features of SN~2021aefx over the
     full wavelength range. However, no single model, or mechanism,
     emerges as a preferred match, and the predicted spectra are
     similar to each other despite the very different explosion
     mechanisms. We discuss possible causes for the mismatch of the
     models, including ejecta asymmetries and ionisation effects. Our
     new calculations of the collisional strengths for Ni\three\ have
     a major impact on the two prominent lines at 7.35\,\micron\ and
     11.00\,\micron, and highlight the need for more accurate
     collisional data for forbidden transitions. Using updated atomic
     data, we identify a strong feature due to
     [Ca\four]\,3.21\,\micron, attributed to [Ni\one] in previous
     studies.  We also provide a tentative identification of a
     forbidden line due to [Ne\two]\,12.81\,\micron, whose peaked
     profile indicates the presence of neon all the way to the
     innermost region of the ejecta, as predicted for instance in violent
     merger models. Contrary to previous claims, we show that the
     [Ar\three] 8.99\,\micron\ line can be broader in sub-\mch\ models
     compared to near-\mch\ models. Last, the total luminosity in
     lines of Ni is found to correlate strongly with the stable nickel
     yield, although ionisation effects can bias the inferred
     abundance.}
   {Our models suggest that key physical ingredients are missing from
     either the explosion models, or the radiative-transfer
     post-processing, or both. Nonetheless, they also show the
     potential of the near- and MIR to uncover new
     spectroscopic diagnostics of \snia\ explosion mechanisms.}


   \maketitle


\section{Introduction}\label{sect:intro}

Current models for Type Ia supernovae (\sneia) invoke variations in
the mass of the exploding carbon-oxygen white dwarf (WD) and in the
conditions of the thermonuclear runaway. These models include delayed
detonations in near-Chandrasekhar-mass (\mch) WDs
\citep{Khokhlov:1991}, double detonations in sub-\mch\ WDs (e.g.
\citealt{Woosley/Weaver:1994}), and violent mergers of two
sub-\mch\ WDs \citep[e.g.][]{Pakmor/etal:2012}. However, due to
numerous degeneracies in \snia\ light curves and spectra,
distinguishing between these various models has been a challenge
\citep[e.g.][]{Maoz/etal:2014}.

The difficulty arises in part from the multi-dimensional nature of the
explosion while most radiative-transfer simulations assume spherical
symmetry (see e.g. \citealt{Gamezo/Khokhlov/Oran:2005},
\citealt{Seitenzahl/etal:2016}, \citealt{Raskin/etal:2009}, and
\citealt{Pakmor/etal:2010} for examples of 3D explosion
models). Another difficulty resides in the intrinsic complexity of
radiative transfer in \snia\ ejecta, including non-local thermodynamic
equilibrium (non-LTE) simulations and non-thermal
effects, or limitations of the atomic data (see e.g.
\citealt{Hoeflich/etal:1998}, \citealt{Pinto/Eastman:2000b},
\citealt{Sim:2007}, \citealt{D14_tech}, \citealt{Shen/etal:2021}, and
\citealt{Blondin/etal:2022b}). These issues are composition dependent
and related to the complicated explosive nucleosynthesis in
\snia\ explosions (see e.g. \citealt{Bravo/Martinez-Pinedo:2012},
\citealt{Bravo:2020}; and \citealt{Seitenzahl/Townsley:2017} for a
review).

\input{tab1.tex}

While the early high-brightness phase of \sneia\ ($\lesssim$50\,d
post explosion) yields constraints on the ejecta mass and kinetic
energy, as well as the yields of intermediate-mass elements (IMEs) and
\nifs, the late nebular phase ($>$100\,d post explosion) can provide
complementary information. At such times, the ejecta is powered by
\cofs\ decay with an increasing fraction of the deposited energy
arising from positrons, in conditions that are close to steady
state. The low ejecta density (electron densities $n_\mathrm{e} \lesssim
10^6$\,cm$^{-3}$) is conducive to forbidden-line emission which
controls the cooling rate. This cooling is dominated by a few strong
lines that provide important nucleosynthetic information.

In particular, the abundance of stable iron-group elements (IGE)
that are synthesised during the explosion has been proposed to distinguish
sub-\mch\ from \mch\ progenitors (e.g. the Ni/Fe ratio;
\citealt{Floers/etal:2020}). Accurate abundance determinations are
however difficult to obtain given the amount of line overlap and the
sensitivity of line strengths to small variations in the ionisation
state \citep{Blondin/etal:2022}.  The advent of JWST has paved
the way for systematic studies of \sneia\ in the mid-infrared (MIR),
where lines are typically less blended (and less sensitive to the
electron temperature) than in the optical and near-infrared
(NIR). This allows for a more secure identification of lines as well
as detailed studies of their morphology, which may constrain the
spatial distribution of the emitting material
\citep[e.g.][]{Gerardy/etal:2007}.

The Type Ia SN~2021aefx was recently observed with JWST using
the Near Infrared Spectrograph (NIRSpec) in the Fixed Slits (FS)
spectroscopy mode and the Mid Infrared Instrument (MIRI) in the Low
Resolution Spectroscopy (LRS) mode \citep{Kwok/etal:2023}. It was
discovered within a few hours post explosion in NGC 1566, located at
an estimated distance of $18\,\pm\,2$\,Mpc (and recession velocity $cz =
1504$\,\kms), with a Galactic reddening $\ebv\approx0.008$\,mag
\citep{Schlafly/Finkbeiner:2011} and an estimated host-galaxy
reddening $\ebv=0.097$\,mag \citep{Hosseinzadeh/etal:2022}. Early-time
observations revealed a UV flux excess within the first few days after
explosion in this otherwise normal event
\citep{Ashall/etal:2022,Hosseinzadeh/etal:2022}. The nebular
JWST spectrum at +255\,d past maximum light ($\sim$270\,d
post explosion) published by \cite{Kwok/etal:2023} covered the full
0.35--14\,\micron\ range and revealed the presence of numerous
forbidden lines of IGEs and a distinct line due to argon
([Ar\three]\,8.99\,\micron), whose flat-topped profile was interpreted
as indicative of a chemically stratified ejecta. Subsequent modelling
by \cite{DerKacy/etal:2023} of a MIR spectrum taken $\sim$2 months
later showed satisfactory agreement with a
near-\mch\ delayed-detonation model with $\sim$0.6\,\msun\ of
\nifs. However, their study did not confront these data to alternative
\snia\ explosion models.

Here we compare a diverse set of state-of-the-art, public
\snia\ explosion models with the full optical-MIR nebular spectrum of
SN~2021aefx, and evaluate which one, if any, fares better. In the next
section, we present our numerical approach. We then present our
theoretical spectra for each \snia\ explosion model in
Sect.~\ref{sect:results}, and study the impact of ejecta asymmetries
in Sect.~\ref{sect:los}. We discuss the implications of our
results in Sect.~\ref{sect:disc} and conclude in
Sect.~\ref{sect:ccl}. All model outputs are publicly available
online\footnote{\url{https://zenodo.org/record/8290155}}.


\section{Numerical approach}\label{sect:num}

We select previously-published models from the Heidelberg Supernova
Model Archive (HESMA; \citealt{HESMA}) with a \nifs\ yield in the
range 0.5--0.8\,\msun\ (see Table~\ref{tab:summary}), compatible with
the amount inferred for SN~2021aefx\footnote{The estimated peak
$B$-band magnitudes for SN~2021aefx range from $M_B=-19.28$
\citep{Ashall/etal:2022}, typical for normal \sneia, to $M_B=-19.62$
\citep{Hosseinzadeh/etal:2022}, similar to the most luminous
events. The difference is largely due to the assumed host-galaxy
extinction. Given the uncertainty on the estimated distance to the
host galaxy NGC 1566 ($18\,\pm\,2$\,Mpc; \citealt{Sabbi/etal:2018}), both
peak $M_B$ values are statistically consistent with one
another. However, they translate into \nifs\ mass estimates ranging
from $\lesssim 0.6$\,\msun\ to $> 0.8$\,\msun\ \citep[see
  e.g.][]{Blondin/etal:2017}.}. These include 17 models but for only
four explosion scenarios, namely: a \mch\ delayed detonation,
\mch\ pulsationally assisted gravitationally-confined detonations,
sub-\mch\ double detonations, and a violent merger of two
sub-\mch\ WDs. In the following section (Sect.~\ref{sect:results}), we
select one representative model for each scenario, namely the
\mch\ delayed-detonation model N100 of \cite{Seitenzahl/etal:2013}
[HESMA name ddt\_2013\_N100, hereafter DDT], the \mch\ pulsationally
assisted gravitationally-confined detonation model r10\_d1.0 of
\cite{Lach/etal:2022} [HESMA name gcd\_2021\_r10\_d1.0, hereafter
  GCD], the sub-\mch\ double-detonation model M1002 of
\cite{Gronow/etal:2021} [HESMA name doubledet\_2021\_M1002\_1,
  hereafter DBLEDET], and the violent merger model of
\cite{Pakmor/etal:2012} [HESMA name merger\_2012\_11+09, hereafter
  MERGER].

All models correspond to spherically-averaged versions of
a 3D simulation. This averaging causes a systematic overestimate of
the total mass by 3--8\% and a difference of up to $\pm$10\% for
several isotopic and elemental yields compared to the original 3D
models (see Table~\ref{tab:3d1d}). For the N100 
model we also obtained non-spherically-averaged radial profiles along
six directions (I.~Seitenzahl, priv. comm.), whose density was
rescaled to match the total mass (1.4\,\msun) of the original 3D
model. Here the yields can vary significantly with respect to the
original 3D model, by up to a factor $\sim$2 in some cases
(Table~\ref{tab:3d1d_n100}). We subsequently discuss these models in
Sect.~\ref{sect:los}.

\begin{figure}
\centering
\includegraphics{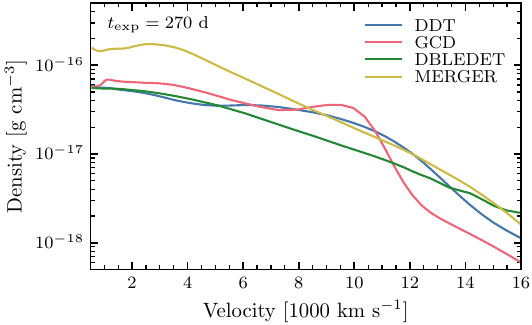}
\caption{\label{fig:dens}
Density profiles at 270\,d post explosion for our reference model set:
the delayed-detonation model ddt\_2013\_N100 (DDT), the pulsationally
assisted gravitationally-confined detonation model
gcd\_2021\_r10\_d1.0 (GCD), the double detonation model
doubledet\_2021\_M1002\_1 (DBLEDET), and the violent merger model
merger\_2012\_11+09 (MERGER).}
\end{figure}

\begin{figure}
\centering
\includegraphics{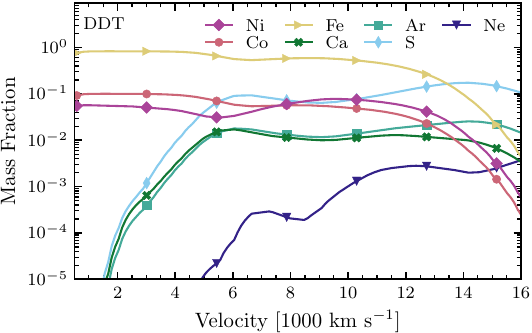}\vspace{-0.55cm}
\includegraphics{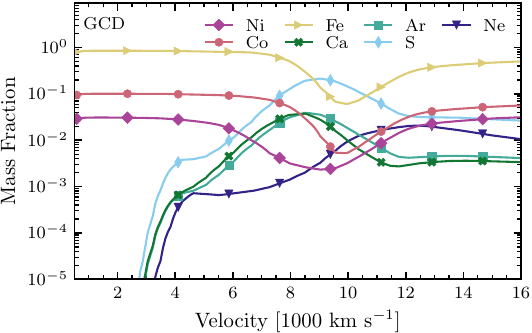}\vspace{-0.55cm}
\includegraphics{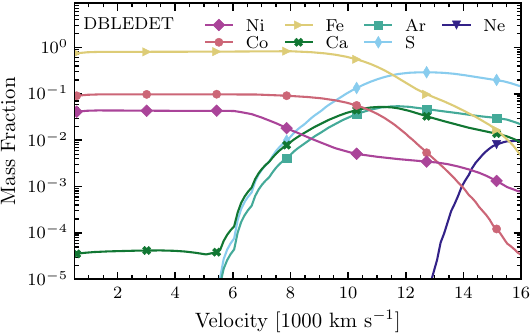}\vspace{-0.55cm}
\includegraphics{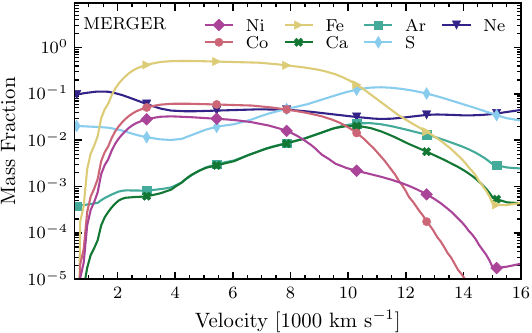}
\caption{\label{fig:comp}
Abundance profiles at 270\,d post explosion for our reference model set.}
\end{figure}

Starting with the spherically-averaged density and abundance profiles
at $t\approx 100$\,s post explosion available on HESMA, we generated
initial conditions at 270\,d post explosion taking into account
changes in composition induced by the decay of radioactive isotopes
(mainly \cofs\ decay at this time) and the decrease in density due to
homologous expansion ($\rho \propto 1/t^3$). We applied a small radial
mixing to the HESMA inputs with a characteristic velocity width
$\Delta \varv_{\rm mix}=300$ or 400\,\kms\ to smooth sharp variations
in composition \citep[see e.g.][]{Blondin/etal:2022b}. The impact of
this mixing on the total yields is below $\sim$0.1\% for all
important species. The initial temperature was set to 5000\,K
throughout the ejecta (this value was found to ease the convergence of
the radiative-transfer calculation and is not too far from the final
value in the inner ejecta layers; see Sect.~\ref{sect:results}). The
density and abundance profiles at 270\,d post explosion for a subset
of four models are shown in Figs.~\ref{fig:dens} and \ref{fig:comp}.

We then solve the 1D non-LTE radiative transfer with CMFGEN
\citep{Hillier/Dessart:2012} assuming steady state. Non-local energy
deposition from radioactive decay ($>99.9$\% of which results from
\cofs\ decay) was treated using a Monte-Carlo approach for
$\gamma$-ray transport. Non-thermal processes are accounted for
through a solution of the Spencer-Fano equation
\citep[see][]{Li/etal:2012}. A new temperature solution is produced by
CMFGEN as part of the full non-LTE solution.

The following ions were included: He\one--\textmc{ii}, C\one--\textmc{iii},
N\one--\textmc{iii}, O\one--\textmc{iii}, Ne\one--\textmc{iii}, Na\one,
Mg\two--\textmc{iii}, Al\two--\textmc{iii}, Si\two--\textmc{iv}, S\two--{\sc
  iv}, Ar\one--\textmc{iii}, Ca\two--\textmc{iv}, Sc\two--\textmc{iii},
Ti\two--\textmc{iii}, Cr\two--\textmc{iv}, Mn\two--\textmc{iii}, Fe\one--{\sc
  v}, Co\one--\textmc{iv}, and Ni\one--\textmc{v}. For all the
aforementioned ions, we also considered ionisations to and
recombinations from the ground state of the next ionisation stage
(e.g. \ion{Fe}{6} in the case of Fe).  The number of levels for all
ions and a description of the sources of atomic data can be found in
Appendix~\ref{sect:atomic_data}.


\section{Ejecta properties and predicted spectra}\label{sect:results}

\begin{figure}
\centering
\includegraphics{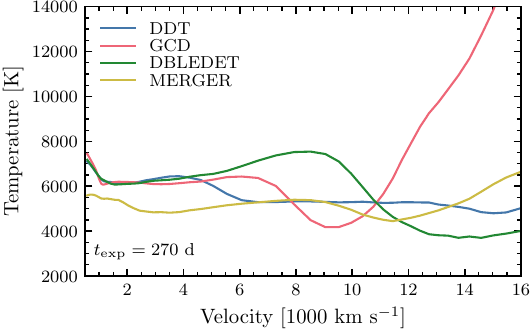}\vspace{-.6cm}
\includegraphics{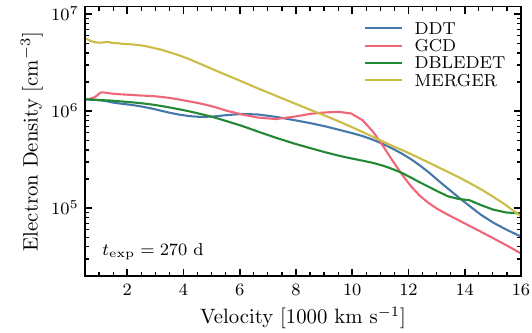}\vspace{-.6cm}
\includegraphics{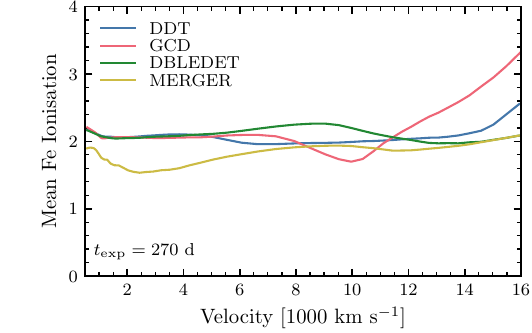}
\caption{\label{fig:tempion}
Characteristic ejecta properties at 270\,d for our reference model set.
\textit{Top:} temperature profiles.
\textit{Middle:} electron density profiles.
\textit{Bottom:} mean Fe ionisation profiles. The mean
ionisation is defined as $\sum_i{in^{i+}}/\sum_i{n^{i+}}$, where
$n^{i+}$ is the number density of ionisation stage $i$ for Fe, such
that a mean ionisation of $\sim$2 indicates that Fe$^{2+}$ is the dominant
stage.}
\end{figure}

\begin{figure*}
\centering
\includegraphics{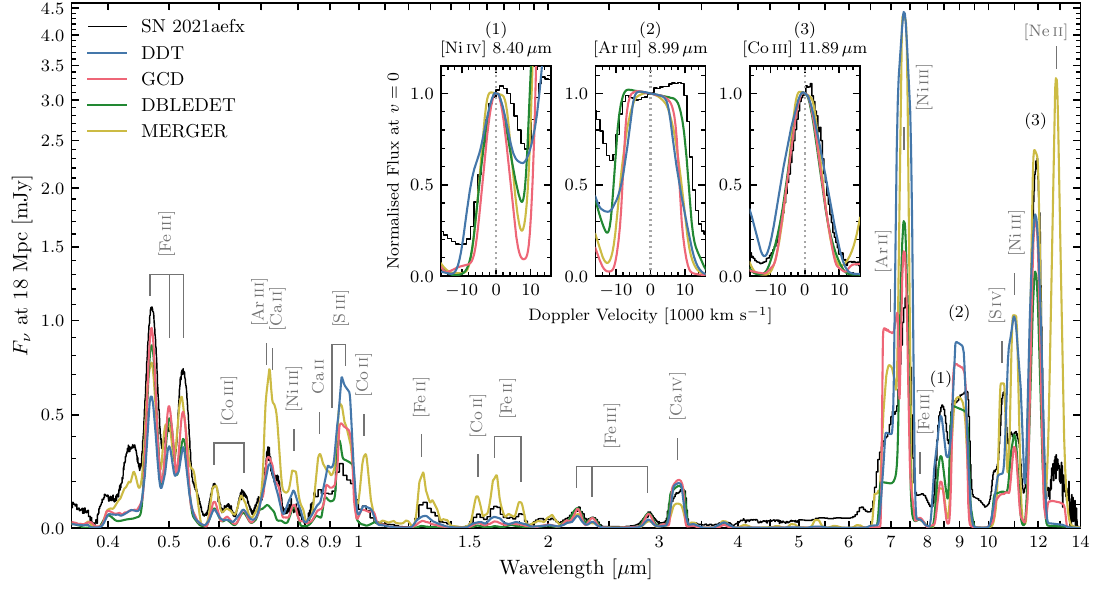}
\caption{\label{fig:comp_spec_overview_full_21aefx}
  Spectra of our reference model set at 270 days post explosion
  compared to SN~2021aefx over the wavelength range
  0.35--14\,\micron\ (logarithmic scale), with flux in $F_\nu$ on a
  non-linear (arcsinh) scale, as in Fig.~1 of
  \cite{Kwok/etal:2023}. The SN~2021aefx spectrum has been corrected
  for redshift and extinction (assuming a host-galaxy reddening of
  0.097\,mag and a MW reddening of 0.008\,mag). The synthetic fluxes
  correspond to the same assumed distance to SN~2021aefx of 18\,Mpc;
  they have not been rescaled or normalised in any way. We include
  selected line identifications based on their maximum Sobolev EW (see
  Table~\ref{tab:lineid}). The insets show normalised line profiles in
  velocity space for selected transitions, illustrating variations in
  line widths and morphology.
}
\end{figure*}


\subsection{Energy deposition, temperature, and ionisation structure}\label{sect:ejecta}

At nebular times, and specifically at 270\,d post explosion in our
model set, only $\sim$5\% of the total decay power is
deposited in the ejecta, of which about 50--70\% comes from
positrons, which we assume are absorbed locally (see
\citealt{Penney/Hoeflich:2014} for a discussion on the validity of
this assumption). Ejecta layers below $\sim$16\,000\,\kms\ capture
99\% of the total volume-integrated decay energy deposition rate (with
the notable exception of the double-detonation model
doubledet\_2021\_M1010\_1, which also has the highest \nifs\ yield of
our model set; see Table~\ref{tab:summary}). This implies that
nebular-phase spectra best probe the regions rich in IGEs but provide
weaker constraints on the outer ejecta where the bulk of IMEs is
located.

All models in our set converge to a roughly uniform temperature of
about 6000\,K below 10\,000\,\kms\ (Fig.~\ref{fig:tempion}, upper
panel), with some diversity at larger velocities that reflect the
variations in the outer density (Fig.~\ref{fig:dens}; a lower outer
density yields a higher temperature for an equivalent power absorbed).
Overall, the higher-mass MERGER model is the cooler one, while the
lower-mass DBLEDET model is the hotter one. The ionisation state of
IGEs, which dominate the composition, reflect
closely this temperature profile (Fig.~\ref{fig:tempion}, lower
panel). In the MERGER model, Fe is a mixture of Fe$^+$ and Fe$^{2+}$,
while Fe$^{2+}$ dominates in all other models (see also
Fig.~\ref{fig:if_fe}). These properties suggest that lines of
Fe\one\ or Fe\four\ should be subdominant. The mean ionisation
profiles for Co and Ni (\textit{not shown}) are qualitatively similar
to those of Fe, in part because of their similar ionisation
potential. However, Co is typically more ionised, with a near-equal
mixture of Co$^{2+}$ and Co$^{3+}$ in the non-MERGER models
(Fig.~\ref{fig:if_co}), and the Ni ionisation level is more uniform
across all models, with Ni$^{2+}$ being the dominant stage
(Fig.~\ref{fig:if_ni}).


\subsection{Model spectra: Overall comparison to SN~2021aefx}\label{sect:spectra}

Figure~\ref{fig:comp_spec_overview_full_21aefx} shows the synthetic
spectra for models DDT, GCD, DBLEDET, and MERGER compared to
SN~2021aefx at 270\,d post explosion in the wavelength range
0.35--14\,\micron\footnote{Similar plots of the full sequence of
pulsationally assisted gravitationally-confined detonation models and
double-detonation models are shown in
Figs.~\ref{fig:comp_spec_gcd_21aefx} and
\ref{fig:comp_spec_doubledet_21aefx}, respectively.}. These data have
been corrected for redshift and reddening \citep{Kwok/etal:2023},
while the models have been scaled to a distance of 18\,Mpc. We report
line identifications in Table~\ref{tab:lineid}, based on their Sobolev
equivalent width (EW)\footnote{Given the weakness of the continuum
flux at this phase, one would normally use the integrated line fluxes
as opposed to their EW for identifying strong lines. However, we found
that using an EW-based criterion provides a better compromise by
limiting the number of line identifications in the optical
($<$0.4\,\micron) while allowing for weaker lines to be identified in
the MIR.}.

The main lesson from this figure is that the models predict all the
observed spectral features of SN\,2021aefx from the blue end of the
optical until the red end of the JWST range. Unfortunately, no
single model produces a perfect match. The modest
difference between the spectral properties of these four distinct
explosion models reflects the similarity in composition for the
dominant coolants in the ejecta. There are, however, numerous offsets
in specific lines or features that result from differences in the
abundance and ionisation of specific elements. The spherical averaging
of the input explosion models will of course impact the predicted
spectra. We attempt to qualitatively evaluate the impact of ejecta
asymmetries later in Sect.~\ref{sect:los}.

Given the similar \nifs\ mass in these models, the deposited decay
power and hence the total radiated energy is also similar, and the
line emission is strongly coupled. Altering the flux in one line will
affect other lines whose flux will adjust to maintain a constant
cooling rate. This effect is most important for the strong IGE lines in
the optical (namely the [Fe\three] complex around 0.5\,\micron) since
they contain most of the flux. The effect is also seen in some IME
lines that can be important coolants in the outer IME-rich regions of
the ejecta (e.g. [S\three--\textmc{iv}], [Ar\two--\textmc{iii}], and
[Ca\two, \textmc{iv}]). Conversely, weak lines that contribute a negligible amount
to the cooling are not affected by this flux redistribution.

\begin{figure}
\centering
\includegraphics{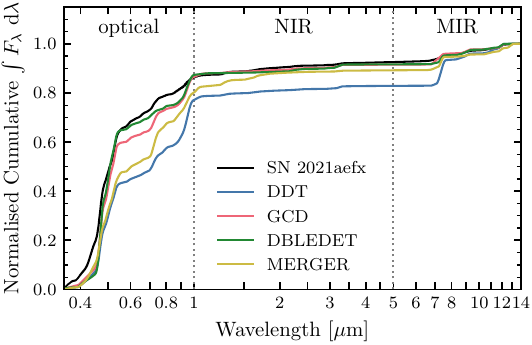}
\caption{\label{fig:fcum}
Normalised cumulative integral of the flux per unit wavelength over
the range 0.35--14\,\micron\ in SN~2021aefx and in our reference model
set. The vertical lines mark the boundaries between the optical-NIR
and NIR-MIR ranges as defined in Sect.~\ref{sect:lamrestrict}.}
\end{figure}

We attempted to quantify the quality of each match by considering
various statistical estimators, such as the mean absolute deviation
(MAD) or the mean fractional error (MFE) with respect to the observed
fluxes over the entire wavelength range (see Appendix~\ref{sect:stats}
and Table~\ref{tab:stats}). To minimise the impact of a mismatch in a
few lines, we also considered a logarithmic flux scale when computing
the MAD or MFE (or the `Score' of
\citealt{Omand/Jerkstrand:2023}). Last, we tried switching to flux per
unit wavelength as opposed to per unit frequency, since the latter
visually enhances the flux in the NIR (1--5\,\micron) and MIR
(5--14\,\micron) ranges compared to the optical range
(0.35--1\,\micron), even though their combined contribution to the
total flux is only $\sim$6\% and $\sim$7\%, respectively (see
Fig.~\ref{fig:fcum}).  However, the results were inconclusive, as each
model was identified in turn as the `best-fit' model depending on the
adopted estimator or flux scale (also when taking into account the
uncertainty in host-galaxy extinction, or after scaling the model
fluxes to match the mean observed flux).  These statistics therefore
only provide a subjective measure of the goodness of fit, and reflect
the difficulty in distinguishing a best-match model for SN~2021aefx.

Throughout the optical and IR ranges, the spectrum is dominated by
forbidden lines. Some exhibit clearly a broad flat top, e.g.
[Ar\three]\,8.99\,\micron\ (except in the MERGER model) or
[Ni\two]\,6.63\,\micron\ (most easily seen in the MERGER model),
indicating a formation starting at large velocities and possibly
extending to even larger velocities (the corresponding models have
little Ar below about 5000\,\kms\ or exhibit an ionisation
stratification).  Numerous lines show a Gaussian-like profile because
they form throughout the ejecta (e.g. [Fe\three]\,0.47\,\micron).
Only a few permitted transitions are predicted in the optical, such as
the Ca\two\ NIR triplet (though significantly weaker in the DBLEDET
model due to the higher ionisation; see Fig.~\ref{fig:if_ca}),
Fe\two\,0.52\,\micron, or Na\one\,D (only noticeable in the MERGER
model).

The `flat'-top profiles in the models exhibit a slight blue excess
which we have identified to arise from the relativistic boost to the
emissivity. When neglecting such a relativistic effect, the line
exhibits a slight flux excess in the red, which is then caused by the
frequency redistribution associated with electron scattering (see
Appendix~\ref{sect:rel}). The observed slant in
[Ar\three]\,8.99\,\micron\ is biased to the red and is probably
dominated by something lacking in our models (e.g. incomplete atomic
data or ejecta asymmetry; \citealt{DerKacy/etal:2023}). We note,
however, that line overlap can also reverse the direction of this tilt
(see Sect.~\ref{sect:mir} below).

All lines from IGEs have typical line widths in the range
$\sim$8000--14\,000\,\kms\ (FWHM values from
\citealt{Kwok/etal:2023}), while those of IMEs, which tend to be
present at large velocities only, exhibit significantly larger widths
($\mathit{FWHM} \gtrsim 20\,000$\,\kms). Nearly all lines appear centred
around their rest wavelength, and the only features exhibiting a
significant offset $\gtrsim 2000$\,\kms\ are affected by line overlap
(e.g. [Fe\three]\,3.23\,\micron, [Ni\two]\,6.63\,\micron,
[Ar\two]\,6.98\,\micron; see \citealt{Kwok/etal:2023}),
with the notable exception of [Ni\three]\,7.35\,\micron\ (see
Sect.~\ref{sect:mir} below). Ejecta
asymmetry is invoked by \citet{DerKacy/etal:2023} but the profile
skewness observed, for example, in individual lines such as
[Co\three]\,11.89\,\micron\ is much weaker than that identified in
multiple lines in nebular phase spectra of some \sneia\ (see
e.g. \citealt{Dong/etal:2015}). The inset in
Fig.~\ref{fig:comp_spec_overview_full_21aefx} gives the typical
profile morphologies in our model spectra, using
[Ni\four]\,8.40\,\micron, [Ar\three]\,8.99\,\micron, and
[Co\three]\,11.89\,\micron.


\subsection{Detailed comparison in restricted wavelength ranges}\label{sect:lamrestrict}

Figure~\ref{fig:comp_spec_overview_opt2mir2_21aefx} shows the spectra
in restricted wavelength ranges (optical: 0.35--1\,\micron; NIR:
1--5\,\micron; lower MIR: 5--14\,\micron), including the
14--28\,\micron\ range (upper MIR) for which there is no data at that
epoch. In what follows we discuss each of these ranges in turn.

\subsubsection{Optical range (0.35--1\,\micron)}\label{sect:opt}

The optical spectrum is dominated by forbidden lines of doubly-ionised
elements (i.e. [Fe\three], [Co\three], [Ar\three], [Ni\three], and
[S\three] lines), with line contributions from singly-ionised Fe, Co,
and Ca (both forbidden and permitted transitions) and neutral Fe and
Na (semi-forbidden and permitted transitions) mostly in the MERGER
model due to the lower ionisation. The (permitted) Ca\two\ NIR triplet
is predicted in all models, except the DBLEDET model due to its higher
ionisation.  This model matches best the [S\three] feature at
0.9--1\,\micron, as both the abundance and ionisation stratification
conspire to confine the line-formation region to a narrower velocity
range (and hence a more restricted volume) compared to the other
models.

As with nebular-phase modelling of \snia\ spectra by other groups
(e.g. \citealt{Mazzali/etal:2015}; \citealt{Shingles/etal:2020}),
none of our models are able to match the flux level in the
$\sim$0.43\,\micron\ feature. In all our models, this feature is
dominated by lines of singly-ionised Fe (forbidden lines in the
non-MERGER models, with contributions from permitted lines in the
MERGER model). The top panel of
Fig.~\ref{fig:comp_spec_overview_opt2mir2_21aefx} seems to suggest
that the semi-forbidden line of Fe\one]\,0.44\,\micron\ dominates this
feature in the MERGER model, however the combined emission of weaker
lines of Fe\two\ is more than a factor of three larger than the
combined emission from neutral Fe. These weaker lines are not reported
in Table~\ref{tab:lineid} because their Sobolev EW is below our
adopted EW cut.

\subsubsection{NIR range (1--5\,\micron)}\label{sect:nir}

The NIR spectral range (second panel of
Fig.~\ref{fig:comp_spec_overview_opt2mir2_21aefx}) is dominated by
forbidden transitions of Fe\two\ and Co\two\ and all four models
predict the observed features with the notable exception of the weak
featureless continuum beyond 4\,\micron\ (which could possibly arise
from molecule emission or dust formation in the outer
ejecta\footnote{\cite{Gerardy/etal:2007} also invoke SiO to explain
the emission at 7.5--8\,\micron\ in the low-luminosity Type Ia
SN~2005df. However, such emission is not expected in higher-luminosity
events (see \citealt{Hoeflich/Khokhlov/Wheeler:1995} and the
discussion in \citealt{Kwok/etal:2023}).}; see e.g.
\citealt{Jerkstrand/etal:2012} in the core-collapse SN context). The
MERGER model indicates that a lower ionisation would improve the
agreement for the Fe\two\ and Co\two\ lines in the three other models.
Higher ionisation features due to [Fe\three] 2.24, 3.01\,\micron,
[Ca\four]\,3.21\,\micron\ and [Ni\three]\,3.39, 3.80\,\micron\ are
also present in all models. We note that the
[Co\three]\,1.55\,\micron\ line identified by \cite{Kwok/etal:2023} is
due to Co\two\ in our models. Moreover, none of our models predict a
contribution from [Ni\one]\,3.12\,\micron\ to the
3.2\,\micron\ feature (see Sect.~\ref{sect:clump}), which is dominated
by [Ca\four]\,3.21\,\micron\ in our models \citep{Nahar23_CaIV}, with
a modest contribution from [Fe\three]\,3.23\,\micron.

\subsubsection{Lower MIR range (5--14\,\micron)}\label{sect:mir}

\begin{figure*}
\centering
\includegraphics{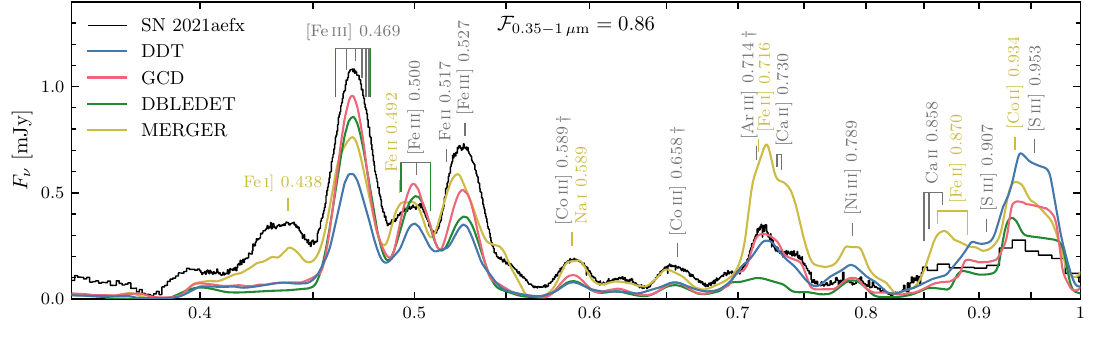}\vspace{-0.475cm}
\includegraphics{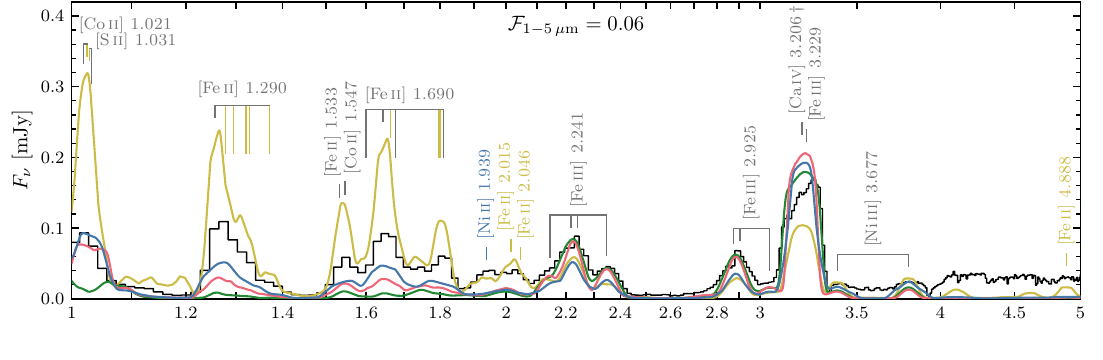}\vspace{-0.475cm}
\includegraphics{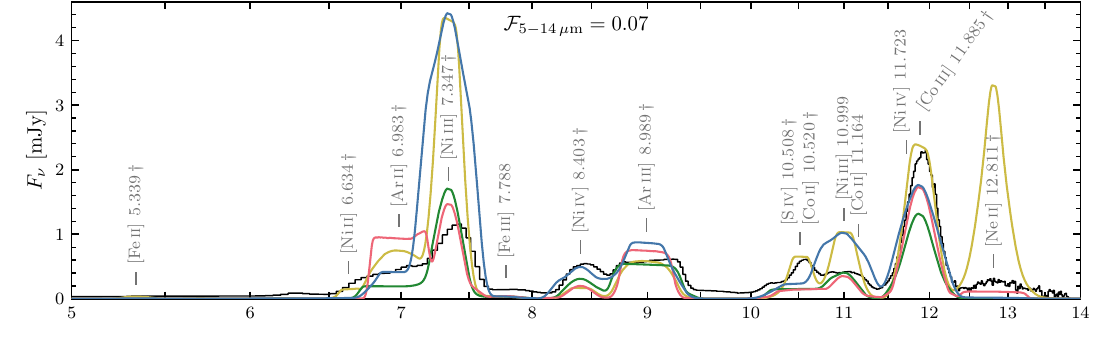}\vspace{-0.475cm}
\includegraphics{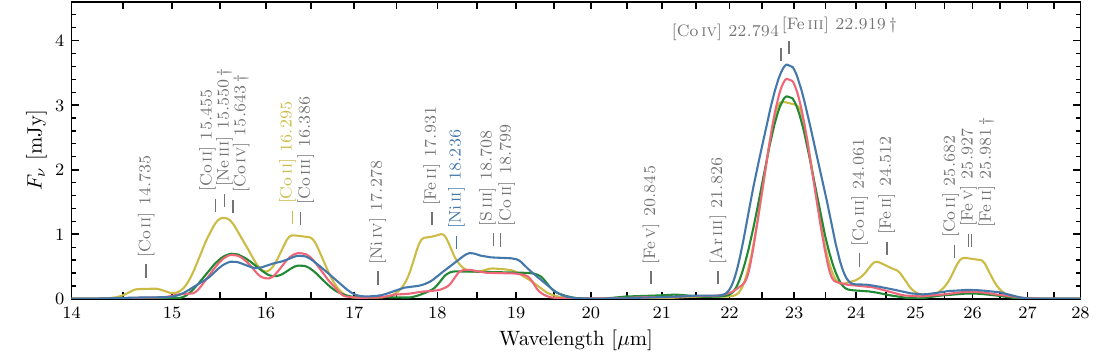}
\caption{\label{fig:comp_spec_overview_opt2mir2_21aefx}
  Same as Fig.~\ref{fig:comp_spec_overview_full_21aefx} but for
  restricted wavelength ranges (from top to bottom):
  optical (0.35--1\,\micron), NIR (1--5\,\micron), lower MIR
  (5--14\,\micron) and upper MIR (14--28\,\micron).
  The $\mathcal{F}_{X-Y\,\micron}$ label gives the fraction of the total
  optical to MIR flux (0.35--14\,\micron) for SN~2021aefx emitted
  within the wavelength range of each plot.
  We include line identifications based on their Sobolev EW
  (see Table~\ref{tab:lineid}). Transitions connected to the ground
  state are marked with a `$\dag$' symbol. For consecutive lines
  within a multiplet (connected by a horizontal line), we give the
  $gf$-weighted mean wavelength of the transitions. Lines that only
  appear in one model class are labelled with the corresponding colour.
}
\end{figure*}

\input{tab2.tex}

The lower MIR range (third panel of
Fig.~\ref{fig:comp_spec_overview_opt2mir2_21aefx}) reveals the
presence of lines from three distinct ionisation stages ({\sc
  ii}--\textmc{iv}) as in the NIR range, and lines from the IMEs argon
and sulfur as in the optical range. We also tentatively identify the
12.8\,\micron\ feature as [Ne\two]\,12.81\,\micron, which is
particularly strong and centrally peaked in the MERGER model due to
the presence of Ne in the inner ejecta layers (see
Fig.~\ref{fig:comp}), combined with the low ionisation. The strongest
lines in this range are low-lying forbidden transitions either
directly connected to the ground state (marked with a `$\dagger$' in
Fig.~\ref{fig:comp_spec_overview_opt2mir2_21aefx} and
Table~\ref{tab:lineid}), such as the prominent
[Ni\three]\,7.35\,\micron\ and [Co\three] 11.89\,\micron\ lines, or
part of a multiplet including the ground state, such as
[Ni\three]\,11.00\,\micron. The [Ni\four] and [S\four] transitions
connect levels at most $\lesssim 0.3$\,eV (equivalent electron
temperature $\lesssim 3500$\,K) above the ground state which are
easily collisionally excited given the ejecta temperature in the
corresponding layers.

In the non-MERGER models, the sharp drop in the IME abundance below
$\sim$3000--4000\,\kms\ ($\sim$8000\,\kms\ in the DBLEDET model)
results in a broad flat-top morphology for lines of Ca\four,
Ar\two--\textmc{iii} and S\four, while the presence of IGEs down to the
centre results in centrally-peaked lines of Fe\three, Co\two--{\sc
  iii}, and Ni\three--\textmc{iv}. The weak [Ni\two]\,6.63\,\micron\ line
is flat-topped in these models owing to an ionisation (as opposed to
abundance) stratification. In the MERGER model, the situation is
reversed as IMEs dominate the innermost regions while IGEs are
present above $\sim$1000-2000\,\kms, resulting in Gaussian-like
profiles for the Ca, Ar and S lines (e.g. [Ar\two]\,6.98\,\micron),
and narrow flat-top profiles for the Ni, Co, and Fe lines (e.g.
[Co\three]\,11.89\,\micron). We note that line overlap can reverse the
direction of the slant of the flat top (from blue to red) in some
profiles, as seen for instance in [Co\two]\,10.52\,\micron\ (overlap
with [Ni\two]\,10.68\,\micron) or [Ni\three]\,11.00\,\micron\ (overlap
with [Co\two]\,11.16\,\micron).

The strength of the [Ni\three]\,7.35\,\micron\ line is largely
overestimated in the DDT and MERGER models (factor of $\sim$4 at
line centre), as is the [Ni\three]\,11.00\,\micron\ line (factor of
$\sim$2 at line centre). These lines are particularly sensitive to
the collisional strengths for Ni\three\ (see Sect.~\ref{sect:nicol}),
which set the collisional de-excitation rate from the upper level and
hence the critical density for each transition ($n_\mathrm{crit}
\approx 1.5\times10^6$\,cm$^{-3}$ and $\sim 6.6\times10^5$\,cm$^{-3}$
at 5000\,K for the 7.35\,\micron\ and 11.00\,\micron\ transitions,
respectively). The DDT model has the largest stable Ni abundance of
our reference model set ($\sim0.08$\,\msun; see
Table~\ref{tab:summary}), with a peak in its abundance profile around
9500\,\kms\ (Fig.~\ref{fig:comp}, top panel) where the electron
density drops below $n_\mathrm{crit}$ (Fig.~\ref{fig:tempion}, middle
panel). The stable Ni abundance is lower in the MERGER model
($\sim0.03$\,\msun) and the electron density is higher, but the offset
of the IGE distribution to larger velocities ensures there is a
sufficient amount of (doubly-ionised) Ni in layers where $n_\mathrm{e}
\lesssim n_\mathrm{crit}$, resulting in strong [Ni\three] lines.

The [Ni\three]\,7.35\,\micron\ line also displays one of the largest
kinematic offsets ($3000\,\pm\,1400$\,\kms; \citealt{Kwok/etal:2023})
that cannot be attributed to line overlap, and which our 1D modelling
approach cannot accommodate. However, \cite{Kwok/etal:2023} note that
the MIRI wavelength solution is more uncertain in this lower-wavelength
range, and the measured offset for the redder isolated
[Ni\four]\,8.40\,\micron\ line is consistent with no offset
($1300\,\pm\,1200$\,\kms).

\subsubsection{Upper MIR range (14--28\,\micron)}\label{sect:mir2}

The upper MIR range (fourth panel of
Fig.~\ref{fig:comp_spec_overview_opt2mir2_21aefx}) is shown for
comparison with future JWST observations extending to higher
wavelengths. The key spectroscopic signatures comprise a double-hump
feature centred on 16\,\micron\ due to [Co\four]\,15.64\,\micron\ (and
[Ne\three]\,15.55\,\micron\ in the MERGER model) and
[Co\three]\,16.39\,\micron, a broad feature around
18--19\,\micron\ dominated by [S\three]\,18.71\,\micron\ with a
contribution on the blue side from [Fe\two]\,17.93\,\micron\ (whose
blue-to-red tilt in the MERGER model is due to line overlap), and a
prominent line at $\sim$23\,\micron\ due to
[Fe\three]\,22.92\,\micron\ (with a small contribution due to
[Co\four]\,22.79\,\micron). This [Fe\three] line constitutes almost
50\% of the total flux in this range for the non-MERGER models
(and $\sim$30\% for the MERGER model). It was speculated to
account for most of the observed flux in the \textit{F2100W} filter
(spanning roughly 18.5--23.5\,\micron) in JWST imaging of
SN~2021aefx taken at +357\,d past maximum by \cite{Chen/etal:2023}. In
our model spectra corresponding to $\sim$100 days earlier, this line
constitutes 75--80\% of the flux in the \textit{F2100W}
bandpass.  Additional lines of Co\two\ and
Fe\two\ contribute to the spectrum of the MERGER model due to the
lower ionisation. The analysis of the differences in line-profile
morphology in the lower MIR range (5--14\,\micron;
Sect.~\ref{sect:mir}) also applies here.


\section{Predicted spectra for different rays of a 3D model}\label{sect:los}

\begin{figure}
\centering
\includegraphics{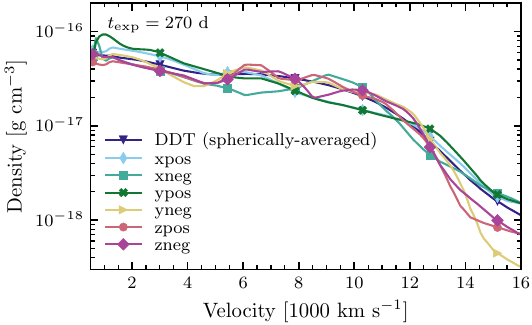}\vspace{-.9cm}
\includegraphics{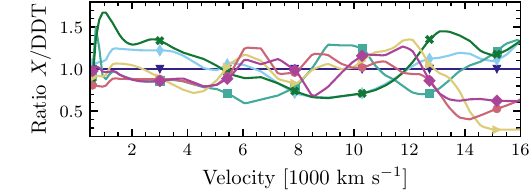}
\caption{\label{fig:dens2}
Density profiles for the spherically-averaged ddt\_2013\_N100 (DDT)
model and along the three orthogonal axes of the original 3D Cartesian
grid, in both positive (\{x,y,z\}pos) and negative (\{x,y,z\}neg) directions.
The bottom panel shows the density ratio with respect to the DDT model.}
\end{figure}

\begin{figure*}
\centering
\includegraphics{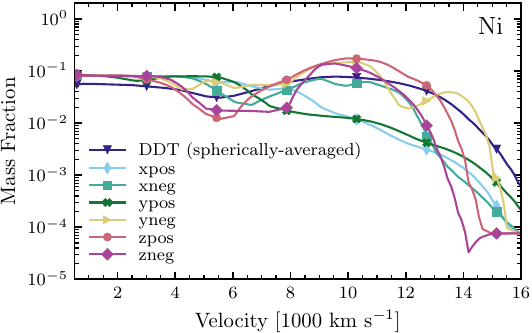}\hspace{0.3cm}
\includegraphics{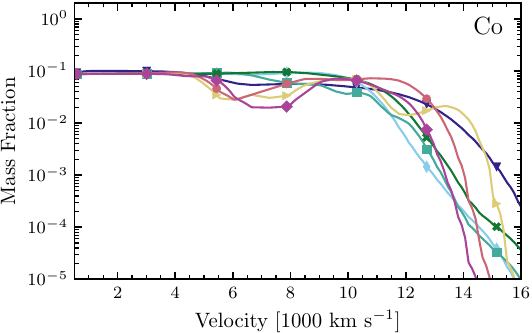}\vspace{-0.8cm}
\includegraphics{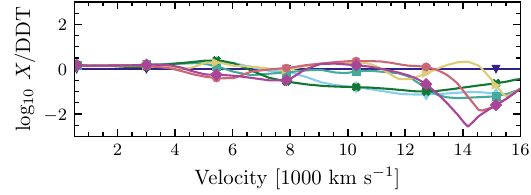}\hspace{0.3cm}
\includegraphics{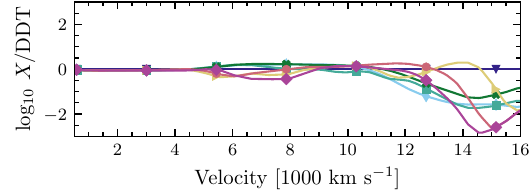}\vspace{-0.5cm}
\includegraphics{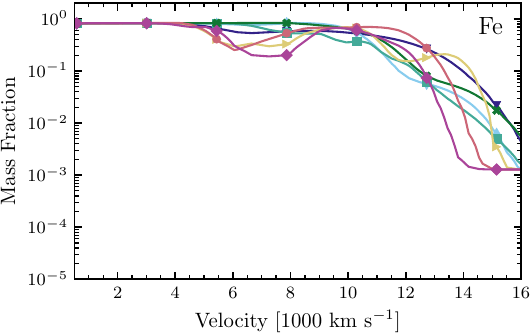}\hspace{0.3cm}
\includegraphics{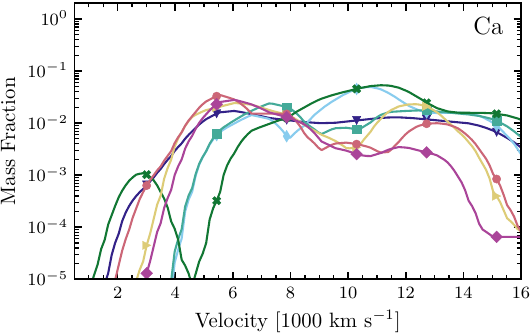}\vspace{-0.8cm}
\includegraphics{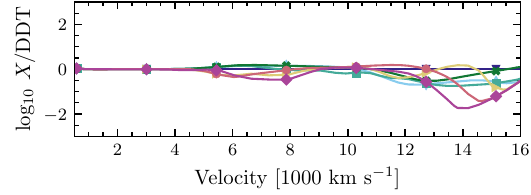}\hspace{0.3cm}
\includegraphics{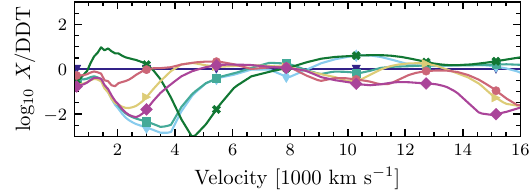}\vspace{-0.5cm}
\includegraphics{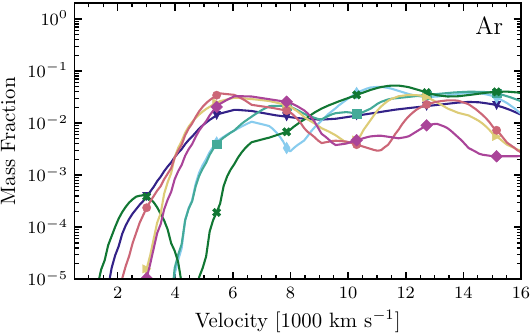}\hspace{0.3cm}
\includegraphics{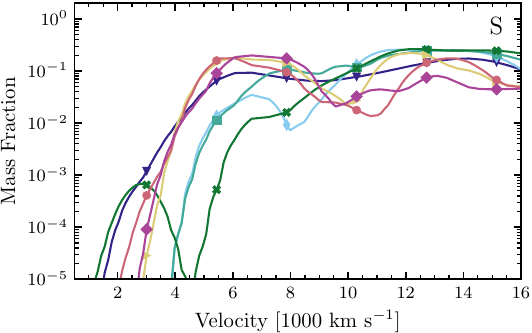}\vspace{-0.8cm}
\includegraphics{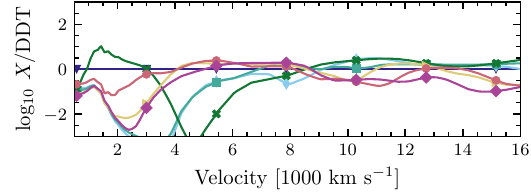}\hspace{0.3cm}
\includegraphics{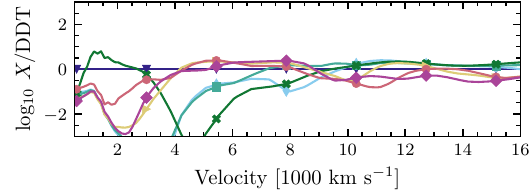}
\caption{\label{fig:comp2}
Abundance profiles for Ni, Co, Fe, Ca, Ar, and S in the
spherically-averaged ddt\_2013\_N100 (DDT) model and along the three
orthogonal axes of the original 3D Cartesian grid, in both positive (\{x,y,z\}pos) 
and negative (\{x,y,z\}neg) directions. The lower panels show the logarithm of the
abundance ratio with respect to the DDT model.
}
\end{figure*}

\begin{figure*}
\centering
\includegraphics{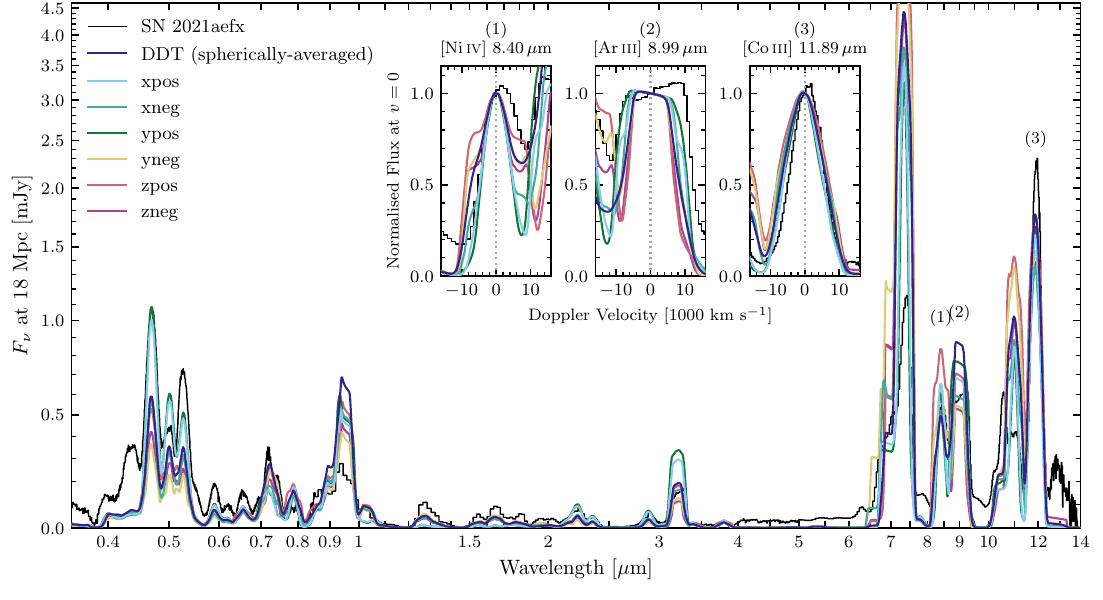}
\caption{\label{fig:comp_spec_n100_3d_21aefx}
Similar to Fig.~\ref{fig:comp_spec_overview_full_21aefx} but for the
ddt\_2013\_N100 (DDT) model along the three
orthogonal axes of the original 3D Cartesian grid, in both positive (\{x,y,z\}pos) 
and negative (\{x,y,z\}neg) directions. We also show the
original (spherically-averaged) DDT model.}
\end{figure*}

One fundamental issue with spherical averaging of multi-dimensional
explosion models is the microscopic mixing that results in otherwise
chemically-segregated zones at a given radius but different
ejecta-centred directions (see e.g. \citealt{Dessart/Hillier:2020}
in the context of core-collapse SNe). In \sneia\ with significant
ejecta asymmetries, with WD-WD mergers and collisions being 
extreme cases, such mixing can alter the composition in the \cofs-rich
zones at nebular epochs, where local energy deposition by positrons
largely governs the plasma emissivity. We attempted to qualitatively
evaluate the impact of such asymmetries on our 1D modelling approach
by considering different directions in the original 3D version of the
DDT model (model N100 in \citealt{Seitenzahl/etal:2013};
I.~Seitenzahl, priv. comm.).  Specifically we considered the three
orthogonal axes of their Cartesian grid $(\hat{\mathbf{x}},
\hat{\mathbf{y}}, \hat{\mathbf{z}})$ in both positive and negative
directions, resulting in six distinct radial profiles.

The density profiles resulted in total ejecta masses in the range
1.15--1.52\,\msun\ (see Table~\ref{tab:3d1d_n100}), so we rescaled the
density in each case to match the total mass of the original 3D model
(1.40\,\msun). The resulting \nifs\ mass is in the range
0.50--0.70\,\msun, corresponding to a $\pm$0.1\,\msun\ ($\pm$17\%)
difference compared to the original 3D model (in which
$M(\nifs)\approx0.60$\,\msun; \citealt{Seitenzahl/etal:2013}). As seen
in Table~\ref{tab:3d1d_n100}, The impact on the yields of specific
isotopes and elements is occasionally larger than 50\%, as is the case
for \nife\ in the model corresponding to the positive
$\hat{\mathbf{z}}$ direction (+81\%), or Ca in the model corresponding
to the positive $\hat{\mathbf{y}}$ direction (+87\%).

We show the density profiles for these six directions in
Fig.~\ref{fig:dens2}. The largest deviations from the
spherically-averaged model occur in the positive y-direction, with a
$\sim$70\% higher density at $\sim$1000\,\kms\ and a $\sim$35\%
lower density at $\sim$9000\,\kms. As expected, the deviation in
composition from the spherically-averaged model at a given velocity
can be far more significant (Fig.~\ref{fig:comp2}). While the mass
fraction of IGEs (Ni, Co, and Fe) are within $\sim$50\%
of one another in the inner ejecta ($\varv < 4000$\,\kms), variations
can span a factor of a few out to $\sim$12\,000\,\kms\ and orders of
magnitude farther out. For the IMEs (illustrated here using Ca, Ar,
and S) the relative variation in mass fraction at a given velocity is
systematically larger, in part due to their underabundance compared to
IGEs.

The variation in spectral properties along the six directions compared
to the spherically-averaged model partly reflects these variations in
abundance (Fig.~\ref{fig:comp_spec_n100_3d_21aefx}). For instance the
width of the prominent [Ar\three]\,8.99\,\micron\ line is set by the
extent of the Ar hole (which determines the width of the flat-top
component\footnote{Due to our 1D modelling approach, the flat top has
the same velocity extent relative to line centre.}; see
Sect.~\ref{sect:ar}) and the peak or global offset of the Ar
distribution (which determines its FWHM). Thus the model corresponding
to the positive $\hat{\mathbf{y}}$ direction has both the broadest
flat-top (approximately $\pm$3000\,\kms\ about line centre) and the
largest $\mathit{FWHM} \approx 21\,200$\,\kms\ (see inset (2) in
Fig.~\ref{fig:comp_spec_n100_3d_21aefx}). Conversely, the model
corresponding to the positive $\hat{\mathbf{z}}$ direction displays a
flat-top extending only $\pm$2000\,\kms\ about line centre and a
$\mathit{FWHM} \approx 13\,600$\,\kms. Modulations in the Ar abundance
distribution also leave an imprint on the line profiles: the dip
around $\sim$8000\,\kms\ in the positive $\hat{\mathbf{x}}$
direction causes a `shoulder' in the [Ar\three]\,8.99\,\micron\ line
profile around $\pm$8000\,\kms\ in this model.

The remaining variation results from differences in the temperature
and ionisation structures due to variations in the \cofs\ distribution
along the different directions, in particular above
$\sim$4000\,\kms\ (see upper-right panel in
Fig.~\ref{fig:comp2})\footnote{This panel in fact shows the total Co
distribution, but it is largely dominated by \cofs\ ($\sim$80--90\%
depending on the direction), the remainder being in the form
of stable $^{59}$Co.}. The impact is most readily seen in the optical
Fe\three-dominated complex around $\sim$0.5\,\micron. It is
strongest in the models corresponding to the positive
$\hat{\mathbf{x}}$ and $\hat{\mathbf{y}}$ directions, where the Fe
ionisation is the largest in the range $\sim$5000--10\,000\,\kms\ due
to the higher \cofs\ abundance in those layers. This ionisation effect
is further enhanced by the larger abundance of Fe in these same
layers, which follows that of Co. By comparing
Figs.~\ref{fig:comp_spec_overview_full_21aefx} and
\ref{fig:comp_spec_n100_3d_21aefx} we see that the spectroscopic
variation along different lines of sight in the DDT model can be as
large as when considering different explosion models. This 
impacts the accuracy of abundance determinations when using
spherically-averaged ejecta as well as our ability to make
quantitative comparisons with observations.

While our approach captures broad trends associated with ejecta
asymmetries, it is inherently 1D and the line-of-sight dependence is
likely overestimated. Moreover, all the predicted line profiles remain
largely symmetric and present no intrinsic Doppler shifts, as inferred
for several lines by \cite{Kwok/etal:2023}.  However, barring a few
outliers strongly affected by line overlap, most lines display a
kinematic offset either consistent with zero or $\lesssim
1000$\,\kms\ (Table~2 in \citealt{Kwok/etal:2023}). Uncertainties in
the wavelength calibration and the low resolution of the MIRI/LRS
spectrograph dominate the measurement uncertainty. For instance, the
strong [Co\three]\,11.89\,\micron\ line is offset by
$500\,\pm\,900$\,\kms\ in the +255\,d past maximum spectrum analysed
here according to \cite{Kwok/etal:2023}, and by
$740\,\pm\,544$\,\kms\ in the +323\,d past maximum spectrum obtained by
\cite{DerKacy/etal:2023}. This latter measurement, only
$\sim$1.4$\sigma$ away from a zero velocity offset, is used as
evidence for an off-centre distribution of \nifs\ in the
ejecta\footnote{In a recent paper, \cite{Ni/etal:2023} also favour an
off-centre explosion for SN~2021aefx due to the presence of
high-velocity features and unburnt carbon in the early-time spectra.}.
Conversely, their model (as does ours) fails to reproduce the much
larger offset observed in [Ni\three]\,7.35\,\micron\ (see
Sect.~\ref{sect:mir}), as the stable Ni is concentrated in the centre
of their off-centre delayed-detonation model.  As noted by the authors
of that study, higher-resolution spectra (with the MIRI
medium-resolution spectrometer, or MRS) and a more accurate wavelength
calibration are needed to reliably measure line shifts in nebular
\snia\ spectra taken with JWST. If confirmed for multiple Ni
lines, such a large kinematic offset would indicate a large asymmetry
in the stable Ni distribution and provide important clues on the
explosion mechanism \citep[see e.g.][]{Maeda/etal:2010a}.


\section{Discussion}\label{sect:disc}

In what follows we discuss in more detail the impact of uncertainties
in the atomic data, focusing on the Ni\three\ collision strengths
which greatly affect the lines at 7.35\,\micron\ and
11.00\,\micron\ (Sect.~\ref{sect:nicol}). We present a case for the
tentative association of the feature at $\sim$12.8\,\micron\ with
[Ne\two]\,12.81\,\micron\ in Sect.~\ref{sect:ne2}. In
Sect.~\ref{sect:clump} we discuss the notable absence of lines from
neutral ions in our model spectra and whether a lower ionisation
through clumping can produce a better match to the observations. Last,
we discuss the constraints on the ejecta (and hence progenitor) mass
that can be inferred from the strength of stable Ni lines
(Sect.~\ref{sect:ni}) or the width of the
[Ar\three]\,8.99\,\micron\ line (Sect.~\ref{sect:ar}).


\subsection{Uncertainties in atomic data: Ni\three\ collision strengths}\label{sect:nicol}

\begin{table}
\centering
\small  
\caption{Ni\three\ transition probabilities ($A_\mathrm{ul}$) and effective collision strengths ($\Upsilon_\mathrm{lu}$) for the three lowest
  levels of the ground configuration (3p$^6$ 3d$^8$).}
\label{tab:ni3_col}
\begin{tabular}{lccccc}
\hline\hline\\[-1.8ex]
\multicolumn{1}{c}{Transition} & \multicolumn{1}{c}{$n_\mathrm{l}\mathrm{-}n_\mathrm{u}$} & \multicolumn{1}{c}{$A_\mathrm{ul}$} & \multicolumn{3}{c}{$\Upsilon_\mathrm{lu}$ ($T=5000$\,K)}        \\
\cline{4-6}\\[-1.8ex]
                               &                                        & (s$^{-1}$)                  & \multicolumn{2}{c}{This paper} & \multicolumn{1}{c}{Priv. comm.} \\
\cline{4-5}\\[-1.8ex]
                               &                                        &                             & \multicolumn{1}{c}{(R07)} & \multicolumn{1}{c}{(S16)} & \multicolumn{1}{c}{(B01)} \\
\hline\\[-1.8ex]
$^3$F$_4$-$^3$F$_3$ & 1-2 & 6.5($-2$) & 2.49 & 2.20 & 2.29 \\
$^3$F$_4$-$^3$F$_2$ & 1-3 & 4.5($-9$) & 0.77 & 0.57 & 0.83 \\
$^3$F$_3$-$^3$F$_2$ & 2-3 & 2.7($-2$) & 1.67 & 1.63 & 2.82 \\
\hline
\end{tabular}
\flushleft
\justifying
\footnotesize
\textbf{Notes.} Numbers in parentheses correspond to powers of ten.
$\Upsilon_\mathrm{lu}$ is the (dimensionless) effective collision strength
averaged over the Maxwellian velocity distribution of the electrons at
a given temperature.\\
\textbf{References.}
B01 = M.~Bautista (priv. comm.),  based on \cite{Bautista:2001};
R07 = This paper, based on \cite{Ramsbottom/etal:2007};
S16 = This paper, based on \cite{Storey/etal:2016}.
\end{table}

Accurate atomic data are of paramount importance to making reliable
predictions with radiative-transfer simulations. The quality of such
data is highly variable, and in some instances no data is
available. This is particularly true of bound-bound collisional
cross-sections between low-lying levels of some ions, relevant to the
formation of forbidden lines in the NIR and MIR ranges. For the
present calculations, we updated the collisional data used by CMFGEN
for the following ions: C\one, O\one, Ar\two, Ca\four, Ti\two--\textmc{iii},
Co\one, Co\three, and Ni\three--\textmc{iv}.

\begin{figure*}
\centering
\includegraphics{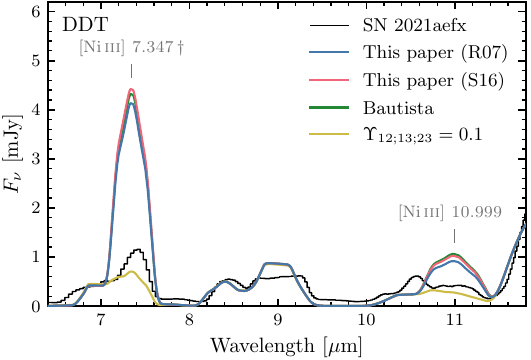}\hspace{0.3cm}
\includegraphics{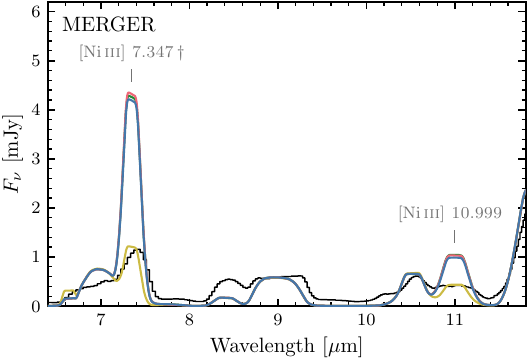}\vspace{-0.5cm}
\includegraphics{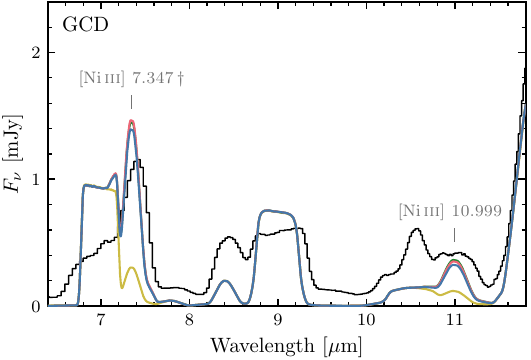}\hspace{0.3cm}
\includegraphics{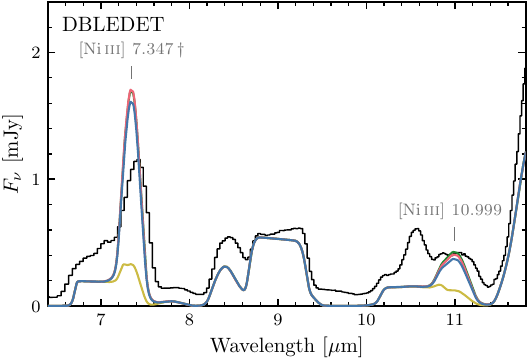}
\caption{\label{fig:nkiiicol}
Impact of different values of the effective collision strengths
amongst the lowest $^3$F term in Ni\three\ (see
Table~\ref{tab:ni3_col}) on the lines at 7.35\,\micron\ and
11.00\,\micron\ in our reference model set. In the top row we show the
DDT and MERGER models where these lines are largely overestimated. In
the bottom row we show the GCD and DBLEDET models where the
predicted strengths for both lines are comparable to those seen in
SN~2021aefx (note the smaller ordinate range compared to the top row).
}
\end{figure*}

Since there are no published collisional strengths for transitions
within the lowest LS term of Ni\three\ (the 3d8 $^3$F configuration, which
gives rise to the 7.35\,\micron\ and 11.00\,\micron\ forbidden lines),
we carried out calculations of collisional cross-sections for
electron-impact excitation of Ni\three\ following the methods of
\cite{Ramsbottom/etal:2007} and \cite{Storey/etal:2016}. We report
effective collision strengths at 5000\,K for transitions among the
three lowest levels in Table~\ref{tab:ni3_col} (a more complete set of
values is shown in Table~\ref{tab:nkiiicol}). We also include values
based on the calculations of \cite{Bautista:2001}
(M.~Bautista, priv. comm.)\footnote{Although
they assume LS coupling, \cite{Bautista:2001} report line intensities
for the fine-structure lines of [Ni\three] by algebraic transformation
of the scattering matrices (their Table~5). They find that this
approximation yields accurate results at the $\sim$1\% level
for the ground 3p$^6$ 3d$^8$ configuration. However, they do not
report collision strengths for these fine-structure transitions.}.
As shown in Fig.~\ref{fig:nkiiicol}, the predicted line strengths for
these three independent sets of collision strengths agree very well
with one another.
In Appendix~\ref{sect:nkiiicol} we compare our values with the
approximations of \cite{Axelrod:1980} for forbidden lines.

As noted previously in Sect.~\ref{sect:spectra}, collisional
excitation of [Ni\three] results in a significant overestimate in the
strength of the two IR lines at 7.35\,\micron\ and 11.00\,\micron\ for
the DDT and MERGER models.  While this visually degrades the match to
the SN~2021aefx observations, this still enables a relative comparison
of the different models.  To illustrate the impact of collisional
excitation on the strengths of both [Ni\three] lines, we recomputed
the spectra for our reference model set by fixing the collision
strengths to a constant $\Upsilon_\mathrm{lu}=0.1$ within the ground
configuration (yellow line in Fig.~\ref{fig:nkiiicol}). The flux in
the [Ni\three] lines decreases significantly, by a factor $\sim$4--7
for the 7.35\,\micron\ line and $\sim$3--4 for the
11.00\,\micron\ line. The match of the DDT and MERGER models to
SN~2021aefx is actually improved with these artificially low collision
strengths, but is significantly degraded for the GCD and DBLEDET
models, for which our computed values yielded [Ni\three] line
strengths comparable to those seen in SN~2021aefx. One cannot
therefore invert the problem and determine collision strengths based
on a goodness-of-fit test of the models to the data, as this requires
a complete control of the underlying systematic uncertainties in the
modelling procedure that we are currently lacking.

\subsection{Possible identification of [Ne\two]\,12.81\,\micron}\label{sect:ne2}

The broad feature at $\sim$12.8\,\micron\ is tentatively associated
with [Ni\two]\,12.73\,\micron\ by \cite{Kwok/etal:2023}. In the MIR
spectrum taken $\sim$2 months later published by
\cite{DerKacy/etal:2023}, the feature has somewhat weakened and is
associated with a combination of [Fe\two]\,12.64\,\micron, [Co\three]
12.68\,\micron, and the aforementioned [Ni\two]\,12.73\,\micron\ line,
though it is difficult to gauge the match of the predicted model flux
to the observed spectrum as both are shown normalised to the peak of
the [Co\three]\,11.89\,\micron\ line (their Fig.~9).  Our DDT and
MERGER models also predict a [Ni\two]\,12.73\,\micron\ line, but it is
at least one order of magnitude too weak compared to the observed
flux. If this feature were entirely due to Ni\two, it should be
matched by our MERGER model which otherwise reproduces the flux levels
of other prominent [Ni\two] lines (at 1.94\,\micron\ and
6.63\,\micron). We therefore propose an alternative identification for
this feature.

\begin{figure}
\centering
\includegraphics{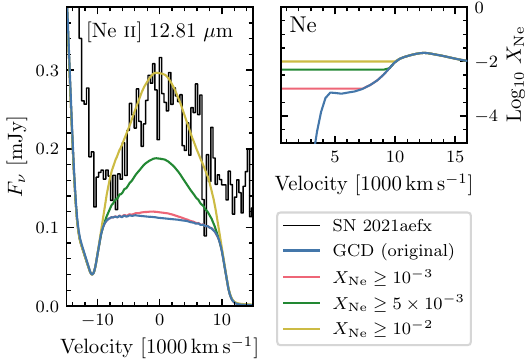}
\caption{\label{fig:ne2}
Impact of the minimum neon mass fraction on the
[Ne\two]\,12.81\,\micron\ line in the GCD model. We can reproduce the
observed 12.8\,\micron\ feature in SN~2021aefx by artificially setting
the minimum Ne mass fraction to $10^{-2}$ throughout the ejecta.}
\end{figure}

Our MERGER model displays a very strong line due to
[Ne\two]\,12.81\,\micron, whose peaked profile reflects the presence
of neon all the way to the innermost region of the ejecta
(Fig.~\ref{fig:comp}). This is a natural expectation of this model as
the central ejecta is dominated by the ashes of the secondary WD which
burns to O and IMEs \citep{Pakmor/etal:2012}. The GCD model also
displays a moderately strong [Ne\two] line. In this model the Ne mass
fraction exceeds $10^{-2}$ above $\sim$10\,000\,\kms\ and Ne$^+$ is
the dominant ionisation stage in the range $\sim$8000--12\,500\,\kms,
resulting in a broad ($\mathit{FWHM}\approx 20\,200$\,\kms) flat-top
profile. This is inconsistent with the 12.8\,\micron\ feature seen in
SN~2021aefx, which is centrally peaked and would suggest that Ne is
present in the inner ejecta. By artificially setting the minimum
Ne mass fraction to $10^{-2}$ all the way to the centre of the GCD
model, we are able to reproduce both the strength and morphology of
the 12.8\,\micron\ feature in SN~2021aefx (Fig.~\ref{fig:ne2}).  Both
models also predict a line due to [Ne\three]\,15.50, which dominates
the $\sim$15.5\,\micron\ feature in the MERGER model. In the other
models the feature results from an overlap of
[Co\two]\,15.46\,\micron\ and [Co\four]\,15.64\,\micron.

The [Ne\two]\,12.81\,\micron\ line was first mentioned in the context
of \sneia\ by \cite{Gerardy/etal:2007} for the low-luminosity
SN~2005df, but it was associated with residual background emission
from the host galaxy. There are hints of narrow emission features at
12.7--12.8\,\micron\ in the MIR spectrum of SN~2014J at $\sim$137\,d
post explosion published by \cite{Telesco/etal:2015}, but the authors
associate them with [Co\three]\,12.68\,\micron\ and
[Ni\two]\,12.73\,\micron. However, the low signal-to-noise ratio and
limited resolution ($\sim$0.15\,\micron, corresponding to
$\sim$3500\,\kms\ at 12.8\,\micron) prevents a thorough
investigation of these features. Interestingly,
\cite{Telesco/etal:2015} do predict a moderately-strong
[Ne\two]\,12.81\,\micron\ line in their reference model (see
their Table~2), but their synthetic spectra are featureless at
this wavelength (their Fig.~4).

Pure deflagration models are also expected to display a prominent
[Ne\two]\,12.81\,\micron\ line, due to the large-scale turbulent
mixing during the explosion that results in the presence of
significant amounts of O, Ne (and unburnt C) in the innermost layers
of the ejecta \citep[see e.g.][]{Fink/etal:2014}. However, the lower 
explosion energy of such models results in significantly narrower line
profiles compared to what is seen in SN~2021aefx.


\subsection{Absence of lines from neutral ions: Impact of clumping}\label{sect:clump}

\begin{figure*}
\centering
\includegraphics{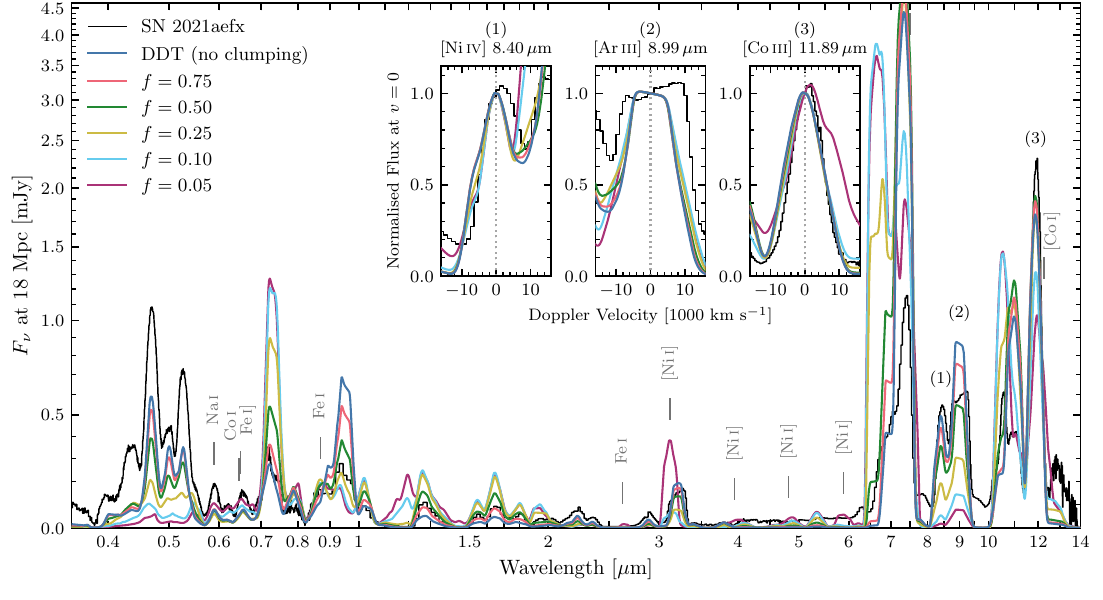}
\caption{\label{fig:comp_spec_n100_clump_21aefx}
Similar to Fig.~\ref{fig:comp_spec_overview_full_21aefx} but for the
ddt\_2013\_N100 (DDT) model with different values for the
volume-filling factor $f$ used to approximate a uniformly-clumped
ejecta \citep[see][]{Wilk/etal:2020}. We show the original
(unclumped) DDT model and clumped models with $f=0.75$, 0.5, 0.25,
0.1, and 0.05. We highlight lines of neutral ions that emerge in the
most clumped models, in particular a strong [Ni\one]\,3.12\,\micron\ line.}
\end{figure*}

Apart from the MERGER model which displays lines of Fe\one\ and
Na\one, none of our other models display lines from neutral ions. In
particular, all our models lack the [Ni\one]\,3.12\,\micron\ line
invoked by \cite{Kwok/etal:2023} to explain the blue edge of the
prominent $\sim$3.2\,\micron\ feature (due to
[Ca\four]\,3.21\,\micron\ with a modest contribution from
[Fe\three]\,3.23\,\micron\ in our models), and the
[Ni\one]\,7.51\,\micron\ line predicted by \cite{DerKacy/etal:2023} to
dominate over the neighbouring [Ni\three]\,7.35\,\micron\ line in
their model for the +323\,d past maximum spectrum\footnote{\cite{DerKacy/etal:2023}
also predict a strong [Ni\one]\,3.12\,\micron\ line and no
[Ca\four]\,3.21\,\micron\ line in their model
(their Table~4), but their observed spectrum of SN~2021aefx only
covers 4--14\,\micron.}. \cite{DerKacy/etal:2023} also predict weaker
forbidden lines due to Co\one\ throughout the NIR-MIR range. For these
lines to show up in our models, the ionisation would have to be
lowered significantly, below that of the MERGER model in the inner
ejecta ($\lesssim 6000$\,\kms, where
$n(\mathrm{Co}^{0+})/n(\mathrm{Co})<10^{-3}$; Fig.~\ref{fig:if_co}).

A clumpy structure may accommodate a broader range of ionisation than
currently predicted in our smooth, quasi-homogeneous ejecta (see
e.g. \citealt{Mazzali/etal:2020}; \citealt{Wilk/etal:2020}). We have
tested the impact of clumping on the DDT model and find that [Ni\one]
lines only start to emerge for a volume-filling factor $f=0.1$, which
results in a ten-fold increase of the density in the clumps
(Fig.~\ref{fig:comp_spec_n100_clump_21aefx}).  The strongest [Co\one]
line at 12.25\,\micron\ is hardly noticeable in the red wing of the
neighbouring [Co\three]\,11.89\,\micron\ line.  The overall fit to the
SN~2021aefx spectrum is significantly degraded, as lines from doubly
and triply-ionised elements become too weak, and those from
singly-ionised elements too strong (although the Fe\two-dominated
feature at $\sim$0.43\,\micron\ remains too weak for all values of the
volume-filling factor; likewise, the Fe\two\,0.517\,\micron\ line
starts to dominate over [Fe\three]\,0.527\,\micron\ for $f<0.5$, but it
contributes at most $\sim$30\% of the flux at that wavelength). This
clearly indicates that if clumping is present, it cannot be uniform as
assumed here. Moreover, it is unclear which physical mechanism, if
any, could lead to such strongly-clumped ejecta in \sneia.

As noted previously in Sect.~\ref{sect:lamrestrict}, the MERGER model
suggests that a lower ionisation would improve the agreement for the
Fe\two\ and Co\two\ lines in the three other models. This
`overionisation problem' in theoretical nebular \snia\ spectra was
investigated in the context of sub-\mch\ models by
\cite{Shingles/etal:2022}. They find that artificially reducing the
non-thermal ionisation rates leads to a better agreement with observed
spectra, although a physical justification for reducing these rates is
yet to be found. We have not investigated this effect in the present
study.


\subsection{Nickel line strengths and stable Ni abundance}\label{sect:ni}

\cite{DerKacy/etal:2023} argue that the strength of the lines of Ni
seen in SN~2021aefx (necessarily resulting from stable isotopes at
this time) are evidence for burning at densities $>5\times10^8$\,\gcc,
which in turn implies a WD mass of at least $\sim$1.2\,\msun. Their
\mch\ model synthesises $\sim$0.06\,\msun\ of stable Ni and
underpredicts the strengths of Ni\two--\textmc{iv} lines in the MIR
spectrum of SN~2021aefx at +323\,d past maximum, which would suggest
an even higher stable Ni mass. The \mch\ DDT model studied here
synthesises $\sim$0.08\,\msun\ of stable Ni and also underpredicts
the strength of Ni\two\ lines in the +255\,d past maximum spectrum of SN~2021aefx,
but overpredicts that of Ni\three\ lines (Sect.~\ref{sect:nicol}),
which points to an overestimation of the ionisation \citep[see
  also][]{Shingles/etal:2022}. The MERGER models also displays strong
Ni lines despite the factor of $\sim$2.4 lower stable Ni abundance
($\sim$0.03\,\msun) compared to the DDT model. This is simply the
result of the lower ionisation in the MERGER model, which leads to
stronger lines of Ni\two\ (and weaker lines of Ni\four) compared to
the DDT model (see also \citealt{Blondin/etal:2022}).

\begin{figure}
\centering
\includegraphics{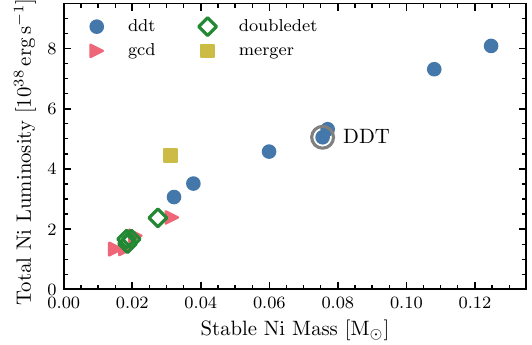}
\caption{\label{fig:fluxni}
Total integrated luminosity in all lines of Ni\one--\textmc{iv} in the
wavelength range 0.35--14\,\micron\ versus
stable Ni mass in our complete model set. There is a strong correlation
between both quantities, with one notable outlier (the MERGER
model). The reference delayed-detonation model (DDT) is highlighted
with a grey circle.
}
\end{figure}

\input{tab4.tex}

When considering the full model set, however, there is a strong
correlation between the total integrated luminosity in Ni\one--\textmc{iv}
lines and the stable Ni mass (Fig.~\ref{fig:fluxni}). The correlation
is mostly driven by the [Ni\three] lines, which account for
$\sim$80--90\% of the total Ni luminosity in our models
(see Table~\ref{tab:lni}). The correlation is significantly degraded when
considering only lines of [Ni\two] (Pearson correlation coefficient
$r=0.61$; cf. last line in Table~\ref{tab:lni}), and highlights the
large uncertainties associated with abundance determinations based on
lines from a single ionisation stage.

The MERGER model remains a notable outlier in Fig.~\ref{fig:fluxni},
and illustrates the impact of ionisation effects when estimating
absolute stable Ni yields. This integrated luminosity is complicated
to determine observationally because of line overlap, which affects
the [Ni\two]\,6.63\,\micron, [Ni\three]\,7.35\,\micron, and
[Ni\three]\,11.00\,\micron\ lines, among others. Despite its weaker
correlation with Ni mass, the
[Ni\four]\,8.40\,\micron\ line is more isolated and could serve as a
reliable tracer of the stable Ni abundance, provided the ionisation balance
is well constrained.

Relating the stable Ni yield to the mass of the exploding WD is
complicated. While \mch\ models tend to synthesise more stable
Ni at a given \nifs\ yield and initial metallicity
\citep{Blondin/etal:2022}, the lower central density of the
\mch\ pulsationally-assisted gravitationally-confined detonation
models of \cite{Lach/etal:2022} results in a stable Ni yield
comparable to sub-\mch\ models ($\sim$0.02--0.03\,\msun; see
Table~\ref{tab:summary}).


\subsection{Width of [Ar\three]\,8.99\,\micron\ line and WD mass}\label{sect:ar}

\cite{Kwok/etal:2023} fit a thick shell emission profile to the broad
($\mathit{FWHM}\approx 23\,700$\,\kms) boxy profile of the [Ar\three]
8.99\,\micron\ line and constrain the shell to be located between
$8700\,\pm\,200$\,\kms\ and $13\,500\,\pm\,300$\,\kms. This suggests a
quasi-absence of Ar in the inner $\sim$9000\,\kms\ of the ejecta, or
`Ar hole', compatible with the near-\mch\ model of
\cite{DerKacy/etal:2023}, who use this as additional evidence in
favour of a high-mass progenitor for SN~2021aefx. The authors argue
that this Ar hole cannot extend to such large velocities in explosions
of lower-mass WDs, based on the sub-\mch\ pure detonation model DET2
of \cite{Hoeflich/Khokhlov:1996} in which this hole only extends to
$\sim$6000\,\kms.

\begin{figure}
\centering
\includegraphics{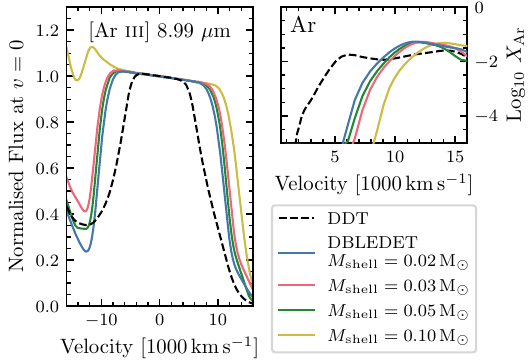}
\caption{\label{fig:ar}
{\it Left:} [Ar\three]\,8.99\,\micron\ line profiles in the \mch\ DDT model
compared to sub-\mch\ double-detonation models with varying He-shell masses (the
reference DBLEDET model has the lowest He-shell mass of 0.02\,\msun).
{\it Right:} Ar distribution, highlighting the larger extent of the Ar
hole in these sub-\mch\ models compared to the \mch\ DDT
model.
}
\end{figure}

In the DBLEDET model from a $\sim$1\,\msun\ progenitor which we
consider here, however, the Ar is located at larger velocities
compared to the \mch\ DDT model (Fig.~\ref{fig:ar}, right panel),
resulting in a broader [Ar\three]\,8.99\,\micron\ line
($\mathit{FWHM}\approx 21\,800$\,\kms\ cf. $\sim$16\,700\,\kms\ for the
DDT model; see left panel in Fig.~\ref{fig:ar}). The same holds for
other sub-\mch\ pure detonation and double detonation models for
1\,\msun\ WDs available in the literature (e.g. 1D models of
\citealt{Shen/etal:2018}; 2D models of \citealt{Townsley/etal:2019};
3D models of \citealt{Gronow/etal:2021}). Therefore, sub-\mch\ models
cannot be excluded based on the large width of the [Ar\three]
8.99\,\micron\ line observed in SN~2021aefx.

The representative DBLEDET model for the class of double-detonation
models consists of a 1\,\msun\ C-O core and a 0.02\,\msun\ He
shell. When considering models with the same core mass but more
massive He shells (0.03--0.1\,\msun), we note that the argon
distribution is indeed shifted to higher
velocities\ (Fig.~\ref{fig:ar}). The FWHM of the
[Ar\three]\,8.99\,\micron\ line increases from
$\sim$21\,800\,\kms\ for $M_\mathrm{shell}=0.02$\,\msun\ to
$\sim$26\,700\,\kms\ for $M_\mathrm{shell}=0.1$\,\msun. The trend is
reversed for the intermediate-mass shells of 0.03\,\msun\ and
0.05\,\msun, but it is worth investigating in a more systematic
manner.


\section{Conclusions}\label{sect:ccl}

We have compared four classes of public state-of-the-art
\snia\ explosion models to nebular observations of SN~2021aefx
covering the full 0.35--14\,\micron\ range \citep{Kwok/etal:2023}. The
input models include \mch\ delayed detonations, pulsationally assisted
gravitationally-confined detonations, sub-\mch\ double detonations and
a violent WD-WD merger. They were selected from the public HESMA
archive to match the \nifs\ yields expected for normal
\sneia\ ($\sim$0.5--0.8\,\msun). The spherically-averaged density
and abundance profiles served as initial conditions to 1D non-LTE
radiative-transfer simulations with CMFGEN at 270\,d post explosion.

Our main result is that no single model emerges as an obvious
candidate for SN~2021aefx based on these data alone. All models
predict the same set of spectroscopic features, all of which have an
observed counterpart. Conversely, all models lack specific
characteristics of the observed spectrum, such as the overall mismatch
in lines of singly-ionised IGEs throughout the optical and infrared.
We tentatively associate the feature at $\sim$12.8\,\micron\ with
[Ne\two]\,12.81\,\micron. If confirmed, this would constitute the
first firm identification of this line in \snia\ ejecta, and would
suggest that neon is present all the way to the innermost region of
the ejecta. We also predict a neon line at longer wavelengths due to
[Ne\three]\,15.55\,\micron, although it overlaps with stronger
neighbouring [Co\two] and [Co\four] lines.

Differences in the abundance structure amongst the different models
(and for different directions in the 3D DDT model) affects the widths
and morphology of some spectral lines, such as the prominent
[Ar\three]\,8.99\,\micron\ line. The predicted blue tilt of its
flat-top profile is due to a relativistic effect in our models. The
observed profile is instead tilted towards the red, indicating a
large-scale asymmetry in the argon distribution
\citep{DerKacy/etal:2023}. We note that line overlap can significantly
skew the line profiles (centred at zero Doppler shift in our 1D
models), and mimic effects normally attributed to ejecta asymmetries.

The largest variations in our model spectra result from differences in
ionisation. A larger density, such as in the inner ejecta of the
MERGER model, results in a lower ejecta temperature and ionisation
state, further enhanced by the increased recombination
rate. Variations in the \cofs\ distribution directly impact the decay
energy deposition rate, of which a large fraction is deposited
(locally in our models) by positrons. Our MERGER model suggest that a
slightly lower ionisation would improve the agreement of the other
(non-MERGER) models with the SN~2021aefx spectrum. However, none of
our models display lines of neutral ions in the NIR or MIR ranges; they
only emerge when a significant amount of clumping is introduced, but
the match to the observations is then severely degraded.

We further show that the large width of the
[Ar\three]\,8.99\,\micron\ line in SN~2021aefx does not invalidate
sub-\mch\ models, contrary to claims made by
\cite{DerKacy/etal:2023}. The double-detonation models from a
1\,\msun\ progenitor we consider here display the largest
[Ar\three]\,8.99\,\micron\ lines, reflecting the larger extent of the
Ar hole in these lower-mass ejecta.  Moreover, while the total
integrated luminosity in lines of stable Ni strongly correlates with the
stable Ni mass, the MERGER model was found to be a clear outlier due
its lower ionisation (see also \citealt{Blondin/etal:2022b}). Provided
the ionisation balance is well constrained, the isolated
[Ni\four]\,8.40\,\micron\ line could be used to estimate the stable Ni
yield in observed \sneia. Connecting this yield to the progenitor mass
requires an accurate knowledge of the explosion model, as some
\mch\ models (e.g. pulsationally-assisted gravitationally-confined
detonations) synthesise similar amounts of stable Ni compared to
sub-\mch\ models.

As for all radiative-transfer simulations, our results are affected by
uncertainties in the atomic data. We could find no published
collisional strengths for low-lying forbidden transitions within the
lowest $^3$F term of Ni\three, and present them here for the first
time. Their values differ significantly from the commonly-used
approximations of \cite{Axelrod:1980}, and the impact on the strength
of the [Ni\three] lines at 7.35\,\micron\ and 11.00\,\micron\ is
significant. Thus despite the low ejecta densities involved,
collisions largely dictate the excitation level and ionisation state
of the plasma (including non-thermal collisional ionisation by
Compton-scattered electrons; see \citealt{Shingles/etal:2022}). It is
clear that more accurate collisional data are needed to correctly
interpret these and future infrared observations of \sneia.

Throughout this study we have implicitly assumed that the input
hydrodynamical models of the explosion provide accurate initial
conditions for our radiative-transfer calculations. It is however
possible that part of the mismatch with the observations can be
attributed to uncertainties in the explosion models themselves, in
particular concerning the predicted nucleosynthetic yields and
chemical abundance profiles in the ejecta \citep[see
  e.g.][]{Bravo:2020}.

Inferring reliable constraints on the progenitor mass and explosion
mechanism of \sneia\ remains a formidable challenge, even with the
extended wavelength coverage and exquisite data quality offered by
JWST. Combined with early-time and optical observations,
nebular infrared spectra of \sneia\ provide an additional validation
criterion when comparing a diverse set of \snia\ models to
observations.


\begin{acknowledgements}
  This paper is dedicated to Prof. Tom Marsh (1961-2022), who was
  S.B.'s lecturer and Bachelor thesis supervisor at the
  University of Southampton during 1998--2002.\\
  S.B. acknowledges support from the Alexander von Humboldt foundation and
  thanks Sherry Suyu and her group at the Technische Universität München
  (TUM) for their hospitality.
  We acknowledge useful discussions with
  Chris Ashall,
  Andreas Floers,
  Peter Hoeflich,
  Griffin Hosseinzadeh,
  Saurabh Jha,
  Anders Jerkstrand,
  Lindsey Kwok, and
  Ruediger Pakmor.
  Many thanks to:
  Manuel Bautista for providing unpublished collisional strengths for [Ni\three] based on \cite{Bautista:2001};
  Lindsey Kwok for sending us the combined spectrum of SN~2021aefx ahead of publication;
  Elena Sabbi and Brent Tully for confirming the error on the estimated distance to NGC 1566;
  Ivo Seitenzahl for extracting radial profiles from the 3D delayed-detonation model N100 of \cite{Seitenzahl/etal:2013};
  Steve Shore, the referee of this paper, for his detailed review and valuable comments that led to the detailed comparison of the 3D vs. spherically-averaged yields in Appendix~\ref{sect:3d1d}.\\
  This work was supported by the `Programme National de Physique
  Stellaire' (PNPS) of CNRS/INSU cofunded by CEA and CNES.
  This research has made use of computing facilities operated by
  CeSAM data centre at LAM, Marseille, France.
  D.J.H. acknowledges partial support for this work through the NASA
  astrophysical theory grant 80NSSC20K0524 and the STScI theory grant
  HST-AR-16131.001-A. STScI is operated by the Association of
  Universities for Research in Astronomy, Inc., under NASA contract
  NAS 5-26555.  
  This research was supported by the Munich Institute for Astro-,
  Particle and BioPhysics (MIAPbP) which is funded by the Deutsche
  Forschungsgemeinschaft (DFG, German Research Foundation) under
  Germany's Excellence Strategy – EXC-2094 – 390783311.
  This work made use of the Heidelberg Supernova Model Archive
  (HESMA; \url{https://hesma.h-its.org}) and of the NIST Atomic
  Spectra Database (\url{https://www.nist.gov/pml/atomic-spectra-database}).
\end{acknowledgements}

\noindent
\textit{Note added in proof.}
Only two days prior to the acceptance of this paper,
\cite{Kwok/etal:2023b} confirmed the presence of a strong
centrally-peaked line due to [Ne\two]\,12.81\,\micron\ in their
nebular spectrum of the Type Ia SN~2022pul, which they associate
with a violent merger event, in line with the predictions of this
paper.


\bibliographystyle{aa} 
\bibliography{ms_21aefx_jwst,atomic} 


\clearpage

\begin{appendix}



\section{3D vs. spherically-averaged yields}\label{sect:3d1d}

In Table~\ref{tab:3d1d} we compare the total mass and yields of
selected isotopes and elements of the original 3D models
($M_\mathrm{3D}$) and their spherically-averaged versions
($M_\mathrm{1D}$) available on HESMA.
The spherical averaging causes a systematic overestimate of
the total mass by 3--8\% and a difference of up to $\pm$10\% for
several isotopic and elemental yields compared to the original 3D
models.

In Table~\ref{tab:3d1d_n100} we repeat this comparison for the 
original 3D model N100 of \cite{Seitenzahl/etal:2013} and 1D models
reconstructed from the three orthogonal axes of the 3D
Cartesian grid in both positive (\{x,y,z\}pos) and negative (\{x,y,z\}neg)
directions. We report two sets of quantities, one based on the original
(unscaled) density profile, the other based on the density profile
rescaled to match the total mass of the original
3D model (1.4\,\msun). Here the impact on the yields of specific
isotopes and elements is in some cases larger than 50\%, as is the case
for \nife\ in the model corresponding to the positive
$\hat{\mathbf{z}}$ direction (+81\%), or Ca in the model corresponding
to the positive $\hat{\mathbf{y}}$ direction (+87\%).

The yields other than $^{56}\mathrm{Ni}_{t=0}$ are given at 270\,d
post explosion. In the spherically-averaged models these yields are
simply integrated over the ejecta that are used as initial conditions
for our 1D radiative-transfer calculations. The 3D yields at 270\,d
were derived from the final decayed abundances at $\sim$2\,Gyr
reported in the HESMA \texttt{abundances.dat} files, corrected for
radioactive decays based on the abundances of radioactive isotopes at
the end of the 3D simulation ($\sim$100\,s post explosion).

\newpage

As an example, we consider the iron abundance in the ddt\_2013\_N100
model, whose final decayed abundance is $7.40 \times
10^{-1}$\,\msun. This includes $^{56}$Fe from the
$^{56}$Ni$\rightarrow ^{56}$Co$\rightarrow ^{56}$Fe decay chain,
$^{57}$Fe from the $^{57}$Ni$\rightarrow ^{57}$Co$\rightarrow ^{57}$Fe
decay chain, and excludes $^{55}$Fe from the $^{55}$Co$\rightarrow
^{55}$Fe$\rightarrow ^{55}$Mn. To obtain the abundance at 270\,d post
explosion, we therefore subtracted the \cofs\ abundance at
270\,d ($5.81 \times 10^{-2}$\,\msun) as well as the $^{57}$Co abundance at
270\,d ($9.48 \times 10^{-3}$\,\msun), and added the $^{55}$Fe abundance at
270\,d ($1.10 \times 10^{-2}$\,\msun), resulting in an iron yield of
$6.83 \times 10^{-2}$\,\msun\ at 270\,d. In principle we would also
need to subtract the \nifs\ and $^{57}$Ni abundances, but these are
essentially zero at 270\,d. Likewise, we added the \cofs\ and $^{57}$Co
abundances at 270\,d to the final decayed Co abundance to obtain its
value at 270\,d.

To compute the stable Ni abundance at 270\,d, however, we would need
to know the initial abundances of several isotopes (such as $^{60}$Zn
and $^{60}$Cu, which decay to $^{60}$Ni) that are not included in the
HESMA \texttt{abundances.dat} files. Instead, we report the
\nife\ abundance, which constitutes the dominant stable isotope of Ni
($>$60\% in the double-detonation models, $>$80\% in the GCD models,
and $>$90\% in the delayed-detonation and violent merger models; see
also \citealt{Blondin/etal:2022}).


\input{tabA1.tex}

\clearpage

\input{tabA2.tex}

\clearpage



\section{Atomic data}\label{sect:atomic_data}

Table~\ref{tab:cmfgen_atoms} give the number of levels (both
super-levels and full levels; see \citealt{Hillier/Miller:1998} for
details) for the model atoms used in the radiative-transfer
calculations presented in this paper.  $N_\mathrm{SL}$ refers to the
number of super levels used for the solution of the rate equations,
and $N_\mathrm{full}$ refers the number of full levels used to solve
the transfer equation and compute the observed spectrum.  We report
the uppermost level for each ion treated in the fourth column.  `W'
refers to states in which higher $\ell$ states (usually f or higher)
have been combined into a single level.  In the last column we give
the number of bound-bound transitions in the model ion taking into
account all $N_\mathrm{full}$ levels.  We considered ionisations to and
recombinations from the ground state of the next ionisation stage for
all elements (i.e. \ion{He}{3}, \ion{C}{4}, \ion{N}{4}, \ion{O}{4},
\ion{Ne}{4}, \ion{Na}{2}, \ion{Mg}{4}, \ion{Al}{4}, \ion{Si}{5},
\ion{S}{5}, \ion{Ar}{4}, \ion{Ca}{5}, \ion{Sc}{4}, \ion{Ti}{4},
\ion{Cr}{5}, \ion{Mn}{4}, \ion{Fe}{6}, \ion{Co}{5}, and \ion{Ni}{6}).

Below we give references to the atomic data used (oscillator and
collision strengths, with a special attention to forbidden transitions)
for all the ions for which we report line identifications in
Table~\ref{tab:lineid}: Ne\two--\textmc{iii}, Na\one, S\two--\textmc{iv},
Ar\two--\textmc{iii}, Ca\two, Ca\four, Fe\one--\textmc{iii}, Fe\five,
Co\two--\textmc{iv}, and Ni\two--\textmc{iv}. We refer the reader to
earlier papers for additional details on the atomic data used by
CMFGEN (e.g. \citealt{Hillier:2011}, \citealt{Blondin/etal:2022b}).

\begin{itemize}
  
\item \ion{Ne}{2}: data for forbidden transitions are from
  \cite{Mendoza1983_col};

\item \ion{Ne}{3}: data for forbidden transitions are very similar
  (mostly within a few per cent or identical) to those reported in
  \cite{Mendoza1983_col};
  
\item \ion{Na}{1}: oscillator strengths are from the Opacity Project
  \citep{Seaton1987_OP}. Collision rates were computed using van
  Regemorter's approximation \citep{vanReg62_col};
  
\item \ion{S}{2}: oscillator and collision strengths for forbidden
  transitions are from \cite{TZ10_SII};
  
\item \ion{S}{3}: data for forbidden and intercombination
  (semi-forbidden) transitions are from \cite{Tay97_SIII_forb},
  \cite{Mendoza1983_col}, and \cite{Hua85_Si_seq}. Collision strengths
  are from \cite{TG99_SIII_col};
  
\item \ion{S}{4}: data for intercombination (semi-forbidden)
  transitions are from the compilation of \cite{Mendoza1983_col}.
  Collision strengths are from \cite{Tay00_SIV_col};
  
\item \ion{Ar}{2}: ground-state fine-structure cross sections are from
  \cite{PB95_Cl_seq}. Collision strengths are from
  \cite{TH96_ArII_col};
  
\item \ion{Ar}{3}: forbidden transition probabilities are from
  \cite{1983MNRAS.202..981M}. Collision strengths are from
  \cite{GMZ95_Si_S_col};
  
\item \ion{Ca}{2}: unfortunately we were not able to trace the origin
  of the forbidden transition probabilities for \ion{Ca}{2}. The
  transition probabilities for the
  $^{2}\text{S}_{1/2}$-$^{2}\text{D}_{3/2}$ (1-2) and
  $^{2}\text{S}_{1/2}$-$^{2}\text{D}_{5/2}$ (1-3)
  transitions making up the 7300\,\AA\ doublet are both set to
  $A_\mathrm{ul}=1.15$\,s$^{-1}$ in our atomic data set, slightly lower than
  the value reported in the National Institute of Standards and Technology (NIST) Atomic
  Spectra Database\footnote{\label{nisturl}\url{https://www.nist.gov/pml/atomic-spectra-database}}
  (ASD; \citealt{NIST_5p10}), namely $A_\mathrm{ul}=1.3$\,s$^{-1}$
  \citep{Osterbrock1951}. More recent calculations result in
  significantly lower values (e.g. \citealt{Kaur2021} report
  $0.805\,\pm\,0.007$\,s$^{-1}$ and $0.827\,\pm\,0.008$\,s$^{-1}$ for the 1-2
  and 1-3 transitions, respectively), in line with experimental
  measurements ($0.833\,\pm\,0.007$\,s$^{-1}$ and
  $0.856\,\pm\,0.005$\,s$^{-1}$; \citealt{Barton2000}); we will use these
  latter experimental values in future work. We ran test calculations
  using the DDT and MERGER models, which showed a very modest impact.
  Collision strengths are from \cite{MBB07_CaII_col};
  
\item \ion{Ca}{4}: ground term transition rates are from
  \cite{Nahar23_CaIV}, as are the collision strengths;
  
\item \ion{Fe}{1}: data for forbidden transitions were downloaded from
  NIST \citep{NIST_V5p2} on 7$^\mathrm{th}$ September 2016 (the
  original data are from \citealt{1995ApJ...441L..97B}). Collision
  strengths between the two lowest terms (10 levels) are from
  \cite{PB_FeI_col};
  
\item \ion{Fe}{2}: forbidden transition probabilities among the first 63
  levels are from \cite{2011A&A...536A..74D}. Collision strengths are
  from \cite{ZP94_FeII_col} and \cite{ZP95_FeII_col};
  
\item \ion{Fe}{3}: data for forbidden transitions are from
  \cite{NP96_FeIII}. Collision strengths are from
  \cite{Zha96_FeIII_col};
  
\item \ion{Fe}{5}: oscillator strengths were computed by R.~Kurucz
  \citep{Kurucz2009_ATD} and obtained through his
  website\footnote{\label{kuruczurl}\url{http://kurucz.harvard.edu}}, complemented with
  data from NIST\cref{nisturl}. Transition probabilities for M1 and E2
  transitions are from \cite{2017ADNDT.114....1A};
  
\item \ion{Co}{2}: oscillator strengths were computed by R.~Kurucz
  \citep{Kurucz2009_ATD} and obtained through his
  website\cref{kuruczurl} (calculation date:
  24$^\mathrm{th}$ November 2006). Oscillator strengths for
  forbidden transitions are from \cite{Qui98_CoII}. Collision
  strengths are from \cite{Storey/etal:2016};
  
\item \ion{Co}{3}: oscillator strengths were computed by R.~Kurucz
  \citep{Kurucz2009_ATD} and obtained through his website\cref{kuruczurl} in
  2009. Transition probabilities for forbidden lines in the 3d7 ground
  configuration are from \cite{HRU83_CoIII_forbid}. Collision
  strengths are from \cite{2016MNRAS.459.2558S};
  
\item \ion{Co}{4}: oscillator strengths were computed by R.~Kurucz
  \citep{Kurucz2009_ATD} and obtained through his website\cref{kuruczurl} in
  2009. Transition probabilities for forbidden lines are scaled values
  for the isoelectronic ion Fe\three\ \citep{Qui96_[FeIII]}. Collision
  strengths are also based on Fe\three\ values from
  \cite{Zha96_FeIII_col};
  
\item \ion{Ni}{2}: oscillator strengths were computed by R.~Kurucz
  \citep{Kurucz2009_ATD} and obtained through his website\cref{kuruczurl} in
  2009. Data for forbidden transitions are from
  \cite{QL96_FeI_FeII_col}. Collision strengths are from
  \cite{Bau04_NiII_col};
  
\item \ion{Ni}{3}: oscillator strengths were computed by R.~Kurucz
  \citep{Kurucz2009_ATD} and obtained through his website\cref{kuruczurl} in
  2000 (newer calculations are available). Forbidden line transition
  probabilities are from \cite{Gar58_nk_forbid}. Collision strengths
  are from this paper (see Sect.~\ref{sect:nicol} and
  Appendix~\ref{sect:nkiiicol});
  
\item \ion{Ni}{4}: data for forbidden transitions among low-lying
  levels are from NIST\cref{nisturl} (\citealt{NIST_5p10}; original
  data from \citealt{HRU83_CoIII_forbid}). Collision strengths are
  from \cite{2019MNRAS.483.2154F}.
  
\end{itemize}


\clearpage

\begin{table}
\centering
\footnotesize
\caption{Model atoms used in the CMFGEN calculations.} 
\label{tab:cmfgen_atoms}
\begin{tabular}{lrrlr}
\hline\hline\\[-1.8ex]
Ion & $N_\mathrm{SL}$ & $N_\mathrm{full}$ & Last level & $N_\mathrm{lines}$  \\
\hline\\[-1.8ex]
\ion{He}{1}    &        40  &          51  &   $n=11$                                                   &             716     \\ 
\ion{He}{2}    &        13  &          30  &   $n=30$                                                   &             813     \\ 
\ion{C}{1}     &        14  &          26  &   2s 2p$^3$ $^3$P$^{\mathrm{o}}$                         &             229     \\ 
\ion{C}{2}     &        14  &          26  &   2s$^2$ 4d $^2$D                                   &             181     \\ 
\ion{C}{3}     &        62  &         112  &   2s 8f$^1$F$^{\mathrm{o}}$                              &          1\,788     \\ 
\ion{N}{1}     &        44  &         104  &   2s$^2$ 2p$^2$($^3$P)5f $^2$F$^{\mathrm{o}}$            &          1\,678     \\ 
\ion{N}{2}     &        23  &          41  &   2s$^2$ 2p 3d $^1$P$^{\mathrm{o}}$                      &             276     \\ 
\ion{N}{3}     &        25  &          53  &   2s 2p($^3$P$^{\mathrm{o}}$)3d $^4$D$^{\mathrm{o}}$          &             523     \\ 
\ion{O}{1}     &        21  &          51  &   2s$^2$ 2p$^3$($^4$S$^{\mathrm{o}}$)4f $^3$F            &             439     \\ 
\ion{O}{2}     &        54  &         123  &   2s$^2$ 2p$^2$($^3$P)4f $^2$F$^{\mathrm{o}}$            &          2\,677     \\ 
\ion{O}{3}     &        44  &          86  &   2s 2p$^2$($^4$P)3p $^3$P$^{\mathrm{o}}$                &          1\,017     \\ 
\ion{Ne}{1}    &        78  &         155  &   2s$^2$ 2p$^5$($^2$P$_{3/2}$)6f $^2$[\sfrac{7}{2}]    &          3\,581     \\ 
\ion{Ne}{2}    &        22  &          91  &   2s$^2$ 2p$^4$($^3$P)4d $^2$P                      &          2\,143     \\ 
\ion{Ne}{3}    &        32  &          80  &   2s$^2$ 2p$^3$($^4$S$^{\mathrm{o}}$)4f $^3$F            &             860     \\ 
\ion{Na}{1}    &        22  &          71  &   30w$^2$W                                          &          3\,128     \\ 
\ion{Mg}{2}    &        31  &          80  &   30w $^2$W                                         &          3\,863     \\ 
\ion{Mg}{3}    &        31  &          99  &   2p$^5$ 7s $^1$P$^{\mathrm{o}}$                         &          1\,526     \\ 
\ion{Al}{2}    &        26  &          44  &   3s 5d$^1$D                                        &             333     \\ 
\ion{Al}{3}    &        27  &          60  &   10z $^2$Z                                         &          1\,191     \\ 
\ion{Si}{2}    &        31  &          59  &   3s$^2$ 7h $^2$H$^{\mathrm{o}}$                       &           1\,059    \\ 
\ion{Si}{3}    &        33  &          61  &   3s 5g$^3$G                                        &             615     \\ 
\ion{Si}{4}    &        37  &          48  &   10f $^2$F$^{\mathrm{o}}$                               &             816     \\ 
\ion{S}{2}     &        56  &         324  &   3s 3p$^3$($^5$S$^{\mathrm{o}}$)4p $^6$P                &         16\,965     \\ 
\ion{S}{3}     &        48  &          98  &   3s 3p$^2$($^2$D)3d $^3$P                          &          1\,714     \\ 
\ion{S}{4}     &        27  &          67  &   3s 3p($^3$P$^{\mathrm{o}}$)4p $^2$D                    &           1\,091    \\ 
\ion{Ar}{1}    &        56  &         110  &   3s$^2$ 3p$^5$($^2$P$_{3/2}$)7p $^2$[\sfrac{3}{2}]   &          3\,030     \\ 
\ion{Ar}{2}    &       134  &         415  &   3s$^2$ 3p$^4$($^3$P$_{1}$)7i $^2$[6]                &         40\,224     \\ 
\ion{Ar}{3}    &        32  &         346  &   3s$^2$ 3p$^3$($^2$D$^{\mathrm{o}}$)8s $^1$D$^{\mathrm{o}}$  &         13\,681     \\ 
\ion{Ca}{2}    &        21  &          77  &   3p$^6$ 30w $^2$W                                  &          3\,365     \\ 
\ion{Ca}{3}    &        16  &          40  &   3s$^2$ 3p$^5$ 5s $^1$P$^{\mathrm{o}}$                  &             210     \\ 
\ion{Ca}{4}    &        18  &          69  &   3s 3p$^5$($^3$P$^{\mathrm{o}}$)3d $^4$D$^{\mathrm{o}}$      &             647     \\ 
\ion{Sc}{2}    &        38  &          85  &   3p$^6$ 3d 4f $^1$P$^{\mathrm{o}}$                      &          1\,905     \\ 
\ion{Sc}{3}    &        25  &          45  &   7h $^2$H$^{\mathrm{o}}$                                &             454     \\ 
\ion{Ti}{2}    &        37  &         152  &   3d$^2$($^3$F)5p $^4$D$^{\mathrm{o}}$                   &         13\,413     \\ 
\ion{Ti}{3}    &        33  &         206  &   3d 6f $^3$H$^{\mathrm{o}}$                             &          9\,485     \\ 
\ion{Cr}{2}    &        28  &         196  &   3d$^4$($^3$G)4p x$^4$G$^{\mathrm{o}}$                  &          8\,249     \\ 
\ion{Cr}{3}    &        30  &         145  &   3d$^3$($^2$D2)4p $^3$D$^{\mathrm{o}}$                  &          4\,996     \\ 
\ion{Cr}{4}    &        29  &         234  &   3d$^2$($^3$P)5p $^4$P$^{\mathrm{o}}$                   &         12\,569     \\ 
\ion{Mn}{2}    &        25  &          97  &   3d$^4$($^5$D)4s$^2$ c$^5$D                        &             464     \\ 
\ion{Mn}{3}    &        30  &         175  &   3d$^4$($^3$G)4p y$^4$H$^{\mathrm{o}}$                  &          6\,292     \\ 
\ion{Fe}{1}    &        44  &         136  &   3d6($^5$D)4s 4p x$^5$F$^{\mathrm{o}}$                  &          3\,934     \\ 
\ion{Fe}{2}    &       228  &      2\,698  &   3d$^5$($^4$F)4s 4p b$^4$G$^{\mathrm{o}}$               &     1\,062\,164     \\ 
\ion{Fe}{3}    &        96  &      1\,001  &   3d$^5$($^6$S)6f $^7$F$^{\mathrm{o}}$                   &        138\,327     \\ 
\ion{Fe}{4}    &       100  &      1\,000  &   3d$^4$($^3$G)4f $^4$P$^{\mathrm{o}}$                   &        144\,005     \\ 
\ion{Fe}{5}    &       139  &      1\,000  &   3d2($^3$P)4s 4p $^3$S$^{\mathrm{o}}$                   &        144\,265     \\ 
\ion{Co}{1}    &        52  &         327  &   3d$^7$ 4s($^5$F)4d f$^4$H                         &         23\,067     \\ 
\ion{Co}{2}    &       112  &      1\,005  &   3d$^7$($^4$P)4f b$^5$G$^{\mathrm{o}}$                  &        205\,747     \\ 
\ion{Co}{3}    &        88  &      1\,075  &   3d$^6$($^3$H)4f $^4$G$^{\mathrm{o}}$                   &        154\,662     \\ 
\ion{Co}{4}    &        56  &      1\,000  &   3d$^5$($^2$D)5s $^1$D                             &        139\,240     \\ 
\ion{Ni}{1}    &        56  &         301  &   3d$^9$($^2$D$_{5/2}$)7d $^2$[\sfrac{7}{2}]           &         22\,065     \\ 
\ion{Ni}{2}    &        59  &      1\,000  &   3d$^8$($^3$F)7f $^4$I$^{\mathrm{o}}$                   &        103\,224     \\ 
\ion{Ni}{3}    &        47  &      1\,000  &   3d$^7$($^2$D)4d $^3$S                               &        132\,677     \\ 
\ion{Ni}{4}    &        54  &      1\,000  &   3d$^6$($^5$D)6p $^6$F$^{\mathrm{o}}$                   &        145\,745     \\ 
\ion{Ni}{5}    &        54  &      1\,000  &   3d$^4$($^1$I)4s$^2$ $^1$I                         &        151\,806     \\ 
\hline\\[-2ex]
\textbf{Total} & \textbf{2\,613}  & \textbf{17\,996}  &                                        & \textbf{2\,739\,134}     \\
\hline
\end{tabular}
\flushleft
\justifying
\textbf{Notes.}
Due to a $gf$ cut (level dependent, $gf > 5\times10^{-4}$) only
672\,614 lines were included in the non-LTE calculations of the level
populations. 1\,373\,191 lines were included when computing the
observed spectrum.
\end{table}

\clearpage


\section{Ionisation fractions}\label{sect:if}

\begin{figure}[h]
\centering
\begin{minipage}{\textwidth}
\includegraphics{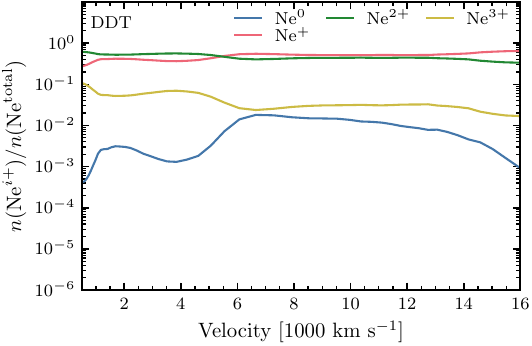}\hspace{0.3cm}
\includegraphics{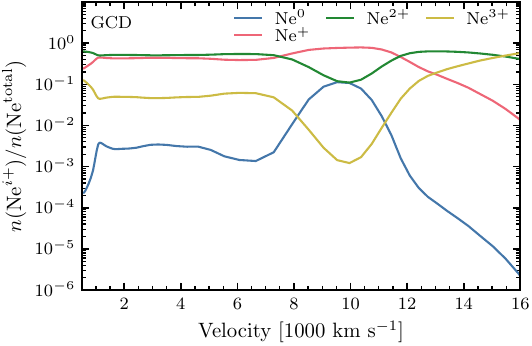}\vspace{-0.5cm}
\includegraphics{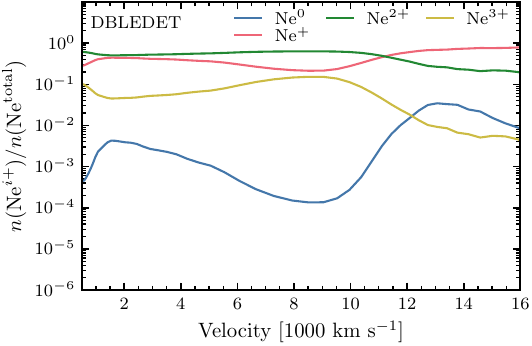}\hspace{0.3cm}
\includegraphics{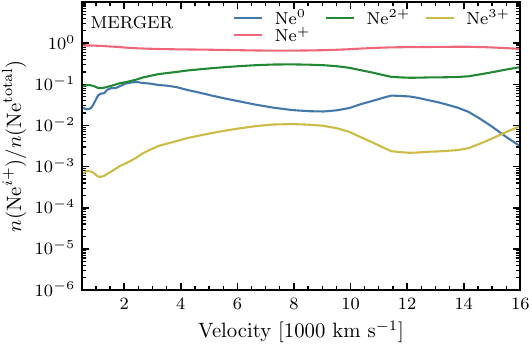}
\caption{
Ratio of the number density of different ionisation states of neon to
the total Ne number density in our reference model set.
}\label{fig:if_ne}
\end{minipage}
\end{figure}

\begin{figure*}
\centering
\includegraphics{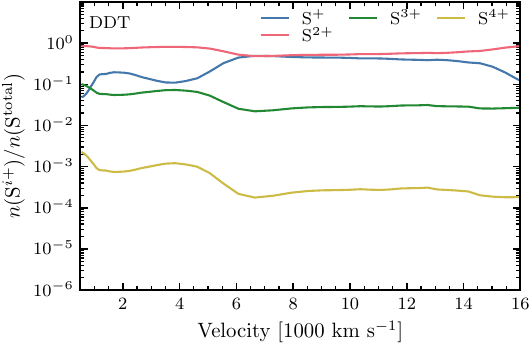}\hspace{0.3cm}
\includegraphics{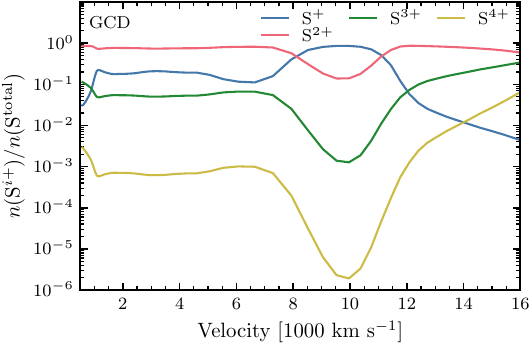}\vspace{-0.5cm}
\includegraphics{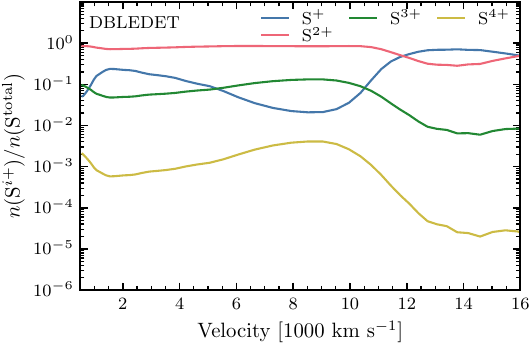}\hspace{0.3cm}
\includegraphics{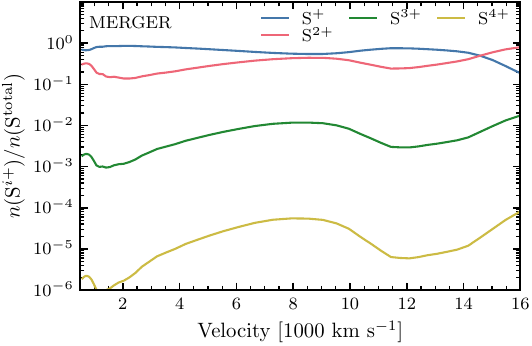}
\caption{\label{fig:if_s}
Same as Fig.~\ref{fig:if_ne} but for sulfur.
}
\end{figure*}

\begin{figure*}
\centering
\includegraphics{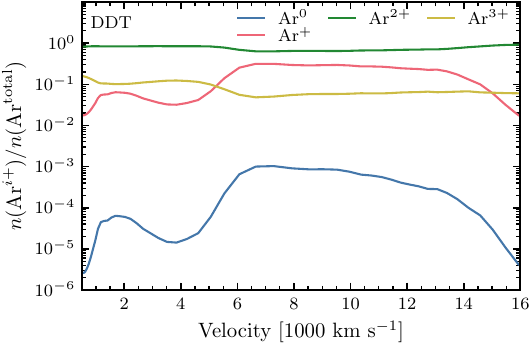}\hspace{0.3cm}
\includegraphics{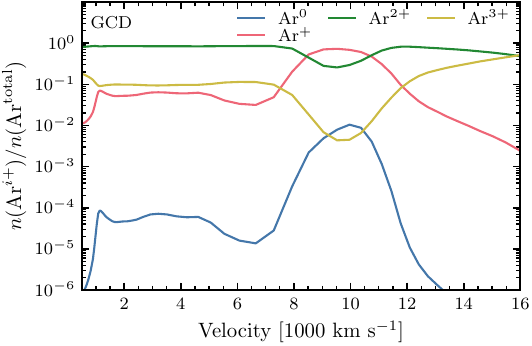}\vspace{-0.5cm}
\includegraphics{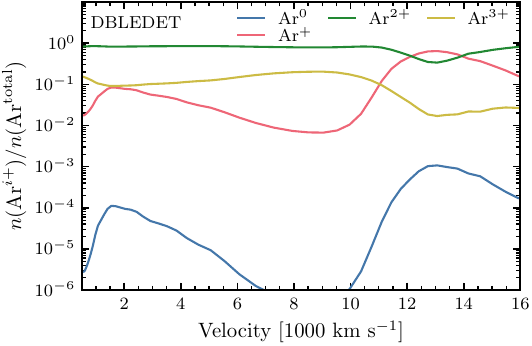}\hspace{0.3cm}
\includegraphics{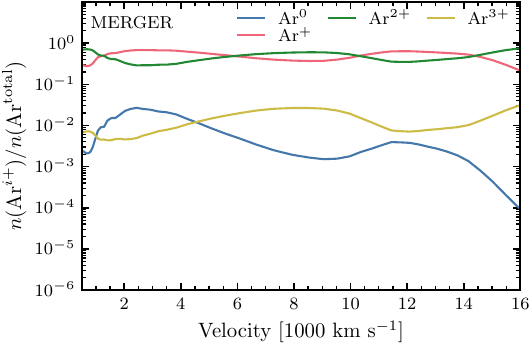}
\caption{\label{fig:if_ar}
Same as Fig.~\ref{fig:if_ne} but for argon.
}
\end{figure*}

\begin{figure*}
\centering
\includegraphics{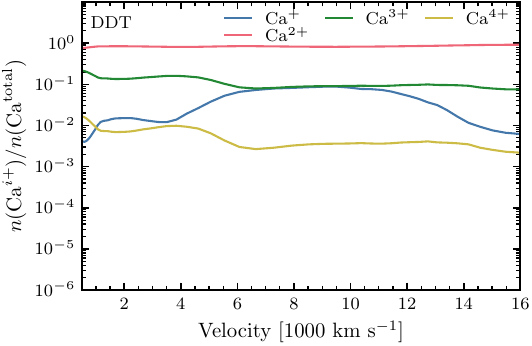}\hspace{0.3cm}
\includegraphics{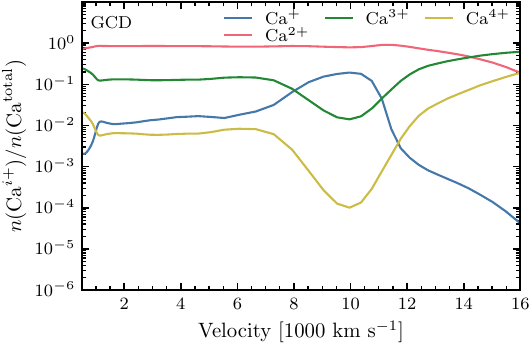}\vspace{-0.5cm}
\includegraphics{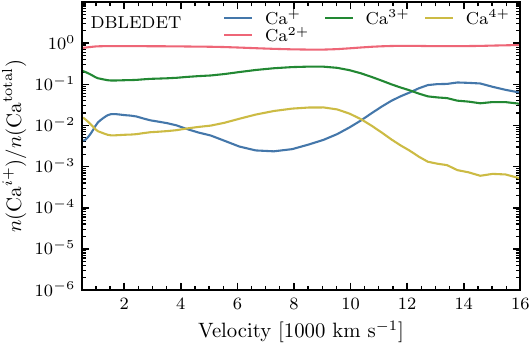}\hspace{0.3cm}
\includegraphics{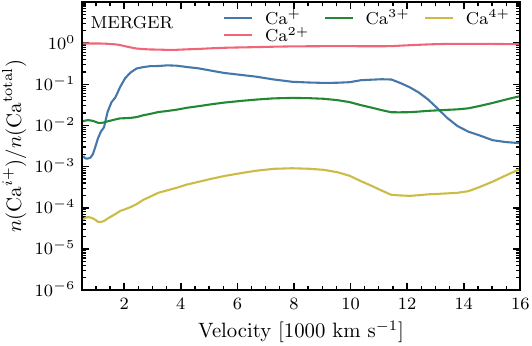}
\caption{\label{fig:if_ca}
Same as Fig.~\ref{fig:if_ne} but for calcium.
}
\end{figure*}

\begin{figure*}
\centering
\includegraphics{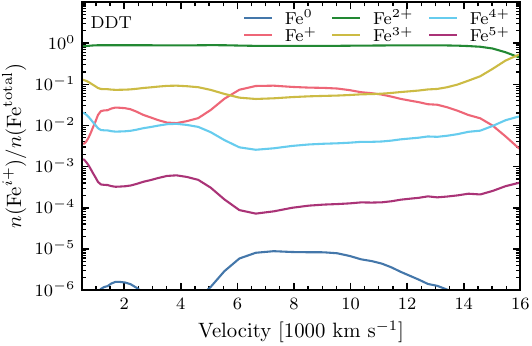}\hspace{0.3cm}
\includegraphics{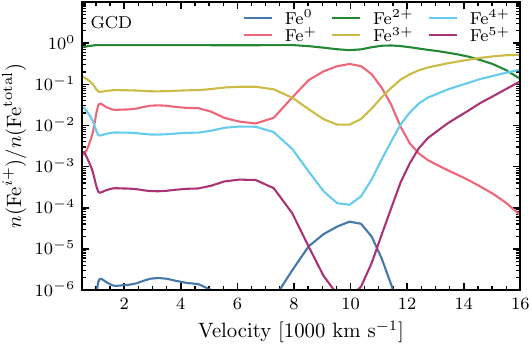}\vspace{-0.5cm}
\includegraphics{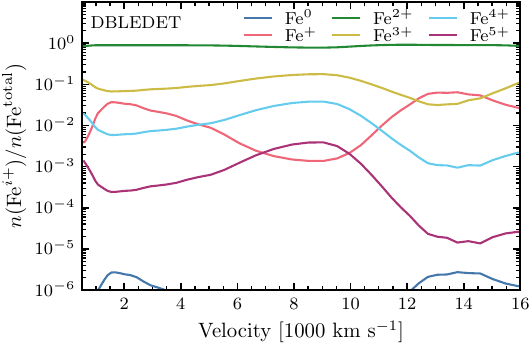}\hspace{0.3cm}
\includegraphics{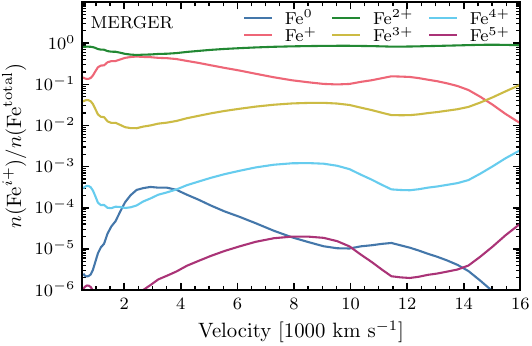}
\caption{\label{fig:if_fe}
Same as Fig.~\ref{fig:if_ne} but for iron.
}
\end{figure*}

\begin{figure*}
\centering
\includegraphics{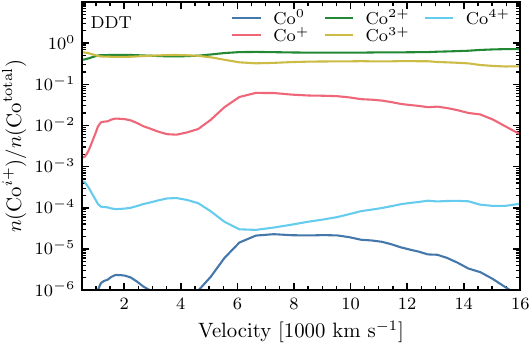}\hspace{0.3cm}
\includegraphics{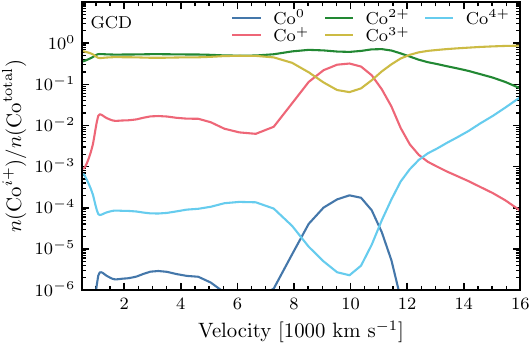}\vspace{-0.5cm}
\includegraphics{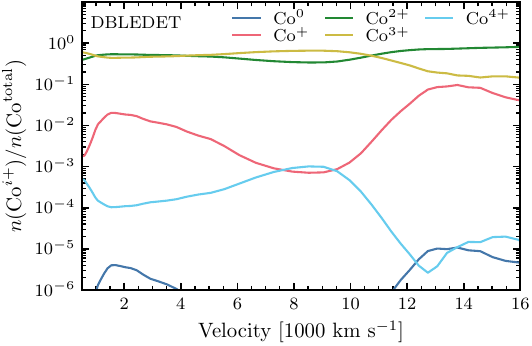}\hspace{0.3cm}
\includegraphics{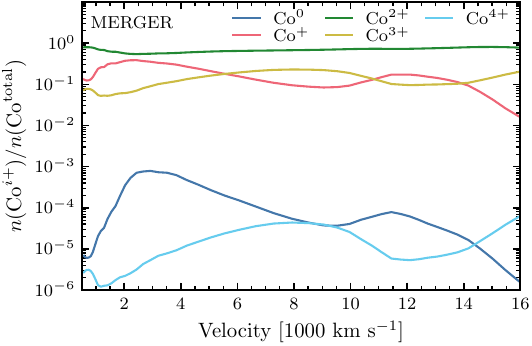}
\caption{\label{fig:if_co}
Same as Fig.~\ref{fig:if_ne} but for cobalt.
}
\end{figure*}

\begin{figure*}
\centering
\includegraphics{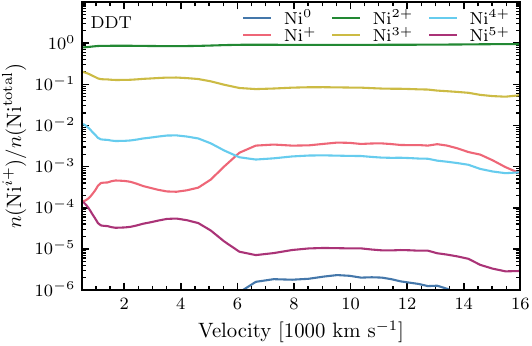}\hspace{0.3cm}
\includegraphics{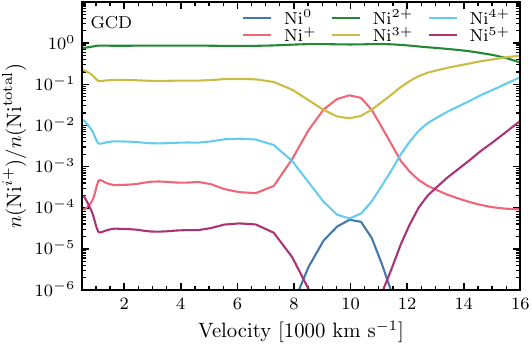}\vspace{-0.5cm}
\includegraphics{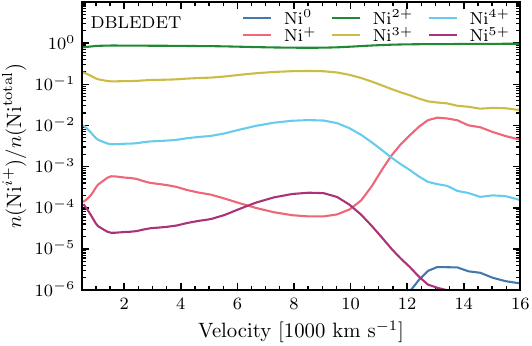}\hspace{0.3cm}
\includegraphics{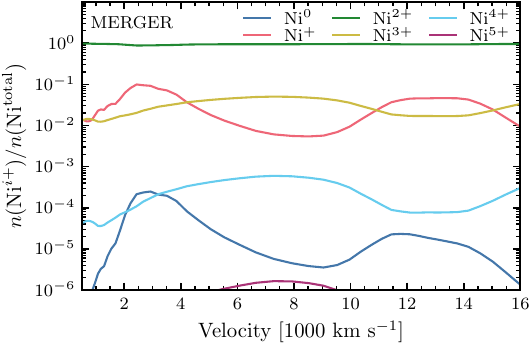}
\caption{\label{fig:if_ni}
Same as Fig.~\ref{fig:if_ne} but for nickel.
}
\end{figure*}

In Figs.~\ref{fig:if_ne}--\ref{fig:if_ni} we show the ratio of the
number density of different ionisation states to the total element
number density in our reference model set for Ne, S, Ar, Ca, Fe, Co,
and Ni.

\clearpage


\section{Statistical estimators}\label{sect:stats}

We used several statistical estimators to gauge the fit quality of
a given model with SN~2021aefx (see Table~\ref{tab:stats}). In what
follows $N_\lambda$ is the number of wavelength bins in the observed
spectrum, $F_{\mathrm{obs},i}$ is the observed flux in the $i$'th
wavelength bin, and $F_{\mathrm{mod},i}$ is the model flux in the same
bin (the model spectra were resampled to match the observed wavelength
scale).

The median absolute deviation (MAD) is calculated as:

\begin{equation}
  \label{eq:mad}
  \mathrm{MAD} = \frac{1}{N_\lambda} \sum_{i=1}^{N_\lambda} |
  F_{\mathrm{mod},i} - F_{\mathrm{obs},i} |.
\end{equation}

\noindent
Since this estimator strongly penalises models that deviate from the
observations in a few lines, we can use the logarithm of the flux
instead, resulting in:

\begin{equation}
  \label{eq:madlog}
  \mathrm{MADlog} = \frac{1}{N_\lambda} \sum_{i=1}^{N_\lambda} \left|
  \log_{10} \left( \frac{F_{\mathrm{mod},i}}{F_{\mathrm{obs},i}}
    \right) \right|.
\end{equation}

The mean fractional error (MFE; see e.g. \citealt{Vogl/etal:2020})
is calculated as:

\begin{equation}
  \label{eq:mfe}  
  \mathrm{MFE} = \frac{1}{N_\lambda} \sum_{i=1}^{N_\lambda}
  \frac{|F_{\mathrm{mod},i} - F_{\mathrm{obs},i}|}{F_{\mathrm{obs},i}}.
\end{equation}

\begin{table}[h]
\begin{minipage}{\textwidth}
\centering
\caption{`Best-fit' model according to different statistical estimators and flux scales.}
\label{tab:stats}
\begin{tabular}{lccccc}
\hline\hline\\[-1.8ex]
\multicolumn{1}{c}{Flux scale}  & \multicolumn{1}{c}{MAD} & \multicolumn{1}{c}{MFE} & \multicolumn{1}{c}{MADlog} & \multicolumn{1}{c}{MFElog} & \multicolumn{1}{c}{Score} \\
\hline\\[-1.8ex]
$F_\nu$                            & GCD    & GCD     & MERGER  & MERGER  & MERGER  \\
$F_{\nu,\mathrm{norm}}$                & GCD    & GCD     & MERGER  & MERGER  & MERGER  \\
$F_{\nu}^{\mathrm{red}}$               & GCD    & GCD     & MERGER  & MERGER  & MERGER  \\
$F_{\nu,\mathrm{norm}}^{\mathrm{red}}$    & GCD    & GCD     & MERGER  & GCD     & MERGER  \\[0.2ex]
\hline\\[-1.8ex]
$F_\lambda$                         & DDT    & DBLEDET & DBLEDET & DBLEDET & DBLEDET \\
$F_{\lambda,\mathrm{norm}}$             & DDT    & GCD     & MERGER  & MERGER  & MERGER  \\
$F_{\lambda}^{\mathrm{red}}$            & DDT    & DBLEDET & DBLEDET & DBLEDET & DBLEDET \\
$F_{\lambda,\mathrm{norm}}^{\mathrm{red}}$ & MERGER & GCD     & MERGER  & MERGER  & MERGER  \\
\hline
\end{tabular}
\flushleft
\justifying
\footnotesize
\textbf{Notes.} The `norm' subscript indicates the model fluxes have
been normalised to the mean observed flux over the full
0.35--14\,\micron\ range, while the `red' superscript indicates the
observed SN~2021aefx spectrum was \textit{not} corrected for
extinction by dust in the host galaxy (i.e. the spectra become redder).
\end{minipage}
\end{table}

\newpage

\noindent
Because of the $F_{\mathrm{obs},i}$ term in the denominator, this
estimator gives almost equal weights to regions of low and high flux
values. As for the MAD estimator above, we can use the logarithm of
the flux to avoid penalising models that deviate from the
observations in a few lines:

\begin{equation}
  \label{eq:mfelog}  
  \mathrm{MFElog} = \frac{1}{N_\lambda} \sum_{i=1}^{N_\lambda} \left|
  \log_{10} \left( \frac{F_{\mathrm{mod},i}}{F_{\mathrm{obs},i}}
    \right) \middle/ \log_{10} (F_{\mathrm{obs},i}) \right|.
\end{equation}

The `Score' estimator is defined in \cite{Omand/Jerkstrand:2023}. Here
we used its mean value:

\begin{equation}
  \label{eq:score}  
  \mathrm{Score} = \frac{1}{N_\lambda} \sum_{i=1}^{N_\lambda} \left[
  \log_{10} \left( \frac{F_{\mathrm{mod},i}}{F_{\mathrm{obs},i}}.
    \right) \right]^2
\end{equation}

\noindent
The logarithm ensures that models that deviate from the observations
in a few lines are not overly penalised, but more so than in the
previous MADlog and MFElog estimators due to the squared exponent.

\clearpage


\section{Relativistic effects}\label{sect:rel}

\begin{figure}[h]
\centering
\includegraphics{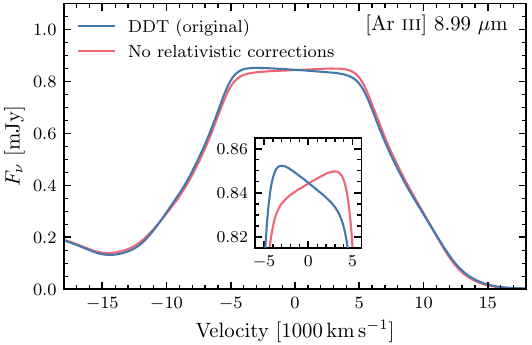}
\caption{\label{fig:rel}
[Ar\three]\,8.99\,\micron\ line profile in our DDT model with (blue)
and without (red) relativistic corrections in the observer-frame
calculation of the spectrum. The inset shows a close-up view of the
`flat'-top portion of the profile, highlighting the reversal of the tilt
direction.
}
\end{figure}

The prominent [Ar\three]\,8.99\,\micron\ line displays a blue-to-red
tilt in its flat-top profile. As noted previously, our `flat'-top line
profiles are tilted the opposite way due to relativistic beaming
(Fig.~\ref{fig:rel}), which results in a $2D\approx 3$\%
contrast between the blue and red edges at $\pm$4500\,\kms\ from the
rest wavelength, where $D$ is the so-called Doppler factor:

\begin{equation}\label{eq:dopfac}
D = \frac{1}{\gamma(1 - \beta \cos \theta)},
\end{equation}

\noindent
where $\beta=v/c$ and $\gamma=1/\sqrt{(1-\beta^2)}$ is the Lorentz
factor. While line overlap can conspire to reverse the direction of
this tilt in some cases, this does not seem to apply to the [Ar\three]
line. The observed $\sim$10\% increase in flux from blue to red
then suggests a pronounced asymmetry in the argon distribution (see
discussion in \citealt{DerKacy/etal:2023}).

\newpage


\section{Collision strengths for Ni\three}\label{sect:nkiiicol}

We report transition probabilities ($A_\mathrm{ul}$) and effective collision
strengths ($\Upsilon_\mathrm{lu}$) in the temperature range 1000--40\,000\,K
among the lowest ten levels of Ni\three\ (see
Table~\ref{tab:nkiiilev}) in Table~\ref{tab:nkiiicol}. Full versions
of both tables (among the lowest 43 levels of Ni\three\ and extending
to $10^5$\,K) are available in electronic form at the CDS.

\input{tabF1.tex}

\begin{figure}[h]
\centering
\includegraphics{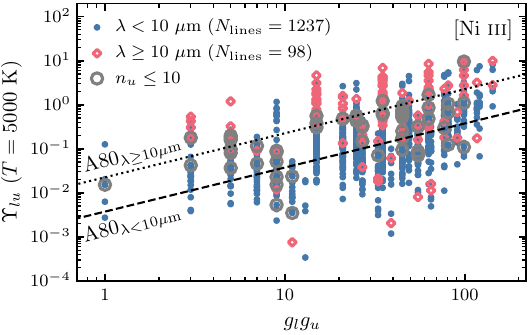}
\caption{\label{fig:axelrod}
Effective collision strength ($\Upsilon_\mathrm{lu}$) at 5000\,K for
forbidden transitions of Ni\three\ versus the product of the
statistical weights of the lower and upper levels. Different symbols
are used for transitions corresponding to wavelengths $<10$\,\micron\ (filled
circles) or $\ge 10$\,\micron\ (open diamonds). We also highlight
transitions among the first ten levels (see
Table~\ref{tab:nkiiilev}). Overplotted are the approximations of
\cite{Axelrod:1980} for the two wavelength regimes. The
$\Upsilon_\mathrm{lu}$ values deviate by more than one order of magnitude for
many transitions (including for some among low-lying levels).
}
\end{figure}

\cite{Axelrod:1980} proposed general approximations for collision
strengths for forbidden-line transitions based on [Fe\three], which
depend on the product of the statistical weights of the lower and
upper levels ($g_\mathrm{l} g_\mathrm{u}$):

\begin{equation}
  \Upsilon_\mathrm{lu} = a_\mathrm{IR} w \frac{g_\mathrm{l} g_\mathrm{u}}{8},
\end{equation}

\noindent
where $w\approx0.03$ and $a_\mathrm{IR}=6$ for infrared transitions
($\gtrsim 10$\,\micron), and $=1$ otherwise. As shown in
Fig.~\ref{fig:axelrod}, these relations only provide a very crude
estimate for most transitions, including among low-lying levels.

\input{tabF2.tex}
\vfill



\section{Spectra for the complete gcd and doubledet model sets}\label{sect:othermod}

\begin{figure*}
\centering
\includegraphics{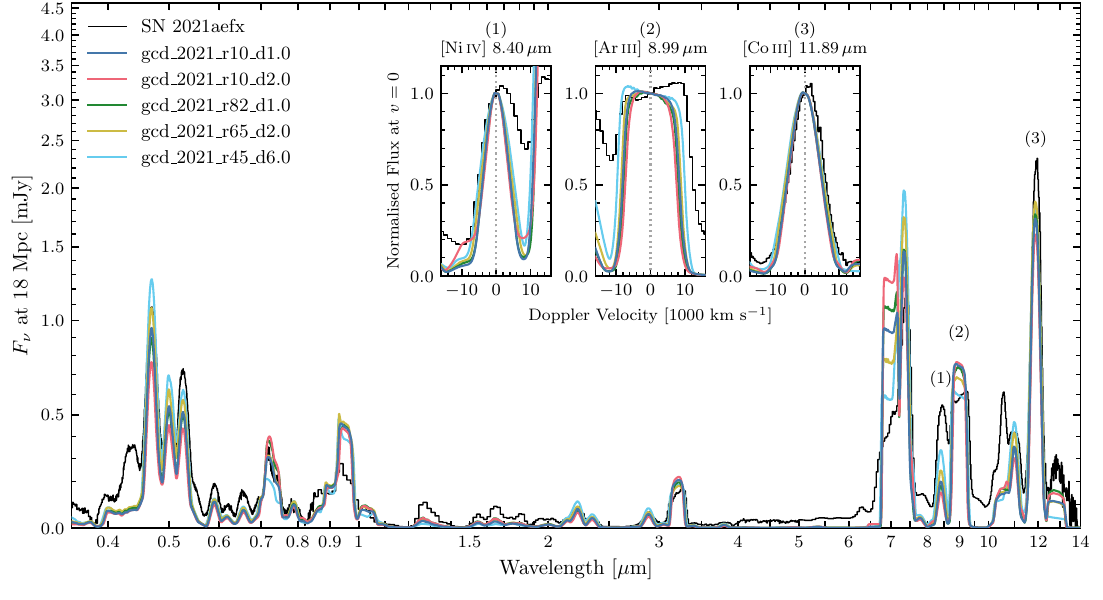}
\caption{
Similar to Fig.~\ref{fig:comp_spec_overview_full_21aefx} but for the full set of
pulsationally assisted gravitationally-confined detonation
models. The gcd\_2021\_r10\_d1.0 (GCD) model is used as the reference
for this class in this paper.
}
\label{fig:comp_spec_gcd_21aefx}
\end{figure*}

\begin{figure*}
\centering
\includegraphics{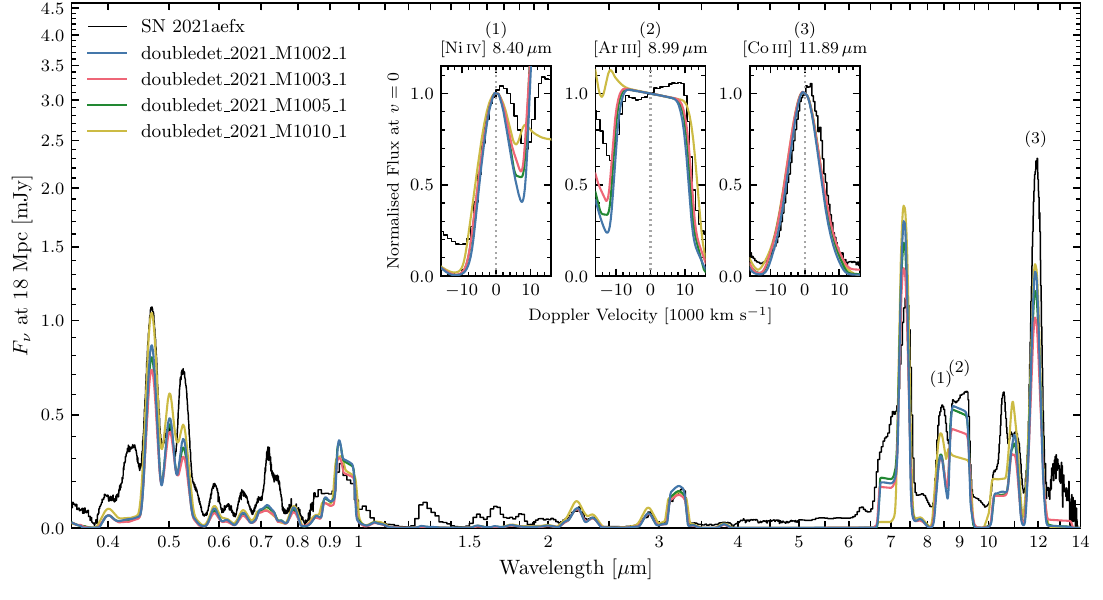}
\caption{\label{fig:comp_spec_doubledet_21aefx}
Similar to Fig.~\ref{fig:comp_spec_overview_full_21aefx} but for the
full set of double-detonation models. The doubledet\_2021\_M1002\_1
(DBLEDET) model is used as the reference for this class in this paper.}
\end{figure*}

We show the synthetic spectra for the complete set of pulsationally
assisted gravitationally-confined detonation models and
double-detonation models compared to SN~2021aefx in the range
0.35--14\,\micron\ in Figs.~\ref{fig:comp_spec_gcd_21aefx} and
\ref{fig:comp_spec_doubledet_21aefx}, respectively.


\end{appendix}

\end{document}

%% file: tab1.tex
\begin{table*}
\centering
\fontsize{8pt}{10pt}
\selectfont
\caption{Basic properties for our complete model set.}\label{tab:summary}
\begin{tabular}{lcccccccccccc}
\hline\hline\\[-1.8ex]
\multicolumn{1}{c}{Model} & \multicolumn{1}{c}{$M_\mathrm{tot}$} & \multicolumn{1}{c}{$^{56}\mathrm{Ni}_{t=0}$} & \multicolumn{1}{c}{Ni} & \multicolumn{1}{c}{Co} & \multicolumn{1}{c}{Fe} & \multicolumn{1}{c}{Ar} & \multicolumn{1}{c}{S} & \multicolumn{1}{c}{$L_\mathrm{dep}$} & \multicolumn{1}{c}{$\mathcal{F}_\mathrm{dep}$} & \multicolumn{1}{c}{$\mathcal{F}_\mathrm{e+}$} & \multicolumn{1}{c}{$\varv_{99, \mathrm{dep}}$} & \multicolumn{1}{c}{Ref.} \\
 & \multicolumn{1}{c}{(M$_\odot$)} & \multicolumn{1}{c}{(M$_\odot$)} & \multicolumn{1}{c}{(M$_\odot$)} & \multicolumn{1}{c}{(M$_\odot$)} & \multicolumn{1}{c}{(M$_\odot$)} & \multicolumn{1}{c}{(M$_\odot$)} & \multicolumn{1}{c}{(M$_\odot$)} & \multicolumn{1}{c}{(erg\,s$^{-1}$)} & \multicolumn{1}{c}{(\%)} & \multicolumn{1}{c}{(\%)} & \multicolumn{1}{c}{(km\,s$^{-1}$)} &  \\
\hline\\[-1.8ex]
\textbf{ddt\_2013\_N100}               &      1.45 &      0.626 &      0.076 &      0.070 &      0.708 &      0.020 &      0.119 & $ 3.92 \times 10^{40}$ &    4.9 &   63.5 & 13\,816 & (1)                            \\
ddt\_2013\_N100\_xpos          &      1.40 &      0.665 &      0.038 &      0.072 &      0.659 &      0.025 &      0.136 & $ 4.29 \times 10^{40}$ &    5.0 &   61.5 & 15\,088 & (2)                            \\
ddt\_2013\_N100\_xneg          &      1.40 &      0.530 &      0.060 &      0.058 &      0.555 &      0.024 &      0.161 & $ 3.37 \times 10^{40}$ &    4.9 &   62.5 & 14\,264 & (2)                            \\
ddt\_2013\_N100\_ypos          &      1.40 &      0.704 &      0.032 &      0.077 &      0.699 &      0.030 &      0.147 & $ 4.53 \times 10^{40}$ &    5.0 &   61.7 & 14\,793 & (2)                            \\
ddt\_2013\_N100\_yneg          &      1.40 &      0.500 &      0.108 &      0.061 &      0.604 &      0.025 &      0.155 & $ 3.20 \times 10^{40}$ &    5.0 &   62.0 & 13\,289 & (2)                            \\
ddt\_2013\_N100\_zpos          &      1.40 &      0.638 &      0.125 &      0.079 &      0.747 &      0.017 &      0.094 & $ 4.06 \times 10^{40}$ &    5.0 &   62.4 & 12\,976 & (2)                            \\
ddt\_2013\_N100\_zneg          &      1.40 &      0.497 &      0.077 &      0.058 &      0.541 &      0.017 &      0.111 & $ 3.14 \times 10^{40}$ &    4.9 &   62.8 & 13\,106 & (2)                            \\
\textbf{gcd\_2021\_r10\_d1.0}           &      1.40 &      0.608 &      0.015 &      0.063 &      0.576 &      0.022 &      0.128 & $ 4.02 \times 10^{40}$ &    5.2 &   59.3 & 15\,957 & (3)                            \\
gcd\_2021\_r10\_d2.0           &      1.42 &      0.542 &      0.018 &      0.057 &      0.533 &      0.023 &      0.135 & $ 3.66 \times 10^{40}$ &    5.3 &   58.6 & 15\,467 & (3)                            \\
gcd\_2021\_r82\_d1.0           &      1.39 &      0.603 &      0.015 &      0.063 &      0.573 &      0.021 &      0.125 & $ 4.05 \times 10^{40}$ &    5.3 &   58.6 & 15\,713 & (3)                            \\
gcd\_2021\_r65\_d2.0           &      1.41 &      0.707 &      0.021 &      0.074 &      0.669 &      0.019 &      0.111 & $ 4.63 \times 10^{40}$ &    5.1 &   59.9 & 15\,833 & (3)                            \\
gcd\_2021\_r45\_d6.0           &      1.45 &      0.775 &      0.032 &      0.082 &      0.754 &      0.018 &      0.106 & $ 4.97 \times 10^{40}$ &    5.0 &   61.3 & 15\,635 & (3)                            \\
\textbf{doubledet\_2021\_M1002\_1}      &      1.07 &      0.571 &      0.018 &      0.061 &      0.552 &      0.021 &      0.108 & $ 3.31 \times 10^{40}$ &    4.5 &   68.5 & 14\,307 & (4)                            \\
doubledet\_2021\_M1003\_1      &      1.10 &      0.538 &      0.019 &      0.058 &      0.518 &      0.021 &      0.105 & $ 2.96 \times 10^{40}$ &    4.3 &   72.2 & 15\,899 & (4)                            \\
doubledet\_2021\_M1005\_1      &      1.10 &      0.576 &      0.020 &      0.062 &      0.553 &      0.020 &      0.106 & $ 3.23 \times 10^{40}$ &    4.4 &   70.6 & 15\,786 & (4)                            \\
doubledet\_2021\_M1010\_1      &      1.16 &      0.802 &      0.027 &      0.085 &      0.741 &      0.015 &      0.072 & $ 4.36 \times 10^{40}$ &    4.3 &   71.2 & 19\,288 & (4)                            \\
\textbf{merger\_2012\_11+09}            &      2.09 &      0.666 &      0.031 &      0.072 &      0.638 &      0.018 &      0.110 & $ 5.84 \times 10^{40}$ &    6.8 &   45.4 & 13\,394 & (5)                            \\

\hline
\end{tabular}
\flushleft
\justifying
\footnotesize
{\bf Notes.}
All quantities other than $^{56}\mathrm{Ni}_{t=0}$ correspond to our
initial conditions at 270\,d post explosion.  Model names are as they
appear in HESMA. Models in boldface correspond to the reference for
each class of explosion mechanism (DDT, GCD, DBLEDET, and MERGER),
which is indicated in the model name (ddt = delayed detonation, gcd =
gravitationally confined detonation, doubledet = double detonation,
merger = violent WD-WD merger). The total mass (\mtot) and initial
\nifs\ mass ($^{56}\mathrm{Ni}_{t=0}$) correspond to our smoothed 1D
models.  The total instantaneous deposited decay power
($L_\mathrm{dep}$) corresponds to a fraction
$\mathcal{F}_\mathrm{dep}$ of the total decay power. The fraction of
this deposited decay power due to positrons (assumed to deposit their
energy locally) is noted $\mathcal{F}_{e+}$. The velocity coordinate
$\varv_{\mathrm{dep},99}$ contains 99\% of the total volume-integrated
specific deposited decay power.\\
\textbf{References.} (1) \cite{Seitenzahl/etal:2013}, (2) Seitenzahl
(priv. comm.), (3) \cite{Lach/etal:2022}, (4) \cite{Gronow/etal:2021},
(5) \cite{Pakmor/etal:2012}.
\end{table*}

%% file: tab2.tex
\begin{table*}
\centering
\caption{Lines in the wavelength range 0.35--28\,$\mu$m in our complete model set (see Table~\ref{tab:summary}) at 270 days post explosion whose absolute Sobolev EW exceeds 5\% of the largest absolute EW in the wavelength range 0.35--1\,$\mu$m.}\label{tab:lineid}
\begin{tabular}{rcc|rcc|rcc}
\hline\hline
&&&&&&&&\\[-1.8ex]
\multicolumn{1}{c}{$\lambda_\mathrm{air}$} & \multicolumn{1}{c}{Ion} & \multicolumn{1}{c|}{Models} & \multicolumn{1}{c}{$\lambda_\mathrm{air}$} & \multicolumn{1}{c}{Ion} & \multicolumn{1}{c|}{Models} & \multicolumn{1}{c}{$\lambda_\mathrm{air}$} & \multicolumn{1}{c}{Ion} & \multicolumn{1}{c}{Models} \\
\multicolumn{1}{c}{($\mu$m)} & & & \multicolumn{1}{c}{($\mu$m)} & & & \multicolumn{1}{c}{($\mu$m)} & & \\
\hline
&&&&&&&&\\[-1.8ex]
0.438\phantom{\,$^{(\dag)}$} & Fe\,\sc{i}]\phantom{ $^{\mathrm{a}}$} & mer & 1.019\,$^{(\dag)}$ & [Co\,\sc{ii}] $^{{\mathrm{{l}}}}$ & ddt, mer & 7.347\,$^{(\dag)}$ & [Ni\,\sc{iii}] $^{{\mathrm{{t}}}}$ & All \\
0.455\phantom{\,$^{(\dag)}$} & Fe\,\sc{ii} $^{{\mathrm{{a}}}}$ & mer & 1.025\phantom{\,$^{(\dag)}$} & [Co\,\sc{ii}] $^{{\mathrm{{l}}}}$ & mer & 7.788\phantom{\,$^{(\dag)}$} & [Fe\,\sc{iii}]\phantom{ $^{\mathrm{a}}$} & All \\
0.458\phantom{\,$^{(\dag)}$} & Fe\,\sc{ii} $^{{\mathrm{{a}}}}$ & mer & 1.029\phantom{\,$^{(\dag)}$} & [S\,\sc{ii}] $^{{\mathrm{{m}}}}$ & mer & 8.403\,$^{(\dag)}$ & [Ni\,\sc{iv}] $^{{\mathrm{{u}}}}$ & All \\
0.461\,$^{(\dag)}$ & [Fe\,\sc{iii}] $^{{\mathrm{{b}}}}$ & ddet & 1.032\phantom{\,$^{(\dag)}$} & [S\,\sc{ii}] $^{{\mathrm{{m}}}}$ & ddt, gcd, mer & 8.989\,$^{(\dag)}$ & [Ar\,\sc{iii}] $^{{\mathrm{{v}}}}$ & All \\
0.466\,$^{(\dag)}$ & [Fe\,\sc{iii}] $^{{\mathrm{{b}}}}$ & All & 1.257\,$^{(\dag)}$ & [Fe\,\sc{ii}] $^{{\mathrm{{n}}}}$ & ddt, gcd, mer & 10.508\,$^{(\dag)}$ & [S\,\sc{iv}]\phantom{ $^{\mathrm{a}}$} & All \\
0.470\phantom{\,$^{(\dag)}$} & [Fe\,\sc{iii}] $^{{\mathrm{{b}}}}$ & All & 1.279\phantom{\,$^{(\dag)}$} & [Fe\,\sc{ii}] $^{{\mathrm{{n}}}}$ & mer & 10.520\,$^{(\dag)}$ & [Co\,\sc{ii}] $^{{\mathrm{{l}}}}$ & All \\
0.473\phantom{\,$^{(\dag)}$} & [Fe\,\sc{iii}] $^{{\mathrm{{b}}}}$ & ddt, gcd, ddet & 1.294\phantom{\,$^{(\dag)}$} & [Fe\,\sc{ii}] $^{{\mathrm{{n}}}}$ & mer & 10.679\phantom{\,$^{(\dag)}$} & [Ni\,\sc{ii}] $^{{\mathrm{{w}}}}$ & ddt, mer \\
0.475\phantom{\,$^{(\dag)}$} & [Fe\,\sc{iii}] $^{{\mathrm{{b}}}}$ & gcd, ddet, mer & 1.321\phantom{\,$^{(\dag)}$} & [Fe\,\sc{ii}] $^{{\mathrm{{n}}}}$ & mer & 10.999\phantom{\,$^{(\dag)}$} & [Ni\,\sc{iii}] $^{{\mathrm{{t}}}}$ & All \\
0.477\phantom{\,$^{(\dag)}$} & [Fe\,\sc{iii}] $^{{\mathrm{{b}}}}$ & gcd, ddet & 1.328\phantom{\,$^{(\dag)}$} & [Fe\,\sc{ii}] $^{{\mathrm{{n}}}}$ & mer & 11.164\phantom{\,$^{(\dag)}$} & [Co\,\sc{ii}] $^{{\mathrm{{l}}}}$ & ddt, mer \\
0.478\phantom{\,$^{(\dag)}$} & [Fe\,\sc{iii}] $^{{\mathrm{{b}}}}$ & ddet & 1.372\phantom{\,$^{(\dag)}$} & [Fe\,\sc{ii}] $^{{\mathrm{{n}}}}$ & mer & 11.723\phantom{\,$^{(\dag)}$} & [Ni\,\sc{iv}] $^{{\mathrm{{u}}}}$ & All \\
0.492\phantom{\,$^{(\dag)}$} & Fe\,\sc{ii} $^{{\mathrm{{c}}}}$ & mer & 1.533\phantom{\,$^{(\dag)}$} & [Fe\,\sc{ii}] $^{{\mathrm{{o}}}}$ & ddt, mer & 11.885\,$^{(\dag)}$ & [Co\,\sc{iii}] $^{{\mathrm{{x}}}}$ & All \\
0.493\phantom{\,$^{(\dag)}$} & [Fe\,\sc{iii}] $^{{\mathrm{{d}}}}$ & ddet & 1.547\phantom{\,$^{(\dag)}$} & [Co\,\sc{ii}]\phantom{ $^{\mathrm{a}}$} & ddt, mer & 12.725\phantom{\,$^{(\dag)}$} & [Ni\,\sc{ii}] $^{{\mathrm{{w}}}}$ & ddt, mer \\
0.501\phantom{\,$^{(\dag)}$} & [Fe\,\sc{iii}] $^{{\mathrm{{d}}}}$ & All & 1.599\phantom{\,$^{(\dag)}$} & [Fe\,\sc{ii}] $^{{\mathrm{{o}}}}$ & ddt, mer & 12.811\,$^{(\dag)}$ & [Ne\,\sc{ii}]\phantom{ $^{\mathrm{a}}$} & All \\
0.502\phantom{\,$^{(\dag)}$} & Fe\,\sc{ii} $^{{\mathrm{{c}}}}$ & mer & 1.644\phantom{\,$^{(\dag)}$} & [Fe\,\sc{ii}] $^{{\mathrm{{o}}}}$ & ddt, gcd, mer & 14.735\phantom{\,$^{(\dag)}$} & [Co\,\sc{ii}] $^{{\mathrm{{y}}}}$ & ddt, gcd, mer \\
0.508\phantom{\,$^{(\dag)}$} & [Fe\,\sc{iii}] $^{{\mathrm{{d}}}}$ & ddet & 1.664\phantom{\,$^{(\dag)}$} & [Fe\,\sc{ii}] $^{{\mathrm{{o}}}}$ & mer & 15.455\phantom{\,$^{(\dag)}$} & [Co\,\sc{ii}] $^{{\mathrm{{l}}}}$ & ddt, gcd, mer \\
0.517\phantom{\,$^{(\dag)}$} & Fe\,\sc{ii} $^{{\mathrm{{c}}}}$ & gcd, mer & 1.677\phantom{\,$^{(\dag)}$} & [Fe\,\sc{ii}] $^{{\mathrm{{o}}}}$ & ddt, mer & 15.550\,$^{(\dag)}$ & [Ne\,\sc{iii}]\phantom{ $^{\mathrm{a}}$} & All \\
0.527\phantom{\,$^{(\dag)}$} & Fe\,\sc{i}\phantom{ $^{\mathrm{a}}$} & mer & 1.797\phantom{\,$^{(\dag)}$} & [Fe\,\sc{ii}] $^{{\mathrm{{o}}}}$ & mer & 15.643\,$^{(\dag)}$ & [Co\,\sc{iv}] $^{{\mathrm{{z}}}}$ & All \\
0.527\phantom{\,$^{(\dag)}$} & [Fe\,\sc{iii}] $^{{\mathrm{{d}}}}$ & All & 1.800\phantom{\,$^{(\dag)}$} & [Fe\,\sc{ii}] $^{{\mathrm{{o}}}}$ & mer & 16.295\phantom{\,$^{(\dag)}$} & [Co\,\sc{ii}] $^{{\mathrm{{l}}}}$ & mer \\
0.589\,$^{(\dag)}$ & [Co\,\sc{iii}]\phantom{ $^{\mathrm{a}}$} & gcd, ddet, mer & 1.809\phantom{\,$^{(\dag)}$} & [Fe\,\sc{ii}] $^{{\mathrm{{o}}}}$ & ddt, mer & 16.386\phantom{\,$^{(\dag)}$} & [Co\,\sc{iii}] $^{{\mathrm{{x}}}}$ & All \\
0.589\,$^{(\dag)}$ & Na\,\sc{i} $^{{\mathrm{{e}}}}$ & mer & 1.939\phantom{\,$^{(\dag)}$} & [Ni\,\sc{ii}]\phantom{ $^{\mathrm{a}}$} & ddt & 17.278\phantom{\,$^{(\dag)}$} & [Ni\,\sc{iv}] $^{{\mathrm{{u}}}}$ & All \\
0.590\,$^{(\dag)}$ & Na\,\sc{i} $^{{\mathrm{{e}}}}$ & mer & 2.015\phantom{\,$^{(\dag)}$} & [Fe\,\sc{ii}]\phantom{ $^{\mathrm{a}}$} & mer & 17.931\phantom{\,$^{(\dag)}$} & [Fe\,\sc{ii}] $^{{\mathrm{{aa}}}}$ & All \\
0.658\,$^{(\dag)}$ & [Co\,\sc{iii}]\phantom{ $^{\mathrm{a}}$} & gcd, ddet & 2.046\phantom{\,$^{(\dag)}$} & [Fe\,\sc{ii}]\phantom{ $^{\mathrm{a}}$} & mer & 18.236\phantom{\,$^{(\dag)}$} & [Ni\,\sc{ii}] $^{{\mathrm{{w}}}}$ & ddt \\
0.714\,$^{(\dag)}$ & [Ar\,\sc{iii}]\phantom{ $^{\mathrm{a}}$} & All & 2.145\phantom{\,$^{(\dag)}$} & [Fe\,\sc{iii}] $^{{\mathrm{{p}}}}$ & All & 18.708\phantom{\,$^{(\dag)}$} & [S\,\sc{iii}]\phantom{ $^{\mathrm{a}}$} & All \\
0.716\phantom{\,$^{(\dag)}$} & [Fe\,\sc{ii}]\phantom{ $^{\mathrm{a}}$} & mer & 2.218\phantom{\,$^{(\dag)}$} & [Fe\,\sc{iii}] $^{{\mathrm{{p}}}}$ & All & 18.799\phantom{\,$^{(\dag)}$} & [Co\,\sc{ii}] $^{{\mathrm{{y}}}}$ & ddt, gcd, mer \\
0.729\,$^{(\dag)}$ & [Ca\,\sc{ii}] $^{{\mathrm{{f}}}}$ & ddt, gcd, mer & 2.242\phantom{\,$^{(\dag)}$} & [Fe\,\sc{iii}] $^{{\mathrm{{p}}}}$ & ddt, gcd, ddet & 20.845\phantom{\,$^{(\dag)}$} & [Fe\,\sc{v}] $^{{\mathrm{{ab}}}}$ & ddt, gcd, ddet \\
0.732\,$^{(\dag)}$ & [Ca\,\sc{ii}] $^{{\mathrm{{f}}}}$ & gcd, mer & 2.348\phantom{\,$^{(\dag)}$} & [Fe\,\sc{iii}] $^{{\mathrm{{p}}}}$ & All & 21.826\phantom{\,$^{(\dag)}$} & [Ar\,\sc{iii}] $^{{\mathrm{{v}}}}$ & All \\
0.789\phantom{\,$^{(\dag)}$} & [Ni\,\sc{iii}] $^{{\mathrm{{g}}}}$ & All & 2.874\phantom{\,$^{(\dag)}$} & [Fe\,\sc{iii}] $^{{\mathrm{{q}}}}$ & All & 22.794\phantom{\,$^{(\dag)}$} & [Co\,\sc{iv}] $^{{\mathrm{{z}}}}$ & All \\
0.850\phantom{\,$^{(\dag)}$} & Ca\,\sc{ii} $^{{\mathrm{{h}}}}$ & gcd, mer & 2.904\phantom{\,$^{(\dag)}$} & [Fe\,\sc{iii}] $^{{\mathrm{{q}}}}$ & ddt, gcd, ddet & 22.896\phantom{\,$^{(\dag)}$} & [Fe\,\sc{ii}]\phantom{ $^{\mathrm{a}}$} & ddt, gcd, mer \\
0.850\phantom{\,$^{(\dag)}$} & [Ni\,\sc{iii}] $^{{\mathrm{{g}}}}$ & ddt, ddet, mer & 3.043\phantom{\,$^{(\dag)}$} & [Fe\,\sc{iii}] $^{{\mathrm{{q}}}}$ & ddt, gcd, ddet & 22.919\,$^{(\dag)}$ & [Fe\,\sc{iii}]\phantom{ $^{\mathrm{a}}$} & All \\
0.854\phantom{\,$^{(\dag)}$} & Ca\,\sc{ii} $^{{\mathrm{{h}}}}$ & ddt, gcd, mer & 3.206\,$^{(\dag)}$ & [Ca\,\sc{iv}]\phantom{ $^{\mathrm{a}}$} & All & 24.061\phantom{\,$^{(\dag)}$} & [Co\,\sc{iii}] $^{{\mathrm{{x}}}}$ & All \\
0.862\phantom{\,$^{(\dag)}$} & [Fe\,\sc{ii}] $^{{\mathrm{{i}}}}$ & mer & 3.229\phantom{\,$^{(\dag)}$} & [Fe\,\sc{iii}] $^{{\mathrm{{q}}}}$ & All & 24.512\phantom{\,$^{(\dag)}$} & [Fe\,\sc{ii}] $^{{\mathrm{{aa}}}}$ & All \\
0.866\phantom{\,$^{(\dag)}$} & Ca\,\sc{ii} $^{{\mathrm{{h}}}}$ & ddt, gcd, mer & 3.393\phantom{\,$^{(\dag)}$} & [Ni\,\sc{iii}] $^{{\mathrm{{r}}}}$ & All & 25.682\phantom{\,$^{(\dag)}$} & [Co\,\sc{ii}] $^{{\mathrm{{y}}}}$ & ddt, gcd, mer \\
0.889\phantom{\,$^{(\dag)}$} & [Fe\,\sc{ii}] $^{{\mathrm{{i}}}}$ & mer & 3.801\phantom{\,$^{(\dag)}$} & [Ni\,\sc{iii}] $^{{\mathrm{{r}}}}$ & All & 25.927\phantom{\,$^{(\dag)}$} & [Fe\,\sc{v}] $^{{\mathrm{{ab}}}}$ & All \\
0.907\phantom{\,$^{(\dag)}$} & [S\,\sc{iii}] $^{{\mathrm{{j}}}}$ & All & 4.888\phantom{\,$^{(\dag)}$} & [Fe\,\sc{ii}] $^{{\mathrm{{s}}}}$ & mer & 25.981\,$^{(\dag)}$ & [Fe\,\sc{ii}]\phantom{ $^{\mathrm{a}}$} & All \\
0.934\phantom{\,$^{(\dag)}$} & [Co\,\sc{ii}] $^{{\mathrm{{k}}}}$ & mer & 5.339\,$^{(\dag)}$ & [Fe\,\sc{ii}] $^{{\mathrm{{s}}}}$ & ddt, mer & 26.124\phantom{\,$^{(\dag)}$} & [Fe\,\sc{iii}]\phantom{ $^{\mathrm{a}}$} & ddet \\
0.953\phantom{\,$^{(\dag)}$} & [S\,\sc{iii}] $^{{\mathrm{{j}}}}$ & All & 6.634\,$^{(\dag)}$ & [Ni\,\sc{ii}]\phantom{ $^{\mathrm{a}}$} & ddt, gcd, mer &   &   &   \\
0.994\phantom{\,$^{(\dag)}$} & [Co\,\sc{ii}] $^{{\mathrm{{k}}}}$ & mer & 6.983\,$^{(\dag)}$ & [Ar\,\sc{ii}]\phantom{ $^{\mathrm{a}}$} & All &   &   &   \\

\hline
\end{tabular}
\flushleft
\justifying
\footnotesize
{\bf Notes.} All wavelengths are given in air. Forbidden and semiforbidden transitions are noted using the appropriate brackets around the ion name. Wavelengths marked with a `$^{(\dag)}$' symbol denote transitions connected to the ground state. Ions with the same superscript correspond to transitions within the same multiplet. The `models' column indicates whether a given line appears in at least one model of a given class (ddt = delayed detonation, gcd = pulsationally-assisted gravitationally-confined detonation, ddet = double detonation, mer = violent WD-WD merger). `All' means the line appears in all four model classes.
\end{table*}

%% file: tab4.tex
\begin{table*}
\centering
\small
\caption{Total line luminosities for (stable) Ni and for individual ionisation stages (including the per cent fraction of the total Ni luminosity).}\label{tab:lni}
\begin{tabular}{lcccrcrcrcr}
\hline\hline\\[-1.8ex]
\multicolumn{1}{c}{Model} & \multicolumn{1}{c}{$M(\text{Ni})$} & \multicolumn{1}{c}{$L(\text{Ni})$} & \multicolumn{2}{c}{$L(\text{\ion{Ni}{1}})$}            & \multicolumn{2}{c}{$L(\text{\ion{Ni}{2}})$}            & \multicolumn{2}{c}{$L(\text{\ion{Ni}{3}})$}            & \multicolumn{2}{c}{$L(\text{\ion{Ni}{4}})$}            \\
\cmidrule(rl{.5em}){4-5}\cmidrule(rl{.5em}){6-7}\cmidrule(rl{.5em}){8-9}\cmidrule(rl{.5em}){10-11}
\multicolumn{1}{c}{}      & \multicolumn{1}{c}{(\msun)}        & \multicolumn{1}{c}{(\ergs)}        & \multicolumn{1}{c}{(\ergs)} & \multicolumn{1}{c}{(\%)} & \multicolumn{1}{c}{(\ergs)} & \multicolumn{1}{c}{(\%)} & \multicolumn{1}{c}{(\ergs)} & \multicolumn{1}{c}{(\%)} & \multicolumn{1}{c}{(\ergs)} & \multicolumn{1}{c}{(\%)} \\
\hline\\[-1.8ex]
ddt\_2013\_N100 & 0.076 & 5.05($+$38) & 2.95($+$36) & 0.6 & 8.15($+$36) & 1.6 & 4.50($+$38) & 89.0 & 4.55($+$37) & 9.0 \\
ddt\_2013\_N100\_xpos & 0.038 & 3.51($+$38) & 3.48($+$35) & 0.1 & 2.79($+$36) & 0.8 & 2.89($+$38) & 82.1 & 6.02($+$37) & 17.1 \\
ddt\_2013\_N100\_xneg & 0.060 & 4.58($+$38) & 2.67($+$37) & 5.8 & 2.97($+$37) & 6.5 & 3.61($+$38) & 78.8 & 4.22($+$37) & 9.2 \\
ddt\_2013\_N100\_ypos & 0.032 & 3.07($+$38) & 3.31($+$35) & 0.1 & 2.39($+$36) & 0.8 & 2.51($+$38) & 81.9 & 5.34($+$37) & 17.4 \\
ddt\_2013\_N100\_yneg & 0.108 & 7.31($+$38) & 3.43($+$37) & 4.7 & 6.59($+$37) & 9.0 & 5.75($+$38) & 78.6 & 6.02($+$37) & 8.2 \\
ddt\_2013\_N100\_zpos & 0.125 & 8.09($+$38) & 9.33($+$36) & 1.2 & 1.51($+$37) & 1.9 & 6.92($+$38) & 85.6 & 9.39($+$37) & 11.6 \\
ddt\_2013\_N100\_zneg & 0.077 & 5.32($+$38) & 1.02($+$37) & 1.9 & 1.40($+$37) & 2.6 & 4.58($+$38) & 86.2 & 5.08($+$37) & 9.6 \\
gcd\_2021\_r10\_d1.0 & 0.015 & 1.34($+$38) & 1.54($+$36) & 1.1 & 3.20($+$36) & 2.4 & 1.12($+$38) & 83.6 & 1.75($+$37) & 13.0 \\
gcd\_2021\_r10\_d2.0 & 0.018 & 1.34($+$38) & 7.16($+$36) & 5.4 & 7.30($+$36) & 5.5 & 1.04($+$38) & 77.5 & 1.60($+$37) & 11.9 \\
gcd\_2021\_r82\_d1.0 & 0.015 & 1.33($+$38) & 1.92($+$36) & 1.4 & 3.73($+$36) & 2.8 & 1.12($+$38) & 84.0 & 1.58($+$37) & 11.9 \\
gcd\_2021\_r65\_d2.0 & 0.021 & 1.78($+$38) & 1.54($+$36) & 0.9 & 3.50($+$36) & 2.0 & 1.47($+$38) & 82.9 & 2.57($+$37) & 14.5 \\
gcd\_2021\_r45\_d6.0 & 0.032 & 2.39($+$38) & 5.43($+$36) & 2.3 & 4.68($+$36) & 2.0 & 1.88($+$38) & 78.9 & 4.06($+$37) & 17.0 \\
doubledet\_2021\_M1002\_1 & 0.018 & 1.67($+$38) & 7.24($+$34) & 0.0 & 9.94($+$35) & 0.6 & 1.36($+$38) & 81.8 & 2.95($+$37) & 17.7 \\
doubledet\_2021\_M1003\_1 & 0.019 & 1.51($+$38) & 5.49($+$34) & 0.0 & 8.48($+$35) & 0.6 & 1.18($+$38) & 77.9 & 3.27($+$37) & 21.6 \\
doubledet\_2021\_M1005\_1 & 0.020 & 1.67($+$38) & 5.96($+$34) & 0.0 & 9.59($+$35) & 0.6 & 1.33($+$38) & 79.6 & 3.33($+$37) & 19.9 \\
doubledet\_2021\_M1010\_1 & 0.027 & 2.37($+$38) & 1.74($+$34) & 0.0 & 8.76($+$35) & 0.4 & 1.84($+$38) & 77.5 & 5.28($+$37) & 22.3 \\
merger\_2012\_11+09 & 0.031 & 4.45($+$38) & 6.11($+$37) & 13.7 & 3.52($+$37) & 7.9 & 3.42($+$38) & 76.7 & 1.31($+$37) & 2.9 \\
\hline\\[-2ex]
\multicolumn{1}{l}{Pearson correlation with $M(\text{{Ni}}$)} & & 0.97 & 0.31 & & 0.61 & & 0.98 & & 0.79 & \\
\hline
\end{tabular}
\flushleft
\justifying
\footnotesize
\textbf{Notes.} Numbers in parentheses correspond to powers of ten. The total line luminosities were determined by integrating the single-element (for Ni) or single-ion (for Ni\one--\textsc{iv}) spectra obtained from an observer-frame calculation based on the converged radiative-transfer solution including all ions (and correcting for continuum emission). The last line give the Pearson correlation coefficient between the various line luminosities and the total (stable) Ni mass.
\end{table*}

%% file: tabA1.tex
\begin{table*}
\centering
\footnotesize
\caption{Comparison of the total mass and selected yields of the original 3D models ($M_\mathrm{3D}$) and their spherically-averaged versions ($M_\mathrm{1D}$) available on HESMA.}\label{tab:3d1d}
\begin{tabular}{llccccccccc}
\hline\hline\\[-1.8ex]
\multicolumn{1}{c}{Model} & \multicolumn{1}{c}{Quantity} & \multicolumn{1}{c}{Unit} & \multicolumn{1}{c}{Total} & \multicolumn{1}{c}{\nifs$_{t=0}$} & \multicolumn{1}{c}{$^{58}$Ni} & \multicolumn{1}{c}{Co} & \multicolumn{1}{c}{Fe} & \multicolumn{1}{c}{Ca} & \multicolumn{1}{c}{Ar} & \multicolumn{1}{c}{S} \\
\hline\\[-1.8ex]
ddt\_2013\_N100                & $M_\mathrm{3D}$ & \msun      &       1.40 &      0.604 &      0.069 &      0.068 &      0.683 &      0.015 &      0.020 &      0.115 \\
                               & $M_\mathrm{1D}$ & \msun      &       1.45 &      0.627 &      0.071 &      0.070 &      0.708 &      0.015 &      0.020 &      0.119 \\
                               & $\Delta M_{\mathrm{1D},\mathrm{3D}}$ & \msun      &      +0.05 &     +0.023 &     +0.002 &     +0.002 &     +0.025 &     +0.000 &     +0.001 &     +0.004 \\
                               & $\Delta M_{\mathrm{1D},\mathrm{3D}}/M_\mathrm{3D}$ &            &     +3.7\% &     +3.8\% &     +3.3\% &     +2.8\% &     +3.6\% &     +3.1\% &     +3.5\% &     +3.7\% \\
 \\
gcd\_2021\_r10\_d1.0           & $M_\mathrm{3D}$ & \msun      &       1.35 &      0.596 &      0.015 &      0.063 &      0.570 &      0.018 &      0.021 &      0.125 \\
                               & $M_\mathrm{1D}$ & \msun      &       1.40 &      0.609 &      0.015 &      0.063 &      0.576 &      0.019 &      0.022 &      0.128 \\
                               & $\Delta M_{\mathrm{1D},\mathrm{3D}}$ & \msun      &      +0.05 &     +0.013 &     +0.000 &     +0.000 &     +0.006 &     +0.000 &     +0.000 &     +0.003 \\
                               & $\Delta M_{\mathrm{1D},\mathrm{3D}}/M_\mathrm{3D}$ &            &     +3.7\% &     +2.1\% &     +1.2\% &     +0.6\% &     +1.0\% &     +1.1\% &     +1.8\% &     +2.5\% \\
 \\
gcd\_2021\_r10\_d2.0           & $M_\mathrm{3D}$ & \msun      &       1.37 &      0.532 &      0.017 &      0.056 &      0.525 &      0.019 &      0.023 &      0.133 \\
                               & $M_\mathrm{1D}$ & \msun      &       1.42 &      0.542 &      0.017 &      0.057 &      0.533 &      0.019 &      0.023 &      0.135 \\
                               & $\Delta M_{\mathrm{1D},\mathrm{3D}}$ & \msun      &      +0.05 &     +0.010 &     +0.001 &     +0.001 &     +0.008 &     +0.000 &     +0.000 &     +0.002 \\
                               & $\Delta M_{\mathrm{1D},\mathrm{3D}}/M_\mathrm{3D}$ &            &     +3.4\% &     +1.9\% &     +3.8\% &     +1.2\% &     +1.5\% &     +0.1\% &     +0.5\% &     +1.7\% \\
 \\
gcd\_2021\_r82\_d1.0           & $M_\mathrm{3D}$ & \msun      &       1.35 &      0.592 &      0.015 &      0.062 &      0.566 &      0.018 &      0.021 &      0.122 \\
                               & $M_\mathrm{1D}$ & \msun      &       1.39 &      0.604 &      0.015 &      0.063 &      0.573 &      0.018 &      0.021 &      0.125 \\
                               & $\Delta M_{\mathrm{1D},\mathrm{3D}}$ & \msun      &      +0.04 &     +0.011 &     +0.000 &     +0.000 &     +0.007 &   $-$0.000 &     +0.000 &     +0.003 \\
                               & $\Delta M_{\mathrm{1D},\mathrm{3D}}/M_\mathrm{3D}$ &            &     +3.2\% &     +1.9\% &     +1.9\% &     +0.8\% &     +1.2\% &   $-$0.2\% &     +0.9\% &     +2.1\% \\
 \\
gcd\_2021\_r65\_d2.0           & $M_\mathrm{3D}$ & \msun      &       1.37 &      0.695 &      0.020 &      0.074 &      0.665 &      0.016 &      0.019 &      0.109 \\
                               & $M_\mathrm{1D}$ & \msun      &       1.41 &      0.707 &      0.021 &      0.074 &      0.669 &      0.016 &      0.019 &      0.111 \\
                               & $\Delta M_{\mathrm{1D},\mathrm{3D}}$ & \msun      &      +0.04 &     +0.012 &     +0.000 &     +0.000 &     +0.004 &     +0.000 &     +0.000 &     +0.002 \\
                               & $\Delta M_{\mathrm{1D},\mathrm{3D}}/M_\mathrm{3D}$ &            &     +3.1\% &     +1.7\% &     +0.9\% &     +0.3\% &     +0.7\% &     +0.3\% &     +1.2\% &     +2.2\% \\
 \\
gcd\_2021\_r45\_d6.0           & $M_\mathrm{3D}$ & \msun      &       1.40 &      0.760 &      0.029 &      0.081 &      0.746 &      0.016 &      0.018 &      0.102 \\
                               & $M_\mathrm{1D}$ & \msun      &       1.45 &      0.776 &      0.030 &      0.082 &      0.754 &      0.016 &      0.018 &      0.106 \\
                               & $\Delta M_{\mathrm{1D},\mathrm{3D}}$ & \msun      &      +0.06 &     +0.016 &     +0.000 &     +0.000 &     +0.008 &     +0.000 &     +0.001 &     +0.004 \\
                               & $\Delta M_{\mathrm{1D},\mathrm{3D}}/M_\mathrm{3D}$ &            &     +4.1\% &     +2.1\% &     +1.0\% &     +0.5\% &     +1.1\% &     +2.9\% &     +3.3\% &     +3.9\% \\
 \\
doubledet\_2021\_M1002\_1      & $M_\mathrm{3D}$ & \msun      &       1.03 &      0.541 &      0.017 &      0.058 &      0.523 &      0.021 &      0.021 &      0.106 \\
                               & $M_\mathrm{1D}$ & \msun      &       1.07 &      0.571 &      0.018 &      0.061 &      0.552 &      0.020 &      0.021 &      0.108 \\
                               & $\Delta M_{\mathrm{1D},\mathrm{3D}}$ & \msun      &      +0.04 &     +0.030 &     +0.001 &     +0.003 &     +0.029 &   $-$0.001 &     +0.000 &     +0.002 \\
                               & $\Delta M_{\mathrm{1D},\mathrm{3D}}/M_\mathrm{3D}$ &            &     +4.0\% &     +5.5\% &     +7.0\% &     +4.8\% &     +5.5\% &   $-$3.8\% &     +1.5\% &     +2.3\% \\
 \\
doubledet\_2021\_M1003\_1      & $M_\mathrm{3D}$ & \msun      &       1.06 &      0.591 &      0.021 &      0.064 &      0.567 &      0.021 &      0.019 &      0.097 \\
                               & $M_\mathrm{1D}$ & \msun      &       1.10 &      0.538 &      0.018 &      0.058 &      0.518 &      0.022 &      0.021 &      0.105 \\
                               & $\Delta M_{\mathrm{1D},\mathrm{3D}}$ & \msun      &      +0.04 &   $-$0.053 &   $-$0.002 &   $-$0.006 &   $-$0.049 &     +0.001 &     +0.001 &     +0.009 \\
                               & $\Delta M_{\mathrm{1D},\mathrm{3D}}/M_\mathrm{3D}$ &            &     +3.4\% &   $-$9.0\% &  $-$10.4\% &   $-$9.9\% &   $-$8.6\% &     +6.3\% &     +7.5\% &     +9.1\% \\
 \\
doubledet\_2021\_M1005\_1      & $M_\mathrm{3D}$ & \msun      &       1.06 &      0.547 &      0.018 &      0.060 &      0.527 &      0.022 &      0.020 &      0.103 \\
                               & $M_\mathrm{1D}$ & \msun      &       1.10 &      0.576 &      0.020 &      0.062 &      0.553 &      0.020 &      0.020 &      0.106 \\
                               & $\Delta M_{\mathrm{1D},\mathrm{3D}}$ & \msun      &      +0.04 &     +0.029 &     +0.001 &     +0.002 &     +0.026 &   $-$0.001 &     +0.000 &     +0.003 \\
                               & $\Delta M_{\mathrm{1D},\mathrm{3D}}/M_\mathrm{3D}$ &            &     +3.7\% &     +5.2\% &     +6.0\% &     +4.0\% &     +4.9\% &   $-$5.5\% &     +2.0\% &     +3.1\% \\
 \\
doubledet\_2021\_M1010\_1      & $M_\mathrm{3D}$ & \msun      &       1.11 &      0.762 &      0.026 &      0.084 &      0.722 &      0.017 &      0.015 &      0.070 \\
                               & $M_\mathrm{1D}$ & \msun      &       1.16 &      0.801 &      0.027 &      0.085 &      0.741 &      0.017 &      0.015 &      0.072 \\
                               & $\Delta M_{\mathrm{1D},\mathrm{3D}}$ & \msun      &      +0.05 &     +0.039 &     +0.001 &     +0.001 &     +0.019 &   $-$0.000 &     +0.000 &     +0.002 \\
                               & $\Delta M_{\mathrm{1D},\mathrm{3D}}/M_\mathrm{3D}$ &            &     +4.1\% &     +5.2\% &     +4.8\% &     +1.1\% &     +2.6\% &   $-$2.0\% &     +1.8\% &     +2.8\% \\
 \\
merger\_2012\_11+09            & $M_\mathrm{3D}$ & \msun      &       1.94 &      0.614 &      0.028 &      0.067 &      0.589 &      0.013 &      0.017 &      0.103 \\
                               & $M_\mathrm{1D}$ & \msun      &       2.09 &      0.666 &      0.031 &      0.072 &      0.638 &      0.014 &      0.018 &      0.110 \\
                               & $\Delta M_{\mathrm{1D},\mathrm{3D}}$ & \msun      &      +0.15 &     +0.052 &     +0.003 &     +0.005 &     +0.049 &     +0.001 &     +0.001 &     +0.007 \\
                               & $\Delta M_{\mathrm{1D},\mathrm{3D}}/M_\mathrm{3D}$ &            &     +7.8\% &     +8.5\% &     +9.2\% &     +8.0\% &     +8.4\% &     +5.8\% &     +6.2\% &     +6.6\% \\
\hline
\end{tabular}
\flushleft
\justifying
\textbf{Notes.} Model names are as they appear on HESMA. Models in boldface correspond to the reference for each class of explosion mechanism (see Table~\ref{tab:summary}).  All yields other than $^{56}\mathrm{Ni}_{t=0}$ correspond to 270\,d post explosion. For (stable) Ni we report the isotopic $^{58}$Ni yield (see text for details); for all other elements (Co, Fe, Ca, Ar, S) we give the total elemental yields.
\end{table*}

%% file: tabA2.tex
\begin{table*}
\centering
\footnotesize
\caption{Comparison of the total mass and selected yields of the original 3D model N100 of \cite{Seitenzahl/etal:2013} [$M_\mathrm{3D}$] and 1D models reconstructed from non-spherically averaged radial profiles along six directions of the 3D cartesian grid ($M_\mathrm{1D}$).}\label{tab:3d1d_n100}
\begin{tabular}{llccccccccc}
\hline\hline\\[-1.8ex]
\multicolumn{1}{c}{Model} & \multicolumn{1}{c}{Quantity} & \multicolumn{1}{c}{Unit} & \multicolumn{1}{c}{Total} & \multicolumn{1}{c}{\nifs$_{t=0}$} & \multicolumn{1}{c}{$^{58}$Ni} & \multicolumn{1}{c}{Co} & \multicolumn{1}{c}{Fe} & \multicolumn{1}{c}{Ca} & \multicolumn{1}{c}{Ar} & \multicolumn{1}{c}{S} \\
\hline\\[-1.8ex]
ddt\_2013\_N100                & $M_\mathrm{3D}$ & \msun      &       1.40 &      0.604 &      0.069 &      0.068 &      0.683 &      0.015 &      0.020 &      0.115 \\
                               & $M_\mathrm{1D}$ & \msun      &       1.45 &      0.627 &      0.071 &      0.070 &      0.708 &      0.015 &      0.020 &      0.119 \\
                               & $\Delta M_{\mathrm{1D},\mathrm{3D}}$ & \msun      &      +0.05 &     +0.023 &     +0.002 &     +0.002 &     +0.025 &     +0.000 &     +0.001 &     +0.004 \\
                               & $\Delta M_{\mathrm{1D},\mathrm{3D}}/M_\mathrm{3D}$ &            &     +3.7\% &     +3.8\% &     +3.3\% &     +2.8\% &     +3.6\% &     +3.1\% &     +3.5\% &     +3.7\% \\
 \\
ddt\_2013\_N100\_xpos          & $M_\mathrm{1D}$ (unscaled) & \msun      &       1.15 &      0.547 &      0.031 &      0.059 &      0.542 &      0.018 &      0.021 &      0.112 \\
                               & $\Delta M_{\mathrm{1D},\mathrm{3D}}$ & \msun      &    $-$0.25 &   $-$0.057 &   $-$0.038 &   $-$0.009 &   $-$0.141 &     +0.003 &     +0.001 &   $-$0.003 \\
                               & $\Delta M_{\mathrm{1D},\mathrm{3D}}/M_\mathrm{3D}$ &            &  $-$17.9\% &   $-$9.5\% &  $-$54.9\% &  $-$13.3\% &  $-$20.7\% &    +22.0\% &     +5.7\% &   $-$3.0\% \\
 \\[-1ex]
                               & $M_\mathrm{1D}$ (scaled) & \msun      &       1.40 &      0.665 &      0.038 &      0.072 &      0.659 &      0.022 &      0.025 &      0.136 \\
                               & $\Delta M_{\mathrm{1D},\mathrm{3D}}$ & \msun      &    $-$0.00 &     +0.061 &   $-$0.031 &     +0.004 &   $-$0.024 &     +0.007 &     +0.006 &     +0.021 \\
                               & $\Delta M_{\mathrm{1D},\mathrm{3D}}/M_\mathrm{3D}$ &            &   $-$0.1\% &    +10.1\% &  $-$45.1\% &     +5.5\% &   $-$3.6\% &    +48.6\% &    +28.8\% &    +18.2\% \\
 \\
ddt\_2013\_N100\_xneg          & $M_\mathrm{1D}$ (unscaled) & \msun      &       1.44 &      0.545 &      0.062 &      0.060 &      0.572 &      0.015 &      0.025 &      0.165 \\
                               & $\Delta M_{\mathrm{1D},\mathrm{3D}}$ & \msun      &      +0.04 &   $-$0.059 &   $-$0.007 &   $-$0.008 &   $-$0.112 &   $-$0.000 &     +0.005 &     +0.050 \\
                               & $\Delta M_{\mathrm{1D},\mathrm{3D}}/M_\mathrm{3D}$ &            &     +2.8\% &   $-$9.7\% &  $-$10.6\% &  $-$12.0\% &  $-$16.4\% &   $-$0.2\% &    +27.0\% &    +43.9\% \\
 \\[-1ex]
                               & $M_\mathrm{1D}$ (scaled) & \msun      &       1.40 &      0.530 &      0.060 &      0.058 &      0.555 &      0.014 &      0.024 &      0.161 \\
                               & $\Delta M_{\mathrm{1D},\mathrm{3D}}$ & \msun      &    $-$0.00 &   $-$0.074 &   $-$0.009 &   $-$0.010 &   $-$0.128 &   $-$0.000 &     +0.005 &     +0.046 \\
                               & $\Delta M_{\mathrm{1D},\mathrm{3D}}/M_\mathrm{3D}$ &            &   $-$0.1\% &  $-$12.2\% &  $-$13.1\% &  $-$14.5\% &  $-$18.7\% &   $-$3.1\% &    +23.4\% &    +39.9\% \\
 \\
ddt\_2013\_N100\_ypos          & $M_\mathrm{1D}$ (unscaled) & \msun      &       1.17 &      0.591 &      0.027 &      0.065 &      0.587 &      0.023 &      0.025 &      0.123 \\
                               & $\Delta M_{\mathrm{1D},\mathrm{3D}}$ & \msun      &    $-$0.23 &   $-$0.013 &   $-$0.042 &   $-$0.004 &   $-$0.096 &     +0.008 &     +0.005 &     +0.008 \\
                               & $\Delta M_{\mathrm{1D},\mathrm{3D}}/M_\mathrm{3D}$ &            &  $-$16.2\% &   $-$2.2\% &  $-$61.0\% &   $-$5.2\% &  $-$14.1\% &    +57.2\% &    +26.6\% &     +6.7\% \\
 \\[-1ex]
                               & $M_\mathrm{1D}$ (scaled) & \msun      &       1.40 &      0.704 &      0.032 &      0.077 &      0.699 &      0.028 &      0.030 &      0.147 \\
                               & $\Delta M_{\mathrm{1D},\mathrm{3D}}$ & \msun      &    $-$0.00 &     +0.100 &   $-$0.037 &     +0.009 &     +0.016 &     +0.013 &     +0.010 &     +0.032 \\
                               & $\Delta M_{\mathrm{1D},\mathrm{3D}}/M_\mathrm{3D}$ &            &   $-$0.2\% &    +16.5\% &  $-$53.5\% &    +12.8\% &     +2.3\% &    +87.4\% &    +51.0\% &    +27.4\% \\
 \\
ddt\_2013\_N100\_yneg          & $M_\mathrm{1D}$ (unscaled) & \msun      &       1.52 &      0.542 &      0.117 &      0.066 &      0.654 &      0.019 &      0.027 &      0.168 \\
                               & $\Delta M_{\mathrm{1D},\mathrm{3D}}$ & \msun      &      +0.12 &   $-$0.062 &     +0.048 &   $-$0.002 &   $-$0.030 &     +0.004 &     +0.008 &     +0.053 \\
                               & $\Delta M_{\mathrm{1D},\mathrm{3D}}/M_\mathrm{3D}$ &            &     +8.2\% &  $-$10.3\% &    +69.5\% &   $-$2.5\% &   $-$4.3\% &    +27.1\% &    +38.3\% &    +45.9\% \\
 \\[-1ex]
                               & $M_\mathrm{1D}$ (scaled) & \msun      &       1.40 &      0.500 &      0.108 &      0.061 &      0.604 &      0.017 &      0.025 &      0.155 \\
                               & $\Delta M_{\mathrm{1D},\mathrm{3D}}$ & \msun      &    $-$0.00 &   $-$0.104 &     +0.039 &   $-$0.007 &   $-$0.079 &     +0.003 &     +0.005 &     +0.040 \\
                               & $\Delta M_{\mathrm{1D},\mathrm{3D}}/M_\mathrm{3D}$ &            &   $-$0.1\% &  $-$17.2\% &    +56.5\% &   $-$9.9\% &  $-$11.6\% &    +17.4\% &    +27.7\% &    +34.7\% \\
 \\
ddt\_2013\_N100\_zpos          & $M_\mathrm{1D}$ (unscaled) & \msun      &       1.52 &      0.691 &      0.135 &      0.086 &      0.810 &      0.014 &      0.018 &      0.102 \\
                               & $\Delta M_{\mathrm{1D},\mathrm{3D}}$ & \msun      &      +0.12 &     +0.087 &     +0.066 &     +0.018 &     +0.127 &   $-$0.001 &   $-$0.001 &   $-$0.013 \\
                               & $\Delta M_{\mathrm{1D},\mathrm{3D}}/M_\mathrm{3D}$ &            &     +8.2\% &    +14.4\% &    +96.1\% &    +25.8\% &    +18.5\% &   $-$7.2\% &   $-$7.3\% &  $-$11.5\% \\
 \\[-1ex]
                               & $M_\mathrm{1D}$ (scaled) & \msun      &       1.40 &      0.638 &      0.125 &      0.079 &      0.747 &      0.013 &      0.017 &      0.094 \\
                               & $\Delta M_{\mathrm{1D},\mathrm{3D}}$ & \msun      &    $-$0.00 &     +0.034 &     +0.056 &     +0.011 &     +0.064 &   $-$0.002 &   $-$0.003 &   $-$0.021 \\
                               & $\Delta M_{\mathrm{1D},\mathrm{3D}}/M_\mathrm{3D}$ &            &   $-$0.1\% &     +5.7\% &    +81.1\% &    +16.0\% &     +9.3\% &  $-$14.1\% &  $-$14.2\% &  $-$18.1\% \\
 \\
ddt\_2013\_N100\_zneg          & $M_\mathrm{1D}$ (unscaled) & \msun      &       1.42 &      0.506 &      0.078 &      0.059 &      0.549 &      0.011 &      0.017 &      0.114 \\
                               & $\Delta M_{\mathrm{1D},\mathrm{3D}}$ & \msun      &      +0.02 &   $-$0.098 &     +0.009 &   $-$0.010 &   $-$0.134 &   $-$0.004 &   $-$0.003 &   $-$0.001 \\
                               & $\Delta M_{\mathrm{1D},\mathrm{3D}}/M_\mathrm{3D}$ &            &     +1.7\% &  $-$16.2\% &    +13.4\% &  $-$14.1\% &  $-$19.6\% &  $-$25.3\% &  $-$12.8\% &   $-$1.0\% \\
 \\[-1ex]
                               & $M_\mathrm{1D}$ (scaled) & \msun      &       1.40 &      0.497 &      0.077 &      0.058 &      0.541 &      0.011 &      0.017 &      0.111 \\
                               & $\Delta M_{\mathrm{1D},\mathrm{3D}}$ & \msun      &    $-$0.00 &   $-$0.107 &     +0.008 &   $-$0.011 &   $-$0.143 &   $-$0.004 &   $-$0.003 &   $-$0.004 \\
                               & $\Delta M_{\mathrm{1D},\mathrm{3D}}/M_\mathrm{3D}$ &            &   $-$0.1\% &  $-$17.7\% &    +11.4\% &  $-$15.4\% &  $-$20.9\% &  $-$26.7\% &  $-$14.6\% &   $-$3.1\% \\
\hline
\end{tabular}
\flushleft
\justifying
\textbf{Notes.} The first entry for model ddt\_2013\_N100 is identical to Table~\ref{tab:3d1d}. Subsequent entries correspond to 1D models reconstructed from the three orthogonal axes of the original 3D cartesian grid, in both positive (\{x,y,z\}pos) and negative (\{x,y,z\}neg) directions. We first report quantities based on the original (unscaled) density profile, then those based on the density profile rescaled to match the total mass of the original 3D model (1.40\,\msun).
\end{table*}

%% file: tabF1.tex
\begin{table}[h]
\centering
\footnotesize
\caption{Level indexing for Ni\three\ used in Table~\ref{tab:nkiiicol}.}\label{tab:nkiiilev}
\begin{tabular}{cl}
\hline\hline\\[-1.8ex]
\multicolumn{1}{c}{Index} & \multicolumn{1}{c}{Level} \\
\hline\\[-1.8ex]
  1 & 3d$^8$ $^3$F$_{4}$                                \\
  2 & 3d$^8$ $^3$F$_{3}$                                \\
  3 & 3d$^8$ $^3$F$_{2}$                                \\
  4 & 3d$^8$ $^1$D$_{2}$                                \\
  5 & 3d$^8$ $^3$P$_{2}$                                \\
  6 & 3d$^8$ $^3$P$_{1}$                                \\
  7 & 3d$^8$ $^3$P$_{0}$                                \\
  8 & 3d$^8$ $^1$G$_{4}$                                \\
  9 & 3d$^8$ $^1$S$_{0}$                                \\
 10 & 3d$^7$($^4$F)4s $^5$F$_{5}$                       \\
\hline
\end{tabular}
\end{table}

%% file: tabF2.tex
\begin{table*}
\centering
\footnotesize
\caption{Transition probabilities ($A_\mathrm{ul}$) and effective collision strengths ($\Upsilon_\mathrm{lu}$) among the lowest ten levels of Ni\three. See Table~\ref{tab:nkiiilev} for level designations.}\label{tab:nkiiicol}
\begin{tabular}{rrcccccccccccccc}
\hline\hline\\[-1.8ex]
\multicolumn{1}{c}{$l$} & \multicolumn{1}{c}{$u$} & \multicolumn{1}{c}{$A_\mathrm{ul}$} & \multicolumn{13}{c}{Temperature (1000\,K)} \\
\cmidrule(l{.5em}){4-16}
 & & (s$^{-1}$)  & \multicolumn{1}{c}{1.0} & \multicolumn{1}{c}{1.5} & \multicolumn{1}{c}{1.8} & \multicolumn{1}{c}{2.0} & \multicolumn{1}{c}{2.5} & \multicolumn{1}{c}{5.0} & \multicolumn{1}{c}{7.5} & \multicolumn{1}{c}{10.0} & \multicolumn{1}{c}{15.0} & \multicolumn{1}{c}{18.0} & \multicolumn{1}{c}{20.0} & \multicolumn{1}{c}{30.0} & \multicolumn{1}{c}{40.0} \\
\hline\\[-1.8ex]
  1 &   2 & 6.54($-$2)\phantom{0} & 3.380 & 3.300 & 3.210 & 3.150 & 2.980 & 2.490 & 2.330 & 2.260 & 2.180 & 2.140 & 2.120 & 2.010 & 1.900 \\
  1 &   3 & 4.45($-$9)\phantom{0} & 1.320 & 1.160 & 1.090 & 1.050 & 0.962 & 0.771 & 0.708 & 0.680 & 0.653 & 0.640 & 0.630 & 0.587 & 0.560 \\
  1 &   4 & 5.56($-$3)\phantom{0} & 1.580 & 1.420 & 1.360 & 1.320 & 1.240 & 1.050 & 0.954 & 0.900 & 0.855 & 0.840 & 0.830 & 0.805 & 0.780 \\
  1 &   5 & 5.85($-$2)\phantom{0} & 1.520 & 1.500 & 1.490 & 1.480 & 1.460 & 1.320 & 1.220 & 1.150 & 1.080 & 1.050 & 1.040 & 0.977 & 0.921 \\
  1 &   6 & 5.10($-$15) & 0.330 & 0.355 & 0.365 & 0.370 & 0.370 & 0.370 & 0.363 & 0.360 & 0.360 & 0.360 & 0.360 & 0.360 & 0.350 \\
  1 &   7 & 1.07($-$17) & 0.060 & 0.070 & 0.070 & 0.070 & 0.080 & 0.090 & 0.090 & 0.100 & 0.100 & 0.100 & 0.100 & 0.100 & 0.100 \\
  1 &   8 & 3.59($-$1)\phantom{0} & 0.490 & 0.500 & 0.505 & 0.510 & 0.530 & 0.620 & 0.696 & 0.770 & 0.882 & 0.926 & 0.950 & 1.020 & 1.020 \\
  1 &   9 & $<10(-10)$ & 0.062 & 0.061 & 0.060 & 0.060 & 0.059 & 0.056 & 0.054 & 0.053 & 0.052 & 0.052 & 0.052 & 0.055 & 0.062 \\
  1 &  10 & 1.22($-$1)\phantom{0} & 1.950 & 1.880 & 1.850 & 1.830 & 1.810 & 1.780 & 1.760 & 1.710 & 1.600 & 1.540 & 1.500 & 1.340 & 1.220 \\
  2 &   3 & 2.68($-$2)\phantom{0} & 1.900 & 1.900 & 1.880 & 1.870 & 1.820 & 1.670 & 1.630 & 1.600 & 1.580 & 1.550 & 1.540 & 1.460 & 1.390 \\
  2 &   4 & 4.95($-$1)\phantom{0} & 1.090 & 1.050 & 1.020 & 1.000 & 0.961 & 0.811 & 0.731 & 0.690 & 0.653 & 0.645 & 0.640 & 0.622 & 0.600 \\
  2 &   5 & 1.23($-$1)\phantom{0} & 0.750 & 0.777 & 0.790 & 0.800 & 0.810 & 0.790 & 0.755 & 0.730 & 0.713 & 0.705 & 0.700 & 0.675 & 0.640 \\
  2 &   6 & 4.46($-$2)\phantom{0} & 0.530 & 0.590 & 0.616 & 0.630 & 0.650 & 0.670 & 0.645 & 0.630 & 0.595 & 0.585 & 0.580 & 0.545 & 0.520 \\
  2 &   7 & 6.69($-$15) & 0.110 & 0.120 & 0.125 & 0.130 & 0.140 & 0.140 & 0.133 & 0.130 & 0.130 & 0.130 & 0.130 & 0.120 & 0.120 \\
  2 &   8 & 1.83($-$1)\phantom{0} & 0.430 & 0.437 & 0.440 & 0.440 & 0.450 & 0.510 & 0.579 & 0.640 & 0.737 & 0.776 & 0.800 & 0.855 & 0.860 \\
  2 &   9 & $<10(-10)$ & 0.137 & 0.131 & 0.128 & 0.126 & 0.123 & 0.183 & 0.329 & 0.458 & 0.599 & 0.631 & 0.639 & 0.614 & 0.557 \\
  2 &  10 & 9.49($-$3)\phantom{0} & 0.012 & 0.011 & 0.010 & 0.010 & 0.009 & 0.007 & 0.008 & 0.009 & 0.009 & 0.009 & 0.009 & 0.009 & 0.008 \\
  3 &   4 & 2.16($-$1)\phantom{0} & 0.730 & 0.747 & 0.745 & 0.740 & 0.720 & 0.610 & 0.541 & 0.510 & 0.483 & 0.480 & 0.480 & 0.472 & 0.460 \\
  3 &   5 & 2.50($-$2)\phantom{0} & 0.460 & 0.460 & 0.460 & 0.460 & 0.460 & 0.450 & 0.443 & 0.440 & 0.440 & 0.445 & 0.450 & 0.442 & 0.430 \\
  3 &   6 & 1.88($-$2)\phantom{0} & 0.370 & 0.412 & 0.430 & 0.440 & 0.460 & 0.480 & 0.463 & 0.450 & 0.433 & 0.425 & 0.420 & 0.395 & 0.380 \\
  3 &   7 & 5.46($-$2)\phantom{0} & 0.180 & 0.205 & 0.220 & 0.230 & 0.240 & 0.250 & 0.250 & 0.240 & 0.230 & 0.225 & 0.220 & 0.210 & 0.200 \\
  3 &   8 & 4.62($-$4)\phantom{0} & 0.340 & 0.340 & 0.340 & 0.340 & 0.340 & 0.380 & 0.432 & 0.470 & 0.540 & 0.566 & 0.580 & 0.625 & 0.620 \\
  3 &   9 & 1.59($-$1)\phantom{0} & 0.025 & 0.025 & 0.024 & 0.024 & 0.023 & 0.020 & 0.018 & 0.017 & 0.015 & 0.015 & 0.014 & 0.013 & 0.012 \\
  3 &  10 & $<10(-10)$ & 0.007 & 0.006 & 0.006 & 0.006 & 0.006 & 0.005 & 0.005 & 0.005 & 0.004 & 0.004 & 0.004 & 0.004 & 0.004 \\
  4 &   5 & 1.02($-$1)\phantom{0} & 0.610 & 0.593 & 0.585 & 0.580 & 0.570 & 0.540 & 0.530 & 0.550 & 0.585 & 0.606 & 0.620 & 0.648 & 0.640 \\
  4 &   6 & 9.43($-$2)\phantom{0} & 0.360 & 0.343 & 0.335 & 0.330 & 0.320 & 0.300 & 0.297 & 0.300 & 0.317 & 0.325 & 0.330 & 0.340 & 0.330 \\
  4 &   7 & 2.87($-$6)\phantom{0} & 0.150 & 0.123 & 0.115 & 0.110 & 0.110 & 0.090 & 0.090 & 0.090 & 0.100 & 0.100 & 0.100 & 0.110 & 0.110 \\
  4 &   8 & 7.22($-$4)\phantom{0} & 0.390 & 0.390 & 0.395 & 0.400 & 0.410 & 0.490 & 0.559 & 0.620 & 0.697 & 0.731 & 0.750 & 0.788 & 0.790 \\
  4 &   9 & 1.27($+$1)\phantom{0} & 0.105 & 0.100 & 0.098 & 0.097 & 0.094 & 0.083 & 0.077 & 0.073 & 0.068 & 0.067 & 0.066 & 0.062 & 0.059 \\
  4 &  10 & $<10(-10)$ & 0.219 & 0.212 & 0.206 & 0.203 & 0.195 & 0.170 & 0.157 & 0.150 & 0.145 & 0.144 & 0.144 & 0.146 & 0.145 \\
  5 &   6 & 5.93($-$4)\phantom{0} & 0.480 & 0.522 & 0.540 & 0.550 & 0.570 & 0.570 & 0.570 & 0.570 & 0.580 & 0.585 & 0.590 & 0.590 & 0.580 \\
  5 &   7 & 3.07($-$9)\phantom{0} & 0.120 & 0.127 & 0.130 & 0.130 & 0.140 & 0.150 & 0.150 & 0.160 & 0.160 & 0.160 & 0.160 & 0.170 & 0.160 \\
  5 &   8 & 3.18($-$5)\phantom{0} & 0.330 & 0.330 & 0.330 & 0.330 & 0.330 & 0.400 & 0.462 & 0.500 & 0.535 & 0.550 & 0.560 & 0.578 & 0.570 \\
  5 &   9 & 9.85($-$1)\phantom{0} & 0.097 & 0.156 & 0.175 & 0.183 & 0.192 & 0.168 & 0.136 & 0.113 & 0.085 & 0.074 & 0.069 & 0.050 & 0.041 \\
  5 &  10 & $<10(-10)$ & 0.060 & 0.065 & 0.065 & 0.064 & 0.062 & 0.052 & 0.046 & 0.043 & 0.039 & 0.038 & 0.037 & 0.035 & 0.034 \\
  6 &   7 & 8.71($-$4)\phantom{0} & 0.300 & 0.300 & 0.295 & 0.290 & 0.290 & 0.260 & 0.243 & 0.230 & 0.230 & 0.225 & 0.220 & 0.212 & 0.210 \\
  6 &   8 & 1.59($-$21) & 0.190 & 0.190 & 0.190 & 0.190 & 0.190 & 0.230 & 0.257 & 0.270 & 0.287 & 0.295 & 0.300 & 0.300 & 0.300 \\
  6 &   9 & 5.96($+$0)\phantom{0} & 0.135 & 0.137 & 0.140 & 0.141 & 0.148 & 0.213 & 0.292 & 0.358 & 0.442 & 0.473 & 0.488 & 0.527 & 0.534 \\
  6 &  10 & $<10(-10)$ & 0.036 & 0.037 & 0.038 & 0.038 & 0.041 & 0.067 & 0.100 & 0.128 & 0.166 & 0.180 & 0.187 & 0.204 & 0.206 \\
  7 &   8 & 2.50($-$25) & 0.070 & 0.070 & 0.070 & 0.070 & 0.070 & 0.080 & 0.087 & 0.090 & 0.100 & 0.100 & 0.100 & 0.100 & 0.100 \\
  7 &   9 & $<10(-10)$ & 0.065 & 0.065 & 0.067 & 0.069 & 0.082 & 0.244 & 0.386 & 0.459 & 0.491 & 0.481 & 0.471 & 0.405 & 0.348 \\
  7 &  10 & $<10(-10)$ & 0.053 & 0.051 & 0.049 & 0.048 & 0.047 & 0.051 & 0.059 & 0.062 & 0.062 & 0.060 & 0.058 & 0.051 & 0.045 \\
  8 &   9 & $<10(-10)$ & 0.249 & 0.242 & 0.238 & 0.236 & 0.232 & 0.221 & 0.217 & 0.214 & 0.211 & 0.209 & 0.208 & 0.204 & 0.201 \\
  8 &  10 & 2.97($-$6)\phantom{0} & 0.361 & 0.320 & 0.300 & 0.290 & 0.268 & 0.216 & 0.194 & 0.179 & 0.159 & 0.151 & 0.147 & 0.132 & 0.121 \\
  9 &  10 & $<10(-10)$ & 0.173 & 0.167 & 0.163 & 0.160 & 0.155 & 0.142 & 0.142 & 0.146 & 0.150 & 0.151 & 0.150 & 0.144 & 0.137 \\
\hline
\end{tabular}
\flushleft
\justifying
{\bf Notes.} Numbers in parentheses correspond to powers of ten. Entries for transitions among the lowest eight levels for temperatures $\le 40000$\,K were computed following the methods outlined in \cite{Storey/etal:2016}.
 All other entries are based on \cite{Ramsbottom/etal:2007}; in this latter approach, transition probabilities $A_\mathrm{ul}<10^{-10}$\,s$^{-1}$ are deemed too small to be significant and we report these as upper limits.
 A complete set of transition probabilities and effective collision strengths among the lowest 43 levels of Ni\three\ and extending to $10^5$\,K is available in electronic format at the CDS.
\end{table*}